\pdfoutput=1
\documentclass[aps,prd,twocolumn,floatfix,amsmath,amssymb,superscriptaddress,notitlepage,10pt,preprintnumbers]{revtex4}

\usepackage{graphicx}
\usepackage{dcolumn}
\usepackage{bm}
\usepackage{url}
\usepackage{color}
\usepackage{comment}
\usepackage{mathrsfs}
\usepackage{hyperref}
\usepackage[dvipsnames]{xcolor}

\newcommand{\hiMpc}{h^{-1}{\rm Mpc}}

\newcommand{\cov}{{\rm Cov}}
\newcommand{\dSigma}{\Delta\!\Sigma}
\newcommand{\wgg}{w_{\rm p}}
\newcommand{\avrg}[1]{\left\langle#1\right\rangle}
\newcommand{\simgt}{\lower.5ex\hbox{$\; \buildrel > \over \sim \;$}}
\newcommand{\sqdeg}{deg$^2$}

\newcommand{\Seight}{0.795^{+0.049}_{-0.042}}
\newcommand{\Seightopt}{0.745^{+0.039}_{-0.031}}
\newcommand{\sigmaeight}{0.718^{+0.044}_{-0.031}}
\newcommand{\Omegam}{0.383^{+0.028}_{-0.053}}

\begin{document}

\preprint{IPMU21-0069, YITP-21-125}

\title{Cosmological inference from the emulator based halo model II: Joint analysis of galaxy-galaxy weak lensing and galaxy clustering from HSC-Y1 and SDSS}

\author{Hironao~Miyatake}
\email{miyatake@kmi.nagoya-u.ac.jp}
\affiliation{Kobayashi-Maskawa Institute for the Origin of Particles and the Universe (KMI),
Nagoya University, Nagoya, 464-8602, Japan}
\affiliation{Institute for Advanced Research, Nagoya University, Nagoya 464-8601, Japan}
\affiliation{Division of Particle and Astrophysical Science, Graduate School of Science, Nagoya University, Nagoya 464-8602, Japan}
\affiliation{Kavli Institute for the Physics and Mathematics of the Universe (WPI), The University of Tokyo Institutes for Advanced Study (UTIAS), The University of Tokyo, Chiba 277-8583, Japan}
\affiliation{Jet Propulsion Laboratory, California Institute of Technology,
Pasadena, CA 91109, USA}

\author{Sunao~Sugiyama}
\affiliation{Kavli Institute for the Physics and Mathematics of the Universe (WPI), The University of Tokyo Institutes for Advanced Study (UTIAS), The University of Tokyo, Chiba 277-8583, Japan}
\affiliation{Department of Physics, The University of Tokyo, Bunkyo, Tokyo 113-0031, Japan}

\author{Masahiro~Takada}
\email{masahiro.takada@ipmu.jp}
\affiliation{Kavli Institute for the Physics and Mathematics of the Universe (WPI), The University of Tokyo Institutes for Advanced Study (UTIAS), The University of Tokyo, Chiba 277-8583, Japan}

\author{Takahiro~Nishimichi}
\affiliation{Center for Gravitational Physics, Yukawa Institute for Theoretical Physics, Kyoto University, Kyoto 606-8502, Japan}
\affiliation{Kavli Institute for the Physics and Mathematics of the Universe
(WPI), The University of Tokyo Institutes for Advanced Study (UTIAS),
The University of Tokyo, Chiba 277-8583, Japan}

\author{Masato~Shirasaki}
\affiliation{National Astronomical Observatory of Japan, Mitaka, Tokyo 181-8588, Japan}
\affiliation{The Institute of Statistical Mathematics,
Tachikawa, Tokyo 190-8562, Japan}

\author{Yosuke~Kobayashi}
\affiliation{Department of Astronomy/Steward Observatory, University of Arizona, 933 North Cherry Avenue, Tucson, AZ 85721-0065, USA}
\affiliation{Kavli Institute for the Physics and Mathematics of the Universe
(WPI), The University of Tokyo Institutes for Advanced Study (UTIAS),
The University of Tokyo, Chiba 277-8583, Japan}

\author{Rachel~Mandelbaum}
\affiliation{Department of Physics, McWilliams Center for Cosmology, Carnegie Mellon University, Pittsburgh, PA 15213, USA}

\author{Surhud~More}
\affiliation{The Inter-University Centre for Astronomy and Astrophysics, Post bag 4, Ganeshkhind, Pune 411007, India}
\affiliation{Kavli Institute for the Physics and Mathematics of the Universe
(WPI), The University of Tokyo Institutes for Advanced Study (UTIAS),
The University of Tokyo, Chiba 277-8583, Japan}

\author{Masamune~Oguri}
\affiliation{Research Center for the Early Universe, The University of Tokyo, Bunkyo, Tokyo 113-0031, Japan}
\affiliation{Department of Physics, The University of Tokyo, Bunkyo, Tokyo 113-0031, Japan}
\affiliation{Kavli Institute for the Physics and Mathematics of the Universe (WPI), The University of Tokyo Institutes for Advanced Study (UTIAS), The University of Tokyo, Chiba 277-8583, Japan}

\author{Ken~Osato}
\affiliation{Center for Gravitational Physics, Yukawa Institute for Theoretical Physics, Kyoto University, Kyoto 606-8502, Japan}
\affiliation{LPENS, D\'epartement de Physique, \'Ecole Normale Sup\'erieure,
Universit\'e PSL, CNRS, Sorbonne Universit\'e, Universit\'e de Paris, 75005 Paris, France}

\author{Youngsoo~Park}
\affiliation{Kavli Institute for the Physics and Mathematics of the Universe
(WPI), The University of Tokyo Institutes for Advanced Study (UTIAS),
The University of Tokyo, Chiba 277-8583, Japan}

\author{Ryuichi~Takahashi}
\affiliation{Faculty of Science and Technology, Hirosaki University, 3 Bunkyo-cho, Hirosaki, Aomori 036-8561, Japan}

\author{Jean~Coupon}
\affiliation{Astronomy Department, University of
Geneva, Chemin d’Ecogia 16, CH-1290 Versoix, Switzerland}

\author{Chiaki~Hikage}
\affiliation{Kavli Institute for the Physics and Mathematics of the Universe
(WPI), The University of Tokyo Institutes for Advanced Study (UTIAS),
The University of Tokyo, Chiba 277-8583, Japan}

\author{Bau-Ching~Hsieh}
\affiliation{Institute of Astronomy and Astrophysics, Academia Sinica, Taipei 10617, Taiwan}

\author{Yutaka~Komiyama}
\affiliation{National Astronomical Observatory of Japan, Mitaka, Tokyo 181-8588, Japan}

\author{Alexie~Leauthaud}
\affiliation{Department of Astronomy and Astrophysics, University of California, 1156 High Street, Santa Cruz, CA 95064, USA}

\author{Xiangchong~Li}
\affiliation{Department of Physics, McWilliams Center for Cosmology, Carnegie Mellon University, Pittsburgh, PA 15213, USA}
\affiliation{Kavli Institute for the Physics and Mathematics of the Universe
(WPI), The University of Tokyo Institutes for Advanced Study (UTIAS),
The University of Tokyo, Chiba 277-8583, Japan}

\author{Wentao~Luo}
\affiliation{Department of Astronomy, School of Physical Sciences, University of Science and Technology of China, Hefei, Anhui 230026, China}
\affiliation{Key Laboratory for Research in Galaxies and Cosmology, School of Astronomy and Space Science, University of Science and Technology of China, Hefei, Anhui 230026, China}

\author{Robert~H.~Lupton}
\affiliation{Department of Astrophysical Sciences, Peyton Hall, Princeton University, Princeton, NJ 08540, USA}

\author{Satoshi~Miyazaki}
\affiliation{National Astronomical Observatory of Japan, Mitaka, Tokyo 181-8588, Japan}

\author{Hitoshi~Murayama}
\affiliation{Department of Physics, University of California, Berkeley, CA 94720, USA}
\affiliation{Kavli Institute for the Physics and Mathematics of the Universe (WPI), The University of Tokyo Institutes for Advanced Study (UTIAS), The University of Tokyo, Chiba 277-8583, Japan}
\affiliation{Ernest Orlando Lawrence Berkeley National Laboratory, Berkeley, CA 94720, USA}

\author{Atsushi~J.~Nishizawa}
\affiliation{Institute for Advanced Research, Nagoya University, Nagoya 464-8601, Japan}

\author{Paul~A.~Price}
\affiliation{Department of Astrophysical Sciences, Peyton Hall, Princeton University, Princeton, NJ 08540, USA}

\author{Melanie~Simet}
\affiliation{University of California Riverside, 900 University Ave, Riverside, CA 92521, USA}
\affiliation{Jet Propulsion Laboratory, California Institute of Technology,
Pasadena, CA 91109, USA}

\author{Joshua~S.~Speagle}
\altaffiliation{Banting \& Dunlap Fellow}
\affiliation{Department of Statistical Sciences, University of Toronto, Toronto, ON M5S 3G3, Canada}
\affiliation{David A. Dunlap Department of Astronomy \& Astrophysics, University of Toronto, Toronto, ON M5S 3H4, Canada}
\affiliation{Dunlap Institute for Astronomy \& Astrophysics, University of Toronto, Toronto, ON M5S 3H4, Canada}

\author{Michael~A.~Strauss}
\affiliation{Department of Astrophysical Sciences, Peyton Hall, Princeton University, Princeton, NJ 08544, USA}

\author{Masayuki~Tanaka}
\affiliation{National Astronomical Observatory of Japan, Mitaka, Tokyo 181-8588, Japan}

\author{Naoki~Yoshida}
\affiliation{Department of Physics, The University of Tokyo, Bunkyo, Tokyo 113-0031, Japan}
\affiliation{Kavli Institute for the Physics and Mathematics of the Universe (WPI), The University of Tokyo Institutes for Advanced Study (UTIAS), The University of Tokyo, Chiba 277-8583, Japan}

\date{\today}

\begin{abstract}
\noindent We present high-fidelity cosmology results from a blinded joint analysis of galaxy-galaxy weak lensing ($\dSigma$) and projected galaxy clustering ($\wgg$) measured from the Hyper Suprime-Cam Year-1 (HSC-Y1) data and spectroscopic Sloan Digital Sky Survey (SDSS) galaxy catalogs in the redshift range $0.15<z<0.7$.  We define luminosity-limited samples of SDSS galaxies to serve as the tracers of $\wgg$ in three spectroscopic redshift bins, and as the lens samples for $\dSigma$.  For the $\dSigma$ measurements, we select a single sample of 4 million source galaxies over 140\,\sqdeg\ from HSC-Y1 with photometric redshifts (photo-$z$) greater than 0.75, enabling a better handle of photo-$z$ errors by comparing the $\dSigma$ amplitudes for the three lens redshift bins. The deep, high-quality HSC-Y1 data enable significant detections of the $\dSigma$ signals, with integrated signal-to-noise ratio $S/N\sim 15$ in the range $3\le R/[h^{-1}{\rm Mpc}]\le 30$ for the three lens samples, despite the small area coverage. For cosmological parameter inference, we use an input galaxy-halo connection model built on the {\tt Dark Emulator} package (which uses an ensemble set of high-resolution $N$-body simulations and enables fast, accurate computation of the clustering observables) with a halo occupation distribution that includes nuisance parameters to marginalize over modeling uncertainties. We model the $\dSigma$ and $\wgg$ measurements on scales from $R\simeq 3$ and $2\,h^{-1}{\rm Mpc}$, respectively, up to $30\,h^{-1}{\rm Mpc}$ (therefore excluding the BAO information) assuming a flat $\Lambda$CDM cosmology, marginalizing over about 20 nuisance parameters and demonstrating the robustness of our results to them. With various tests using mock catalogs described in \citet{2021arXiv210100113M}, we show that any bias in the clustering amplitude $S_8\equiv \sigma_8(\Omega_{\rm m}/0.3)^{0.5}$ due to uncertainties in the galaxy-halo connection is less than  $\sim50$\% of the statistical uncertainty of $S_8$, {\it unless} the assembly bias effect is unexpectedly large. Our best-fit models have $S_8=\Seight$ (mode and 68\% credible interval) for the flat $\Lambda$CDM model; we find tighter constraints on the quantity $S_8(\alpha=0.17)\equiv\sigma_8(\Omega_{\rm m}/0.3)^{0.17} =\Seightopt$. 
\end{abstract}

\maketitle

\section{Introduction}
\label{sec:introduction}
Wide-area imaging galaxy surveys offer exciting opportunities to address fundamental questions in cosmology such as the nature of dark matter and the origin of cosmic acceleration \citep{Weinbergetal:13}. Current-generation imaging surveys such as the Subaru Hyper Suprime-Cam \footnote{\url{https://hsc.mtk.nao.ac.jp/ssp/}} \citep[HSC][]{HSCoverview:17,2019PASJ...71...43H,2020PASJ...72...16H}, the Dark Energy Survey \footnote{\url{https://www.darkenergysurvey.org}} \citep[DES][]{2018PhRvD..98d3526A,DES-Y3,Amon:2021, Secco:2021, Porredon:2021, Pandey:2021}, and the Kilo-Degree Survey \footnote{\url{http://kids.strw.leidenuniv.nl}} \citep[KiDS][]{Heymansetal:2021,Asgari:2021} have used accurate measurements of weak gravitational lensing effects to obtain tight constraints on cosmological parameters. Intriguingly, the cosmological models inferred from these large-scale structure probes consistently (albeit at low significance) exhibit a lower value of $\sigma_8$ or $S_8$, which characterizes the clustering amplitude in the late universe \citep[e.g.][]{2019PASJ...71...43H,Park:2020}, than do cosmological models inferred from the {\it Planck} cosmic microwave background (CMB) measurement \citep{2020A&A...641A...6P}, hinting at the possibility of new physics beyond the standard cosmological model, i.e. the flat $\Lambda$CDM model with adiabatic, Gaussian initial conditions \citep[e.g.][]{Park:2020}.

The main challenge of large-scale structure probes lies in the uncertainty in galaxy bias; that is, the unknown relation between the distributions of matter and galaxies  \citep{2014Natur.509..177V,2018MNRAS.475..676S}. Since the physical processes inherent in the formation and evolution of galaxies are still difficult to accurately model from first principles, we need both observational and theoretical approaches to tackle the galaxy bias uncertainty in order for us to obtain ``unbiased'' and ``precise'' inference of the underlying cosmological parameters from large-scale structure observables. On  scales large enough to be described by linear perturbation theory, galaxy bias is expected to have a simple form \cite{Kaiser:1984}, however, there is considerable statistical power to be gained by exploiting the information in the mildly non-linear regime \citep[e.g.][]{2005PhRvD..71j3515S,Mandelbaumetal:13,2021arXiv210100113M}. 

Combining multiple observational probes offers a promising way to mitigate the impact of galaxy bias uncertainty on cosmology inference \citep{2005PhRvD..71j3515S,OguriTakada:11,2013MNRAS.430..767C,Mandelbaumetal:13,Miyatakeetal:15,2015ApJ...806....2M,2018PhRvD..98d3526A,Heymansetal:2021,Sugiyama:2020}. In particular, galaxy-galaxy weak lensing, obtained by cross-correlating the positions of foreground (lens) galaxies with shapes of background galaxies, can be used to infer the average mass distribution around lens galaxies. Combining the galaxy-galaxy weak lensing with the auto-correlation function of galaxies in the same sample as the lens galaxies can be used to observationally disentangle the galaxy bias and the correlation function of the underlying matter distribution.  

The halo model prescription of large-scale structure \citep{Seljak:00,PeacockSmith:00,MaFry:00,Scoccimarroetal:01} is a useful theoretical method to make model predictions of galaxy clustering quantities. Halos are locations where galaxies likely form, and the clustering properties of halos are relatively well understood, both from analytical approaches and $N$-body simulations \citep[][]{CooraySheth:02}. An empirical model such as the halo occupation distribution (HOD) method \cite{1998ApJ...494....1J,Zhengetal:05} can be used to connect galaxies to halos. An advantage of this method is that it allows one to use small-scale information in cosmology inference, thereby yielding tighter constraints on cosmological parameters \citep[see e.g.,][]{Cacciato:2009,More:2013b,vandenBosch:2013,2015ApJ...806....2M,Kobayashi:2021}. However, if the model is not sufficiently accurate nor flexible enough to capture the complicated galaxy-scale physics or marginalize over uncertainties in the galaxy-halo connection, the method might lead to a significant bias in cosmological parameters, more than the statistical credible interval \citep[see e.g.,][]{HutererTakada:05,Zentner:2014,McEwen:2018,2019MNRAS.488.1652H}. A worst-case scenario is that one might claim a wrong cosmology, e.g. a time-varying dark energy model, from a given dataset due to inaccurate theoretical templates.

In this paper we estimate cosmological parameters by comparing halo model predictions to the clustering observables, galaxy-galaxy weak lensing and the projected auto-correlation function of galaxies, measured from the Subaru HSC Year 1 datasets \citep[hereafter HSC-Y1;][]{HSCoverview:17} and the spectroscopic LOWZ and CMASS galaxy samples of the Sloan Digital Sky Survey (SDSS; \citet{SDSS}). In this paper we use luminosity-limited, rather than flux-limited, samples of LOWZ and CMASS \citep{Alam:2015} galaxies because those samples are nearly volume-limited in each of the redshift bins we use, and display weaker redshift evolution of the clustering properties in each bin than do the flux-limited samples, allowing us to use redshift-independent model predictions in the parameter inference \citep{Miyatakeetal:15, 2015ApJ...806....2M}. In addition, the luminosity-limited samples allow us to rather straightforwardly model the magnification bias effect on the galaxy-galaxy weak lensing measurement \cite{2020A&A...638A..96U}. For the source sample of HSC-Y1 galaxies used in the galaxy-galaxy weak lensing measurement, we employ a single sample of source galaxies for the three lens samples of LOWZ and CMASS galaxies, following the method described in \citet{OguriTakada:11}. This method allows for a calibration of photo-$z$ errors by comparing the relative strengths of the galaxy-galaxy weak lensing at different lens redshifts for the same source sample. This calibration allows us to mitigate photo-$z$ errors, one of the most important systematic effects in weak lensing cosmology.

On the modeling side, we use the publicly available code {\tt Dark Emulator} developed in \citet{2018arXiv181109504N}, which enables fast, accurate computations of halo clustering quantities (the halo mass function, the halo-matter cross-correlation function and the halo auto-correlation function) as a function of redshift, separation and halo mass(es) for an input model within the flat $w$CDM cosmology framework. This emulator is flexible in the sense that we can combine it with a user-specified prescription of the galaxy-halo connection, the HOD prescription used in this paper, to make model predictions of galaxy-galaxy weak lensing and galaxy auto-correlation function for a target sample of galaxies for an input cosmological model \citep{2021arXiv210100113M}. With this emulator, we can perform cosmology inference in a multi-dimensional parameter space, which is equivalent to comparing the measured signals with model predictions from mock catalogs of galaxies in $N$-body simulations. In doing this, we include a sufficient number of nuisance parameters to account for uncertainties in the galaxy-halo connection and other observational effects, and then estimate cosmological parameters after marginalizing over the nuisance parameters. 

We perform the cosmology analysis in a blind manner to avoid confirmation biases affecting our results. After unblinding, we compare our results with those from other cosmological experiments such as other weak lensing surveys and {\it Planck}. We aim to address the question of whether the cosmological parameters, especially $S_8$ or $\sigma_8$ inferred from our joint probes analysis, are consistent with those of the {\it Planck} cosmology. Note that our companion paper \cite{Sugiyama:2021} performs a cosmology analysis limited to large scales using a theoretical template motivated by the perturbation theory, which is complementary to our analysis.

This paper is organized as follows. In Section~\ref{sec:data}, we briefly describe the HSC first-year shear catalog and the spectroscopic SDSS galaxy catalog that are used in the galaxy-galaxy weak lensing and galaxy clustering measurements. In Section~\ref{sec:measurements}, we describe the measurements of our clustering observables. In Section~\ref{sec:analysis}, we describe our analysis method: the theoretical templates based on the halo model and the likelihood analysis. In Section~\ref{sec:blinding} we describe our strategy for the blind cosmology analysis. In Section~\ref{sec:results} we show the main results of this paper: our cosmological constraints and their robustness to different systematics. Finally we give our conclusions in Section~\ref{sec:summary}. We include seven appendices giving technical details of our model and tests of systematic effects.

Throughout this paper, unless stated otherwise, we quote the central value of a parameter from the mode value of parameter that has the highest probability for the marginalized 1-d posterior probability density function in the chain: ${\cal P}(p_{\rm mode})={\rm maximum}$. We quote the 68\% credible interval(s) for parameter(s) from the highest density interval 
of parameter(s) \footnote{E.g., see \url{https://www.sciencedirect.com/topics/mathematics/highest-density-interval}} satisfying 
\begin{align}
\int_{{\bf p}\in{\cal P}>{\cal P}_{68}}\!\mathrm{d}{\bf p}~
{\cal P}({\bf p})=0.68, 
\label{eq:hdi}
\end{align}
where ${\cal P}({\bf p})$ is the 1-d or 2-d marginalized posterior distribution. 
The 95\% credible interval is similarly defined.

\section{Data}
\label{sec:data}
\subsection{HSC-Y1 data: source galaxies for galaxy-galaxy weak lensing}
\label{sec:HSC-Y1}
HSC is a wide-field prime focus camera on the 8.2m Subaru Telescope \citep{2018PASJ...70S...1M,2018PASJ...70S...2K,2018PASJ...70S...3F,2018PASJ...70...66K}. The HSC Subaru Strategic Program (HSC SSP) survey started in 2014, and is using 330 Subaru nights to conduct a five-band ($grizy$) wide-area imaging survey \citep{HSCoverview:17}. The combination of HSC's wide field-of-view (1.77~\sqdeg), superb image quality (typically $0.6^{\prime\prime}$ seeing FWHM in $i$ band), and large photon-collecting power makes it one of the most powerful instruments for weak lensing measurements. The HSC SSP survey consists of three layers; Wide, Deep, and Ultradeep. The Wide layer, which is designed for weak lensing cosmology, plans to cover about 1,200~\sqdeg\ of the sky with a $5\sigma$ depth of $i\sim26$ ($2^{\prime\prime}$ aperture for a point source). Since the $i$-band images are used for galaxy shape measurements in weak lensing analyses, they are preferentially taken under good seeing conditions.

In this paper, we use the HSC first-year (hereafter HSC-Y1) galaxy shape and photo-$z$ catalogs \citep{HSCDR1_shear:17,2018MNRAS.481.3170M} constructed from about 90~nights of HSC Wide data taken between March 2014 and April 2016 \footnote{The shape catalog is publicly available in the HSC database at \url{https://hsc-release.mtk.nao.ac.jp/doc/index.php/database/}}. Both catalogs are based on the object catalog produced by the data reduction pipeline \cite{2018PASJ...70S...5B}. In the following subsections, we describe details of the shape and photo-$z$ catalogs.

\subsubsection{HSC-Y1 galaxy shape catalog}
\label{sec:HSC-Y1_shape}
We apply a number of cuts to construct the shape catalog of HSC galaxies \citep[see][for details]{HSCDR1_shear:17}. For instance, we restrict our analysis to survey regions with approximately full depth in all five filters (the ``full-depth full-color", or FDFC region), to ensure the homogeneity of the galaxy sample. We also adopt a {\tt cmodel} magnitude cut of $i<24.5$ \citep[see][for the definition of {\tt cmodel} magnitude]{2018PASJ...70S...5B}, which is conservative given the depth of the HSC Wide layer. We apply the ``Sirius'' star mask to remove regions affected by bright stars \cite{Coupon:2018}. We remove galaxies located in disconnected regions and regions where PSF modeling fails. The resulting HSC-Y1 shear catalog has more than 12 million galaxies, covering 136.9~\sqdeg\ spread over six distinct fields: XMM, GAMA09H, WIDE12H, GAMA15H, HECTOMAP, and VVDS \citep[see Fig.~1 in Ref.][]{HSCDR1_shear:17}. The HSC Wide survey footprint overlaps fully with the SDSS sky coverage. \citet{HSCDR1_shear:17} and \citet{2018PASJ...70S..26O} carried out various null tests to show that the shear catalog satisfies the requirements of HSC-Y1 science for both cosmic shear and galaxy-galaxy weak lensing analyses \footnote{Note that the requirement set in \citet{HSCDR1_shear:17} was much more stringent than that for our analysis because they assumed a larger lens sample and a wider radial range of lensing signal to be used for cosmological analysis.}. In Appendix~\ref{sec:lens_systematics}, we give further null tests that are specific for cosmology analysis with the galaxy-galaxy lensing measurements based on the HSC-Y1 and SDSS datasets.

The shape catalog includes the following quantities relevant to our weak lensing analysis. The shape catalog has the PSF-corrected galaxy ellipticity $(e_1, e_2)=(e\cos2\phi, e\sin2\phi)$, where $\phi$ is the position angle. Since the ellipticity is defined in terms of {\it distortion}, i.e., $e = (a^2-b^2)/(a^2+b^2)$, where $a$ or $b$ is the major or minor axis, one needs to apply the appropriate responsivity factor when estimating weak lensing shear from galaxy shapes (see Section~\ref{sec:gglensing_estimator} for details). The shape catalog contains, for each galaxy, an estimate of the rms intrinsic ellipticity per component $e_{\rm rms}$, from which the ellipticity measurement noise is already subtracted, and  contains the calibration factors derived from the image simulations \cite{2018MNRAS.481.3170M}. The calibration factors consist of the shear multiplicative bias $m$ and the additive bias $(c_1, c_2)$ which relate a measured shape to a true shape as $\gamma_{{\rm meas}, i}=(1+m)\gamma_{{\rm true}, i}+c_i$. The shape catalog also contains the inverse-variance weight $w_s$ which takes into account the intrinsic shape 
and measurement noise.

\subsubsection{Source galaxy catalog for galaxy-galaxy weak lensing}
\label{sec:source_galaxies}
Thanks to the depth of the HSC-Y1 data, we can define a secure sample of source galaxies behind lens galaxies, for galaxy-galaxy weak lensing measurements. In this paper we select three distinct samples of lens galaxies from the database of spectroscopic SDSS galaxies up to $z=0.7$. To select background galaxies, we use photometric redshift (hereafter photo-$z$) estimates of each HSC galaxy. Multiple photo-$z$ catalogs using different algorithms are available \citep[see][for details]{Tanaka:2018}. For our fiducial analysis, we use the {\tt MLZ} method \cite{2014MNRAS.438.3409C} as for our fiducial catalog; in Section \ref{sec:lcdm} we will discuss how
the choice of different photo-$z$ algorithms affects
the cosmological results.

In this paper, following the method in \citet{OguriTakada:11}, we use a {\it single} sample of source galaxies for the galaxy-galaxy weak lensing measurements for all three samples of lens galaxies in different redshift bins (see below). This method enables us to mitigate the impact of photo-$z$ uncertainties on cosmological constraints, as we will explicitly demonstrate later. We define a sample of background galaxies by requiring that the posterior that the galaxy has redshift less than 0.75 be less than 1\% \citep{2014MNRAS.444..147O,2018PASJ...70...30M,2019ApJ...875...63M}:
\begin{align}
\int_{z_{{\rm l,max}}+0.05}^{7}\!\!\mathrm{d}z_{\rm s}~P_i(z_{\rm s})\ge 0.99, 
\label{eq:sourcegalaxy_cut}
\end{align}
where $P_i(z_{\rm s})$ is the posterior distribution of photo-$z$ estimation for the $i$-th HSC galaxy.  Note that we use a lower bound of $z_{\rm s} = 0.75$, comfortably larger than the upper bound of the SDSS lens galaxy sample, $z_{{\rm l,max}}= 0.7$. With this cut, the sample includes 4,308,983 HSC galaxies over about 140~\sqdeg, corresponding to a net number density of $\bar{n}_{\rm s}\simeq 8.74\,{\rm arcmin}^{-2}$ and a weighted number density (defined in \citealt{2013MNRAS.434.2121C}) of $\bar{n}_{\rm s}\simeq 7.95\,{\rm arcmin}^{-2}$. The mean redshift of the sample is $\langle z_{\rm s}\rangle\simeq 1.34$.

\subsection{SDSS spectroscopic galaxy catalog for lens galaxies: LOWZ and CMASS}
\label{sec:SDSS}
\begin{figure}
    \includegraphics[width=\columnwidth]{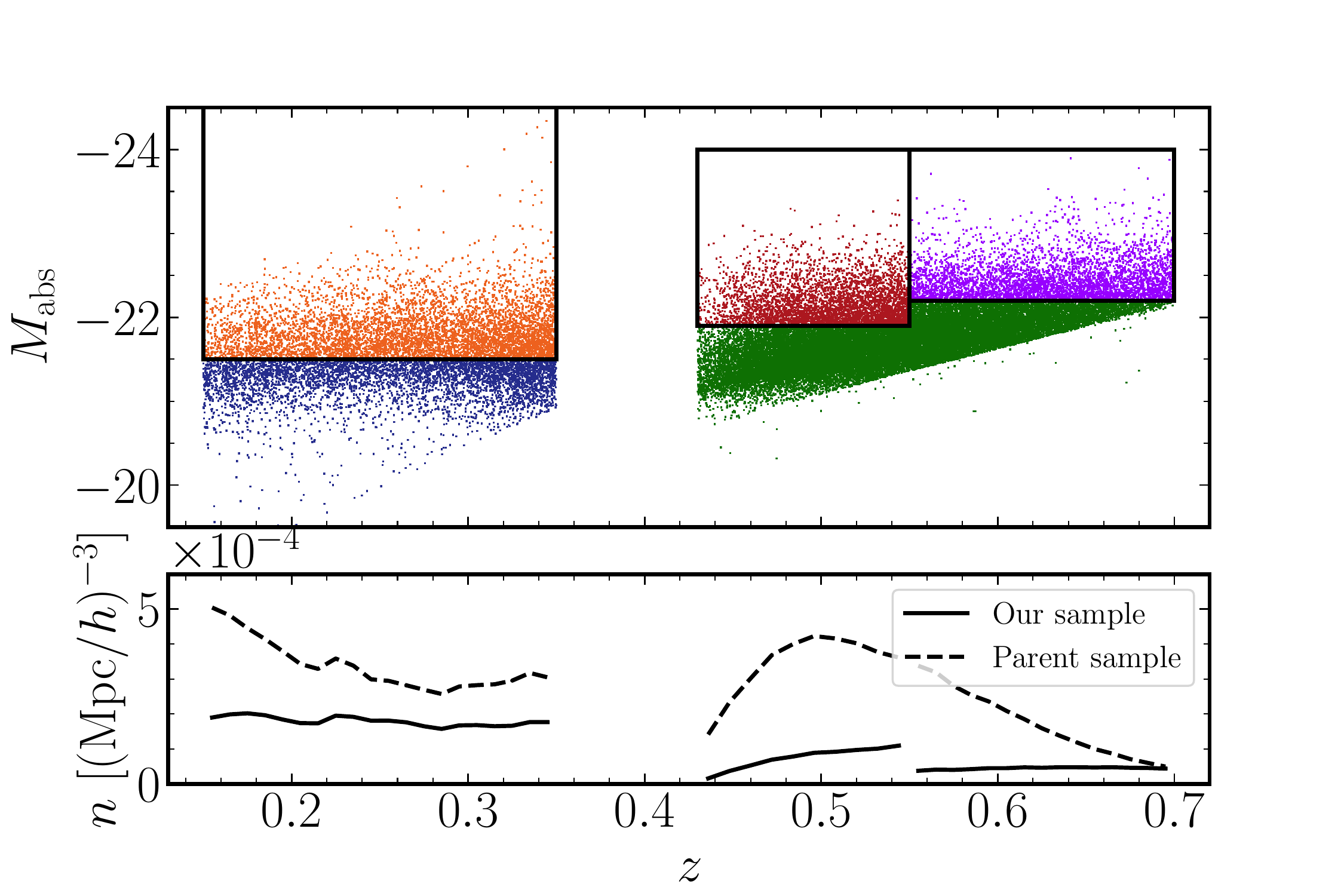}
    \caption{{\it Upper panel}: The LOWZ and CMASS galaxy samples in the plane of redshift and absolute magnitude. In this study we use the three samples denoted by the black boxes: the LOWZ sample in the redshift range $z\in [0.15,0.35]$ with absolute magnitudes $M<-21.5$, the ``CMASS1'' sample in $z\in [0.43,0.55]$ with $M<-21.9$, and the ``CMASS2'' sample in $z\in [0.55,0.70]$ with $M<-22.2$. The figure includes only 5\% of the galaxies to avoid crowding. {\it Lower panel}: The solid (dashed) lines show the redshift dependence of the comoving number density of galaxies in each of our galaxy samples (the parent LOWZ and CMASS sample), obtained assuming the {\it Planck} cosmology.}
    \label{fig:boss_selection}
\end{figure}
To trace the large-scale structure, we use the large-scale structure sample compiled in Data Release 11 (DR11) \footnote{\url{https://www.sdss.org/dr11/}} \cite{Alam:2015} of the SDSS-III (Baryon Oscillation Spectroscopic Survey) project \citep{2013AJ....145...10D}. The  SDSS-III  is  a  spectroscopic  follow up of galaxies and quasars selected from the imaging data obtained  by  the  SDSS-I/II \citep{SDSS} covering  about $11,000$ deg$^2$ \cite{2009ApJS..182..543A} using the dedicated 2.5m SDSS Telescope \cite{2006AJ....131.2332G}. Imaging data obtained in five photometric bands ($ugriz$) in the SDSS I/II surveys \cite{1996AJ....111.1748F,2002AJ....123.2121S,2010AJ....139.1628D} were augmented with an additional 3,000~deg$^2$ of imaging data  by the SDSS-III BOSS project \citep{2011AJ....142...72E,2012ApJS..203...21A,2013AJ....145...10D,2011ApJS..193...29A}. These data were processed by a series of image processing pipelines \citep{2001ASPC..238..269L,2003AJ....125.1559P,2008ApJ...674.1217P} and corrected for Galactic extinction \citep{1998ApJ...500..525S} to obtain a reliable photometric catalog which serves as an input to select targets for spectroscopy \citep{2013AJ....145...10D}. Targets are assigned to tiles using an adaptive tiling algorithm designed to maximize the number of targets that can be successfully observed \cite{2003AJ....125.2276B}. The resulting spectra were processed by an automated pipeline to perform redshift  determination and spectral classification \cite{2012AJ....144..144B}. The BOSS large-scale structure (LSS) samples are selected using algorithms focused on galaxies in different redshifts:  $0.15<z<0.35$ (LOWZ) and $0.43<z<0.7$ (CMASS). In addition to the galaxies targeted by the BOSS project, we also use galaxies which pass the target selection but have already been observed as part of the SDSS-I/II project (legacy galaxies). These legacy galaxies are subsampled in each sector so that they obey the same completeness as that of the LOWZ/CMASS samples in their respective redshift ranges \citep{2014MNRAS.441...24A}.

To perform measurements of the clustering and lensing signals, we create various subsamples, cutting on redshift and absolute magnitude, of the parent LSS catalog provided with DR11. To estimate the $i$-band absolute magnitudes for individual SDSS galaxies, we employ the method in \citet{2006MNRAS.372..537W} to make k-corrections (using  ``passive plus star-forming galaxies'' in \citet{2006MNRAS.372..537W} constructed using templates from the stellar population synthesis model in \citet{Bruzual_Charlot:2003}) of individual galaxies based on {\tt cmodel} photometry.  We k-correct the photometry LOWZ galaxies to a redshift of 0.20 and that of CMASS galaxies to a redshift of 0.55.

We use three galaxy subsamples in three redshift bins: ``LOWZ'' galaxies in the redshift range $z=[0.15,0.35]$ and two subsample of ``CMASS'' galaxies, hereafter called ``CMASS1'' and ``CMASS2'', respectively, which are obtained from subdivision of CMASS galaxies into two redshift bins, $z=[0.43,0.55]$ and $z=[0.55,0.70]$, respectively. As shown in Fig.~\ref{fig:boss_selection}, we define each of the subsamples by selecting galaxies having the absolute magnitudes $M_i-5\log {\rm h}<-21.5$, $-21.9$ and $-22.2$ for the LOWZ, CMASS1 and CMASS2 samples, respectively. The comoving number densities for the {\it Planck} cosmology $\bar{n}_{\rm g}/[10^{-4}\,(h^{-1}{\rm Mpc})^{-3}] \simeq 1.8, 0.74$ and 0.45, respectively, which are a few times smaller than those of the entire parent (color-cut and flux-limited) LOWZ and CMASS samples. As shown in Fig.~\ref{fig:boss_selection}, the number density depends only weakly on redshift within each sample. The CMASS1 sample does show a somewhat stronger redshift dependence,  but we will show later that our cosmological constraints remain almost unchanged when we exclude the CMASS1 sample from the cosmological analysis.

In Appendix~\ref{sec:lens_systematics}, we quantify the sensitivity of our lensing analysis to the choice of k-correction method we use. We find that the differences are smaller than the statistical uncertainties.  

The SDSS DR11 large-scale structure catalogs \citep{2014MNRAS.441...24A} provide weights to account for various systematic effects, including (i) the inverse correlation between the number density of galaxies and that of stars \citep{Ross:2012} and issues related to seeing ($w_\ast$), (ii) fiber collided galaxies that do not have a spectroscopic redshift, and (iii) systematic failures to obtain the spectroscopic redshifts of galaxies, respectively.

The last two weights correct for the full LOWZ and CMASS sample, not the absolute-magnitude-limited subsamples we use here, and thus are not applicable for our purposes. Instead we assign the redshift of the nearest neighbor to all fiber collided or redshift-failure galaxies, and compute their absolute magnitudes and include or exclude them depending upon our selection criteria. In summary, in our clustering analysis, we set the weights to $w_{\rm l}=w_\ast$ if the galaxies satisfy our absolute magnitude threshold criteria given below \citep[also see][for a similar treatment for a stellar-mass selected sample]{Miyatakeetal:15}. Detailed tests on mock galaxy catalogs of the use of nearest neighbor redshifts can be found in \citet{Guo:2012}; they demonstrated the nearest-neighbor correction achieves sub-percent accuracy in the projected galaxy auto-correlation function for scales used in this paper. 

For simplicity, throughout this paper we ignore redshift evolution of the clustering observables within the redshift bin of each sample, as is usually done in galaxy clustering analyses. The volume-limited samples should have weaker redshift evolution than does a flux-limited sample, because the volume limited sample would tend to reside in host halos of similar masses over the redshift range of the bin. In fact, \citet{Miyatakeetal:15} verified that both the lensing and clustering signals of volume-limited samples, defined in the range $0.43<z<0.59$, have weak redshift dependence. In addition the luminosity-limited sample allows a simpler treatment of the magnification bias effect on 
the galaxy-galaxy weak lensing than does the flux-limited sample, as we will describe below. 

\section{Measurements}
\label{sec:measurements}
In this paper we use the galaxy-galaxy weak lensing $\dSigma(R)$ and the projected correlation function $\wgg(R)$ as clustering observables.  This section describes details of the measurement methods of these two quantities. 

\subsection{Galaxy-galaxy weak lensing: $\Delta\Sigma(R)$}
\label{sec:gglensing}
Cross-correlating the positions of spectroscopic galaxies (spectroscopic SDSS galaxies in our study) with shapes of background galaxies (HSC galaxies) enables us to probe the averaged mass distribution around lens galaxies -- galaxy-galaxy weak lensing \citep{Mandelbaumetal:05}. Throughout this paper we use the average excess surface mass density profile, $\dSigma(R)$, as the  galaxy-galaxy weak lensing observable, where $\dSigma$ has dimensions of $[hM_\odot~{\rm pc}^{-2}]$ and is given as a function of the projected comoving separation $R$ with units of $[\hiMpc]$. An alternative choice of for the weak lensing observable is the tangential shear profile $\gamma_{+}(\theta)$ (a dimensionless quantity) as a function of angular separation $\theta$. As shown in \citet{2018MNRAS.478.4277S}, $\dSigma(R)$ is typically measured with higher signal-to-noise ratio than is $\gamma_+$, because it upweights source galaxies at higher redshifts that therefore have higher lensing efficiency for a given lens sample. In this section we describe the measurement method of galaxy-galaxy weak lensing and show the signal-to-noise ratio of the measurements from the HSC-Y1 dataset.

\begin{figure*}
\includegraphics[width=\columnwidth]{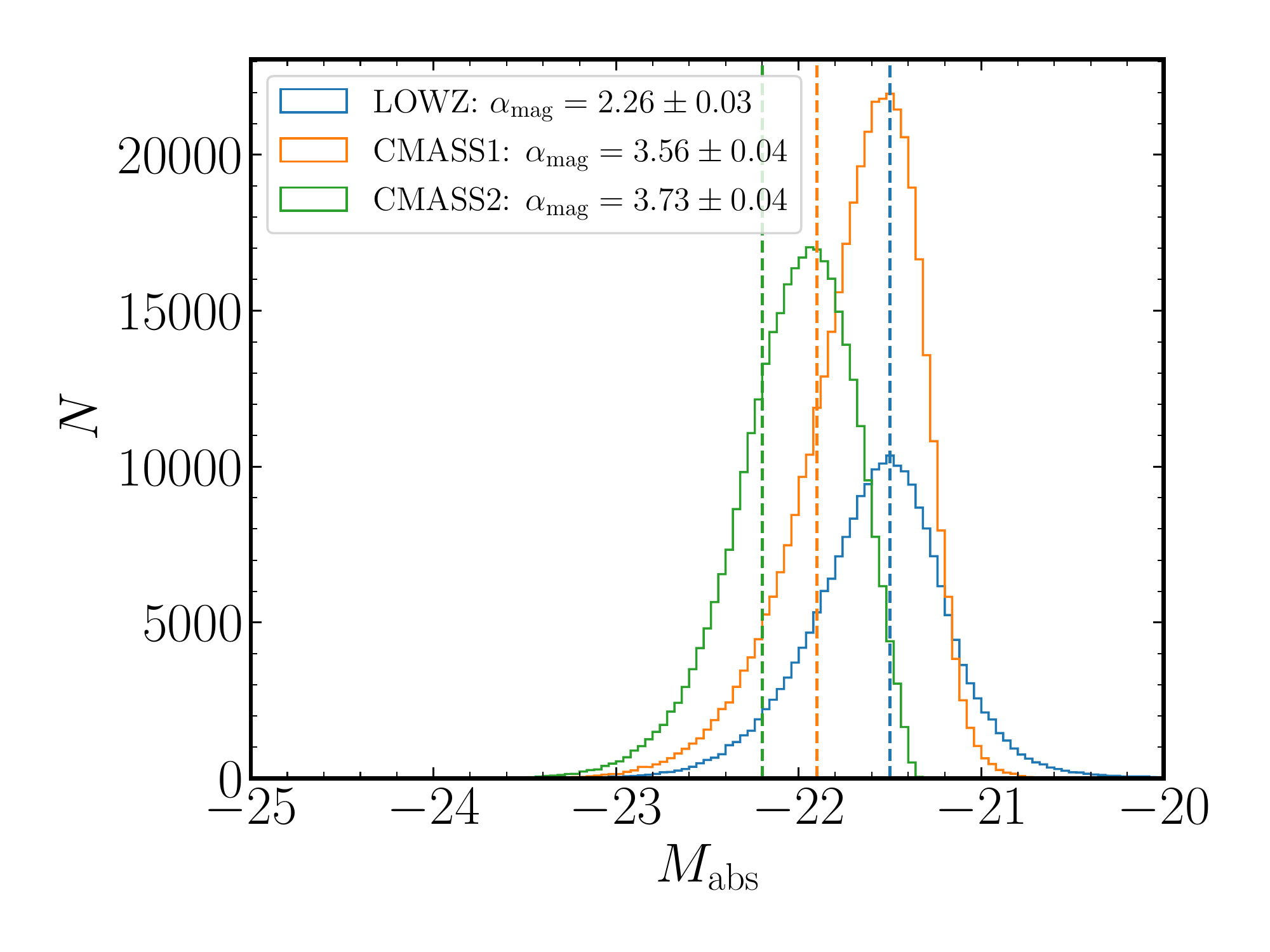}
\includegraphics[width=\columnwidth]{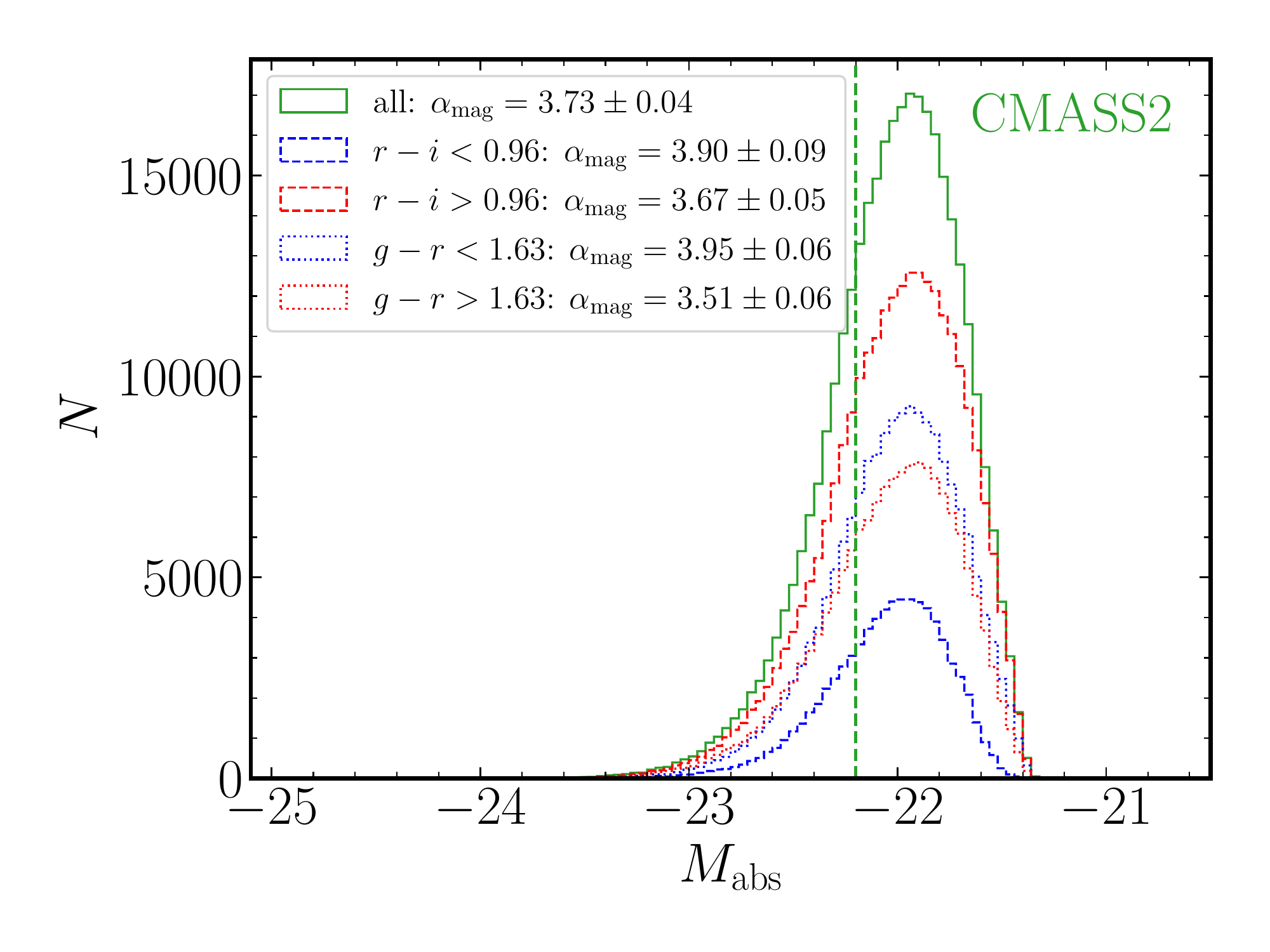}
\caption{{\it Left panel}: The differential number counts of galaxies as a function of absolute magnitude, integrated over redshift, for each of the LOWZ, CMASS1 and CMASS2 subsamples (see Fig.~\ref{fig:boss_selection}). The vertical line for each sample denotes the magnitude cut that is used to define the sample in the clustering analysis. {\it Right panel}: The number counts for subsamples of galaxies divided based on the different color cuts for the CMASS2 sample.}
\label{fig:magnification}
\end{figure*}
\begin{figure*}
\includegraphics[width=\textwidth]{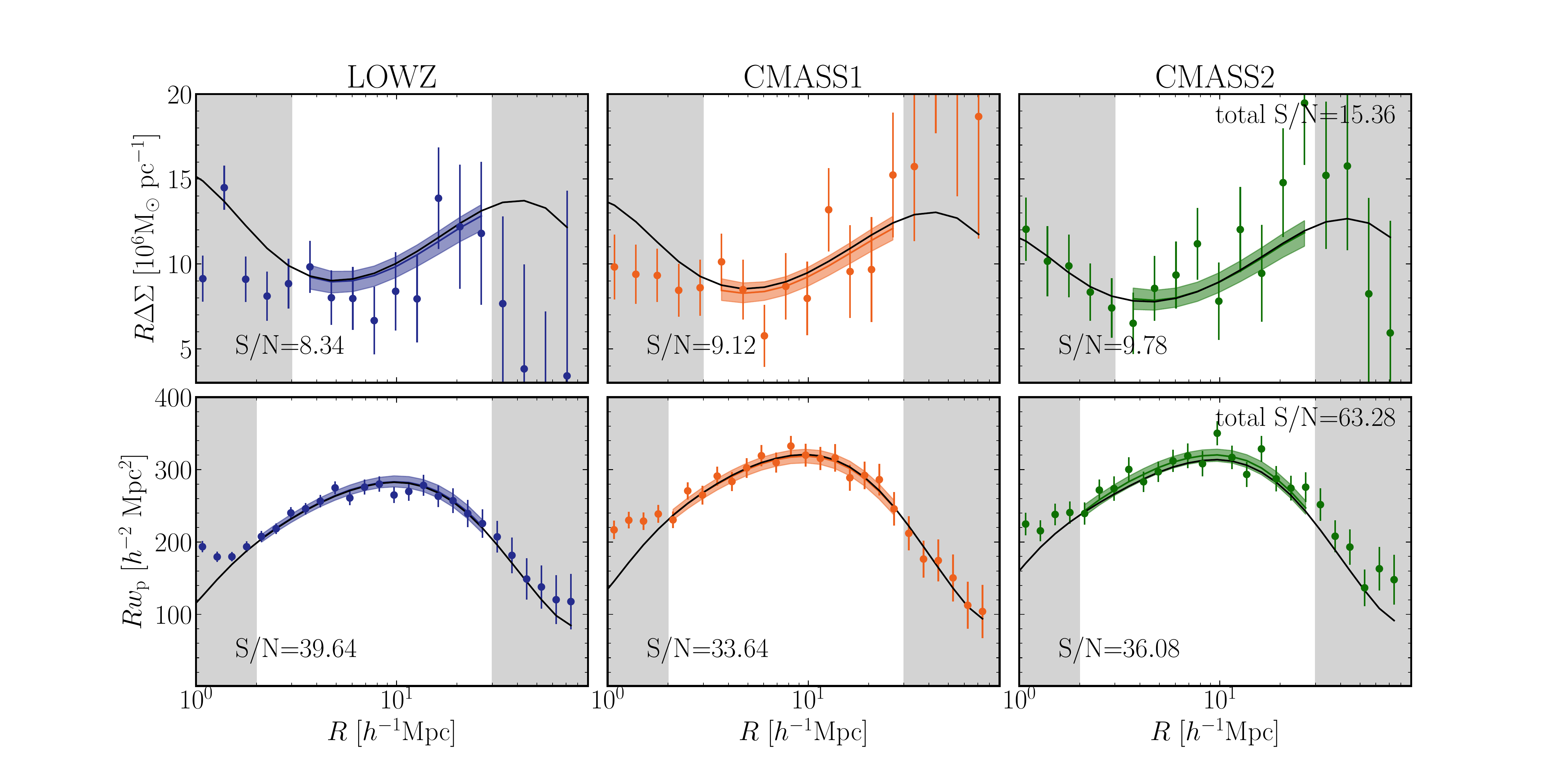}
\caption{{\it Upper panels}: The galaxy-galaxy weak lensing signals, $R\times \dSigma(R)$, measured by combining spectroscopic SDSS galaxies and HSC-Y1 galaxies for lens and source galaxy samples, respectively. Here we consider the LOWZ, CMASS1 and CMASS2 lens samples in the redshift range, $z=[0.15,0.35], [0.43,0.55]$ and $[0.55,0.70]$, respectively, and we employ a single source population from the HSC-Y1 shape catalog (see text for details). The error bars at each bin are computed from the diagonal components of the covariance matrix. The regions which are not grayed out display the range of separations, $3\le R/[h^{-1}{\rm Mpc}]\le 30$, which we use for our baseline cosmology analysis. The legend denotes the cumulative signal-to-noise ratio ($S/N$) over the range $3\le R/[h^{-1}{\rm Mpc}]\le 30$. The total $S/N$ over the three samples is given in the upper right corner, taking into account the cross-covariances. The colored band and line over the fitting range of separations in each panel denote the 68\% credible interval and mode of the posterior distribution of the model predictions in each separation bin, obtained from the Bayesian cosmology inference. The black line in each panel denotes the model prediction at {\it maximum a posteriori} (MAP). {\it Lower panels}: Similarly, the results for the projected correlation function, $R\times \wgg(R)$. The region which is not grayed out displays the range of separations, $2\le R/[h^{-1}{\rm Mpc}]\le 30$, which we use for our cosmology analysis. Note that we employed $\Omega_{\rm m,fid}=0.279$ in the measurements of $\dSigma(R)$ and $\wgg(R)$, which is needed for computations of 
$R$ and $\avrg{\Sigma_{\rm cr}^{-1}}$.}
\label{fig:lens_signal}
\end{figure*}
\subsubsection{Galaxy-galaxy weak lensing estimator}
\label{sec:gglensing_estimator}
An estimator of $\dSigma(R_i)$ for the $i$-th radial bin $R_i$ is given \citep[e.g. see][]{Miyatakeetal:15} by
\begin{widetext}
\begin{align}
    \widehat{\Delta\Sigma}(R_i) = \left.\frac{\sum_{{\rm ls} \in R_i} w_{\rm ls} \left\langle \Sigma_{\rm cr}^{-1} \right\rangle^{-1}_{\rm ls} \left[e_{t, {\rm ls}}/2\mathcal{R}(R_i) - c_{t, {\rm ls}}\right]}{\left[1+K_{\rm sel}(R_i)\right]\left[1+K(R_i)\right]\sum_{{\rm ls} \in R_i}w_{\rm ls}}\right|_{R_i=\chi(z_{\rm l})\Delta\theta_{\rm ls}} - \mbox{(signal around random points)}_{R_i} \, ,
    \label{eq:lens_measurement}
\end{align}
\end{widetext}
where the summation ``${\rm ls}$'' runs over all lens-source pairs that lie in the $i$-th radial bin $R_i\equiv \chi(z_{\rm l})\Delta \theta_{\rm ls}$; $\chi(z_{\rm l})$ is the comoving angular diameter distance to the l-th SDSS lens galaxy at the spectroscopic redshift $z_{\rm l}$, and $\Delta\theta_{\rm ls}$ is the angular separation between lens and source in each pair; $e_{t, {\rm ls}}$ is the ``tangential component'' of ellipticity of the s-th HSC source galaxy \footnote{Here we denote the dependence of each lens-source paper, i.e. ``${\rm ls}$'' in the subscript, because the tangential shear component of the s-th HSC source galaxy shape is defined with respect to the line connecting the source and lens galaxies on the sky}; $c_{t,{\rm ls}}$ is the additive shear calibration factor given in the HSC shape catalog \cite{2018MNRAS.481.3170M}; $\langle \Sigma_{\rm cr}^{-1} \rangle_{\rm ls}$ is the average of the inverse critical surface mass density given by the lensing efficiency averaged over the photo-$z$ posterior distribution function of each source galaxy, $P_{\rm s}(z_{\rm s})$, behind the l-th lens galaxy:
\begin{equation}
\left\langle \Sigma_{\rm cr}^{-1} \right\rangle_{\rm ls}=\frac{\int_0^\infty\!\mathrm{d}z_{\rm s} P_{\rm s}(z_{\rm s})\Sigma_{\rm cr}^{-1}(z_l, z_{\rm s}) }{\int_0^\infty\!\mathrm{d}z_{\rm s}~  P_{\rm s}(z_{\rm s})}, 
\label{eq:ave_Sigmacr_def}
\end{equation}
with 
\begin{align}
\Sigma_{\rm cr}\!(z_{\rm l},z_{\rm s})\equiv \frac{\chi(z_{\rm s})}{4\pi G\chi_{\rm ls}(z_{\rm l},z_{\rm s})\chi(z_{\rm l})(1+z_{\rm l})}. 
\label{eq:Sigma_cr}
\end{align}
Here $\chi_{\rm ls}(z_{\rm l},z_{\rm s})$ is the angular comoving distance between lens and source, given as $\chi_{\rm ls}=\chi(z_{\rm s})-\chi(z_{\rm l})$ for a flat-geometry universe assumed throughout this paper. Note that the factor $(1+z_{\rm l})$ arises from our use of comoving coordinates in the projected separation. Also note that we set $\Sigma_{\rm cr}^{-1}=0$ when $z_{\rm s}<z_{\rm l}$ in Eq.~(\ref{eq:ave_Sigmacr_def}). The factor ``$w_{\rm ls}$'' in Eq.~(\ref{eq:lens_measurement}) denotes the ``weight'' for which we employ an inverse-variance weighting that is nearly optimal in the shape-noise-dominated regime, following \citet{Mandelbaumetal:13} \citep[also see][]{2018MNRAS.478.4277S}:
\begin{equation}
w_{\rm ls}=w_{\rm l} w_{\rm s} \left\langle \Sigma_{\rm cr}^{-1} \right\rangle^2_{\rm ls},
\label{eq:wls_def}
\end{equation}
where $w_{\rm l}$ and $w_{\rm s}$ are the weights given in the BOSS catalog and the HSC shape catalog, respectively (see Section~\ref{sec:SDSS} for details). 

To compute $\langle \Sigma_{\rm cr}^{-1} \rangle$ (Eq.~\ref{eq:ave_Sigmacr_def}), we use the photo-$z$ posterior distribution of individual galaxies, $P_{\rm s}(z_{\rm s})$. Since the posterior distribution of source galaxies, even after averaging, differs from the underlying true redshift distribution, there is  a bias in the estimation of $\langle \Sigma_{\rm cr}^{-1} \rangle$ \citep{2012MNRAS.420.3240N}. As we will discuss below, to quantify this possible bias, we also use the ``re-weighting'' method in \citet{2019PASJ...71...43H} to estimate the intrinsic redshift distribution for the source sample, by matching the populations of the COSMOS 30-band photo-$z$ sample \cite{Ilbert:2009, Laigle:2016} to that of our background galaxy sample in the color space, because the COSMOS 30-band photo-$z$'s are much more accurate than ours. 

The shear responsivity $\mathcal{R}(R)$ in Eq.~(\ref{eq:lens_measurement}), which accounts for conversion of ``distortion'' ($[a^2-b^2]/[a^2+b^2]$) to ``shear'' ($[a-b]/[a+b])$ \cite{BernsteinJarvis:02}, is given by  
\begin{equation}
\mathcal{R}(R_i)=1-\frac{\sum_{{\rm ls} \in R_i}w_{\rm ls}e_{{\rm rms}, {\rm s}}^2}{\sum_{{\rm ls} \in R_i}w_{\rm ls}},
\label{eq:responsibity_def}
\end{equation}
where $e_{{\rm rms}, {\rm s}}$ is the rms {\it intrinsic} ellipticity of the s-th source per component. The factor $\left[1+K(R_i)\right]$ is the multiplicative shear calibration factor that is given in the HSC shape catalog, defined as 
\begin{equation}
\label{eq:multiplicative_bias}
    1+K(R_i)=1+\frac{\sum_{{\rm ls} \in R_i}w_{\rm ls}m_{\rm s}}{\sum_{{\rm ls} \in R_i}w_{\rm ls}},
\end{equation}
where $m_{\rm s}$ is the multiplicative calibration factor for the s-th source galaxy that is estimated per object using the simulations of HSC images \citep{2018MNRAS.481.3170M}. This calibration factor not only accounts for both noise bias \cite[e.g.,][]{BernsteinJarvis:02,Hirata:2004,Kacprzak:2012} and model bias \cite{Voigt:2010, Melchior:2010}, but also the effect of image blending expected in HSC including the impact of unrecognized blends to some degree.

We also need to correct for the effect of selection bias, because our sample of source galaxies is based on the specific photo-$z$ cuts (Eq.~\ref{eq:sourcegalaxy_cut}) in addition to the fiducial cuts used for the source catalog in the cosmic shear analysis \cite{2019PASJ...71...43H,2020PASJ...72...16H}. The factor $[1+K_{\rm sel}(R)]^{-1}$ corrects for the selection bias. We follow the method described in  Section~5.6.2 of \citet{2018MNRAS.481.3170M} to compute the correction factor for the source galaxy sample in each separation bin $R$. In this method, we include the selection bias effect by modifying the multiplicative bias factor $m_{\rm s}$, by computing the probability of galaxies in the sample at the edge of the resolution factor cut, where the resolution factor is given by the ratio of the PSF size to the observed size of the galaxy. We have confirmed that this selection bias effect is very small. Furthermore, in the cosmological inference, we will introduce a nuisance parameter $\Delta m_\gamma$ (see below) to model any residual systematic error in the shear calibration and study its impact on the cosmological results. 

Finally, the second term on the right-hand side of Eq.~(\ref{eq:lens_measurement}) denotes the signal around the random points, which is measured by replacing lens galaxies with random points. We need to subtract this random signal to correct for observational systematic effects such as residual systematics in shape measurements due to an imperfect correction of optical distortions across the field-of-view. The number of random points is 20 times larger than that of lens galaxies, where the random catalogs are generated mimicking the redshift distribution of galaxies in the LOWZ, CMASS1 or CMASS2 sample. We found that the signal around random points starts to deviate from zero at $R\simgt15\hiMpc$, but the deviations are still smaller than the statistical uncertainties, as shown in Fig.~\ref{fig:random_signal} in Appendix~\ref{sec:lens_systematics}.

As can be found from Eq.~(\ref{eq:lens_measurement}), the estimation of $\dSigma(R)$ involves conversion of the observed angular separation between source and lens, $\Delta\theta$, to the comoving separation $R$ and the multiplicative factor of $\langle\Sigma_{\rm cr}^{-1}\rangle_{\rm ls}^{-1}$. To do these, we need to assume a ``fiducial'' cosmology, which generally differs from the underlying true cosmology. For the flat $\Lambda$CDM model which we use throughout this paper, the only relevant free parameter is $\Omega_{\rm m}$ (because we use units such as $\hiMpc$ and $h M_\odot~{\rm pc}^{-2}$ in which the $h$ dependence is made explicit, we are insensitive to the value of $h$). In this paper we use the method in \citet{2013ApJ...777L..26M} to take into account the geometrical dependence of $\Omega_{\rm m}$ in the computations of $\langle\Sigma_{\rm cr}^{-1}\rangle_{\rm ls}$ and $R$  (also see \citet[][]{2021arXiv210100113M}). Throughout this paper we employ $\Omega_{\rm m, fid}=0.279$ for the fiducial cosmology in the measurements of $\dSigma(R)$ and $\wgg(R)$.

\begin{figure}
\includegraphics[width=\columnwidth]{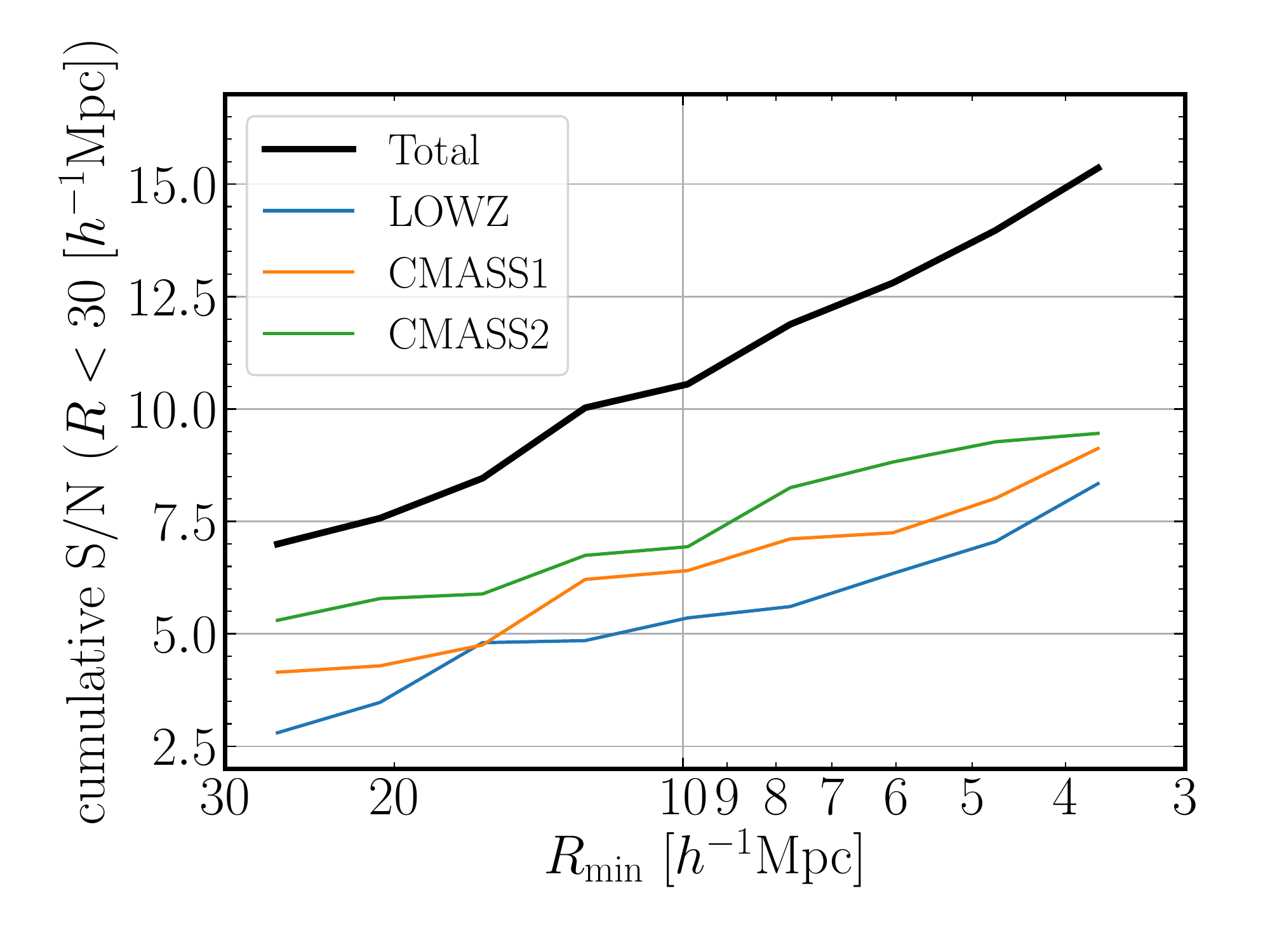}
    \caption{The cumulative signal-to-noise ratio of $\dSigma$, integrated over $R_{\rm min}<R<30\,h^{-1}{\rm Mpc}$, where we vary the minimum separation $R_{\rm min}$. Note the reversed $x$ axis in this plot. We fix the maximum separation to $30\,h^{-1}{\rm Mpc}$. We use the same binning of separations as in Fig.~\ref{fig:lens_signal}. The lines show the results for each of the LOWZ, CMASS1, and CMASS2 samples, and the bold line
    shows the total $S/N$ obtained by combining the $\dSigma$ measurements for all the samples.}
    \label{fig:lens_cumulative_sn_from_largescales}
\end{figure}
\begin{figure}
    \includegraphics[width=\columnwidth]{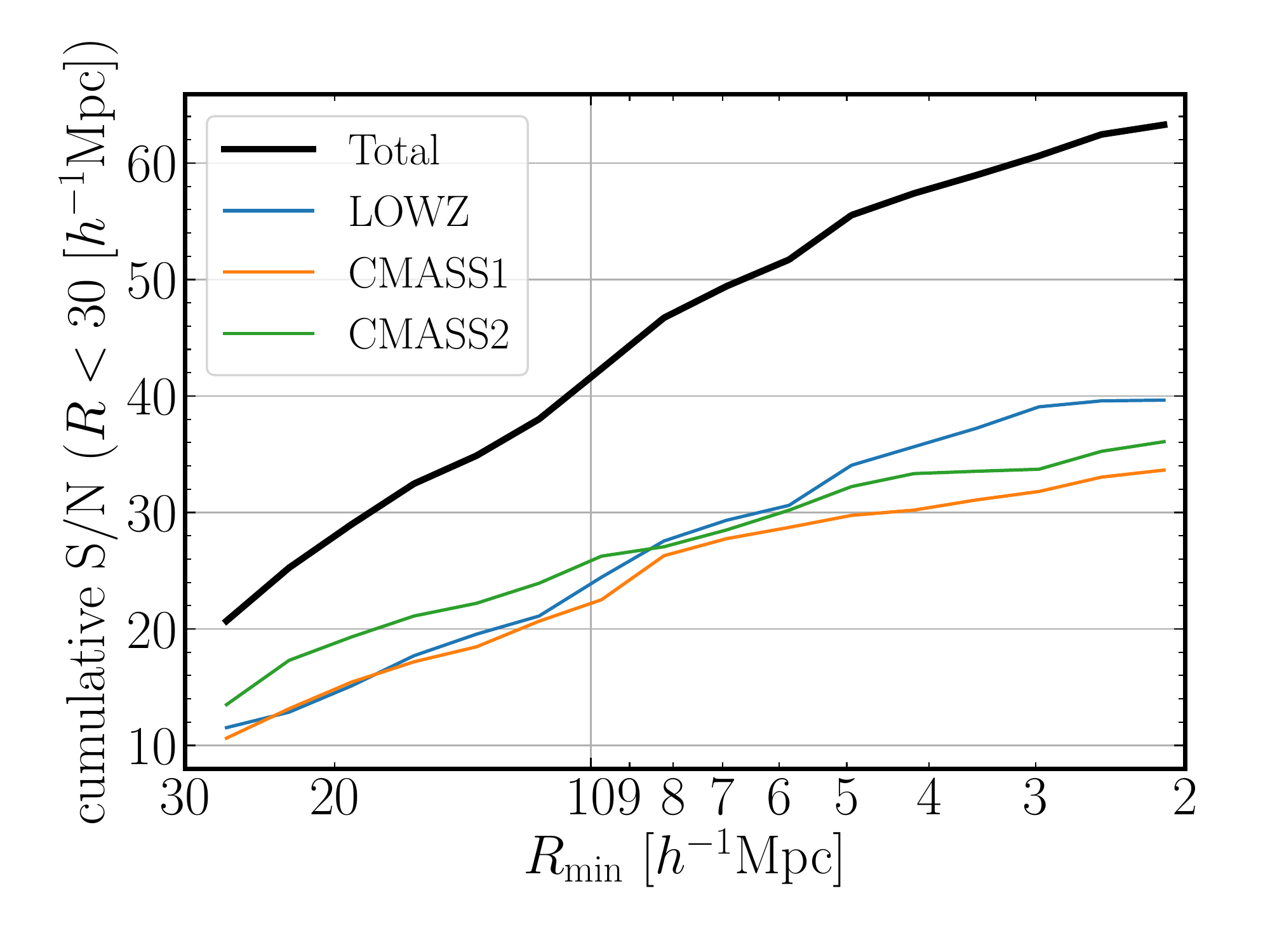}
    \caption{Similarly to Fig.~\ref{fig:lens_cumulative_sn_from_largescales}, the cumulative signal-to-noise ratio of $\wgg(R)$ integrated over $R_{\rm min}\le R\le 30~h^{-1}{\rm Mpc}$ as a function of $R_{\rm min}$.}
\label{fig:clustering_clumulative_sn_from_largescales}
\end{figure}
The large-scale structure which lies between us and the lens galaxies causes distortions of the shapes of the background source galaxy sample. It also modulates the number densities of both the source and lens galaxies due to lensing magnification \citep{2020A&A...638A..96U, vonWietersheim-Kramsta:2021}. This complicates the interpretation of the galaxy-galaxy lensing signal. As shown in \citet{2020A&A...638A..96U}, the effect of the magnification of the source galaxy sample is small and can be neglected. However, the correlated effect of the magnification of the lens sample, and the associated imprints on the shapes of the source galaxies can be a significant source of systematic error \citep{2020A&A...638A..96U}. 
The number density fluctuations of lens galaxies caused by the magnification is given by
\begin{eqnarray}
\delta^{\rm mag}_{\rm g}(\chi_{\rm l},\chi_{\rm l}\boldsymbol{\theta})&\equiv& {\displaystyle \frac{N-N_0}{N_0}}
\nonumber\\
&=&\mu^{\alpha_{\rm mag}-1} -1
\nonumber\\
&\simeq& 2(\alpha_{\rm mag}-1)\kappa(\chi_{\rm l},\chi_{\rm l}\boldsymbol{\theta}),
\end{eqnarray}
where $\kappa$ is the lensing convergence, i.e. the projected mass density field up to $z_{\rm l}$, in the direction $\boldsymbol{\theta}$,  and we have assumed the weak lensing regime, $|\kappa| \ll 1$. Here we approximate the intrinsic number counts of lens galaxies by a power law with respect to magnitudes, and the slope of galaxy counts around a given magnitude cut $\alpha_{\rm mag}$ is defined as
\begin{align}
\alpha_{\rm mag}\equiv -\frac{\mathrm{d}\log N(>f_{\rm lim})}{\mathrm{d}\log f_{\rm lim}},
\end{align}
for which we use the flux corresponding to the absolute magnitude cut, $f_{\rm lim}\propto 10^{-0.4M_{\rm ab,lim}}$. The same foreground large-scale structure causes a weak lensing distortion of the HSC source galaxies. In turn the magnification bias causes an additive contamination to the standard galaxy-galaxy weak lensing as described below.

Fig.~\ref{fig:magnification} displays the number counts of galaxies in the LOWZ, CMASS1 and CMASS2 samples. The estimated slope around the magnitude cut is $\alpha_{\rm mag}\simeq 2.26\pm 0.03$, $3.56\pm 0.04$ or $3.73\pm 0.04$ for the LOWZ, CMASS1 and CMASS2 sample, respectively, where the $1\sigma$ error is estimated assuming the Poisson errors in the number counts in each magnitude bin around the magnitude cut. For the cosmological inference, we use the estimated $\alpha_{\rm mag}$ for the central value and employ a Gaussian prior with width given by $\sigma(\alpha_{\rm mag})=0.5$. We use a significantly larger width than the estimated error on these slopes to be conservative and in order to take into account variations in the slope for different color-cuts of galaxies in the parent SDSS samples in each redshift bin (see the right panel of Fig.~\ref{fig:magnification}). We find that even with such wide Gaussian priors, the estimated cosmological parameters do not change significantly from the results obtained by fixing $\alpha_{\rm mag}$ to the central value. We comment that our luminosity-limited samples are better suited for a treatment of the magnification bias than are the parent LOWZ and CMASS galaxy samples. The parent LOWZ and CMASS galaxies are selected by the color-dependent flux cuts \citep{Eisensteinetal:01,2013AJ....145...10D}. Hence the magnification bias cannot be characterized by a single slope of the number counts in the original sample, and one would have to carefully estimate the effect using the effective slope as done in Ref.~\cite{vonWietersheim-Kramsta:2021}.

Our use of conservative cuts based on the photo-$z$ posterior distribution of the source galaxies also mitigates any contamination of intrinsic alignments of source galaxies to the galaxy-galaxy weak lensing measured that could occur if some of the source galaxies were at the same redshift as the lens galaxies and therefore are physically associated with the same large-scale structure in which the lens galaxies reside \cite[e.g.,][]{Blazek:2012}. As we show in Appendix~\ref{sec:lens_systematics}, we do not see any excess clustering of source galaxies around our lenses. Therefore we do not explicitly model intrinsic alignments in this paper.

We use mock catalogs of HSC- and SDSS-like galaxies to determine the covariance matrix of statistical errors for the $\dSigma$ measurement, as described in Appendix~B of \citet{2021arXiv210100113M}. In Appendix~\ref{sec:covariance} we briefly describe the details of the mock catalogs and our method for the covariance calibration. The correlation matrix is shown in the right panel of Fig.~\ref{fig:lens_signal_covariance}, which shows significant off-diagonal components at $R\simgt 10\,\hiMpc$. The covariance matrix includes cross-correlation between the $\dSigma(R)$ signals of different lens galaxies, which arise from the shape noise of the same source galaxies and the cosmic shear due to the shared foreground large-scale structures. In addition, our companion paper \cite{Sugiyama:2021} derived an additional contribution to the covariance matrix arising from the magnification bias. While this contribution does not significantly affect the cosmological parameter estimation, we include it  for completeness. 

In Fig.~\ref{fig:lens_signal} we show the measured signals of $\dSigma(R)$ for each of the LOWZ, CMASS1 and CMASS2 samples, respectively. We define the radial bins by dividing $0.05 < R/[\hiMpc] < 80$ into 30 evenly-spaced logarithmic bins. The region which is not grayed out displays the range of $R$ bins which we use for our cosmological analysis: $3\le R/[h^{-1}{\rm Mpc}]\le 30$. To be more precise, the smallest bin in this range includes the lens-source pairs in the separation range $3.27\le R/[h^{-1}{\rm Mpc}]\le 4.18$, while the largest bin is in the range $23.4\le R/[h^{-1}{\rm Mpc}]\le 29.9$. As is clear from the figure, the HSC-Y1 data yields a significant detection of $\dSigma$ over this full range of separations. 

To quantify the significance of the lensing measurements, we can define the {\it cumulative} signal-to-noise ratio ($S/N$) as
\begin{align}
\left(\frac{S}{N}\right)^2\equiv \sum_{3\le R_i, R_j\le 30}~ \dSigma(R_i) [{\bf C}_{\dSigma,{\rm sub}}^{-1}]_{ij} \dSigma(R_j),
\label{eq:sn_lens_def}
\end{align}
where ${\bf C}_{\dSigma,{\rm sub}}$ is a sub-matrix of the full covariance matrix including only the elements in the $R$ range $\in [3,30]~h^{-1}{\rm Mpc}$ and $({\bf C}_{\dSigma,{\rm sub}})^{-1}$ is its inverse. The legend of Fig.~\ref{fig:lens_signal} shows the $S/N$ values for each of the  LOWZ, CMASS1 and CMASS2 samples, and the total $S/N$ denotes the total $S/N$ value combining the three samples  taking into account the cross-covariances. Even though it covers only 140 \sqdeg, the HSC-Y1 data give a significant detection of the weak lensing signal, with total $S/N\simeq 15.4$. These $S/N$ values are consistent with those we obtained from the mock catalogs for the {\it Planck}-like cosmology \citep[see Table~III of Ref.][]{2021arXiv210100113M}.

Fig.~\ref{fig:lens_cumulative_sn_from_largescales} shows the $S/N$ values as a function of the minimum separation $R_{\rm min}$ over which the sum in Eq.~(\ref{eq:sn_lens_def}) extends.

\subsubsection{A model for the residual systematic photo-$z$ errors: $\Delta z_{\rm ph}$}
\label{sec:systeamtics_nuisance}
In Appendix~\ref{sec:lens_systematics} we show the results for various systematics tests such as the $B$-mode signal and the ``boost'' factor. The boost factor might arise from contamination by galaxies physically associated with lens galaxies to the source galaxy sample due to imperfect determination of photo-$z$. In brief, we did not find any evidence for such residual systematic effects in our $\dSigma$ measurements, reflecting the high quality of the HSC-Y1 data and the appropriateness of our source galaxy cuts. In our cosmological analysis, we introduce nuisance parameters $\Delta z_{\rm ph}$ (described in this subsection) and $\Delta m_\gamma$ (the following subsection) to model possible residual systematic errors in the photo-$z$ determination and multiplicative shear calibration, and treat those parameters as free parameters in the cosmology inference. Hence, even if we have residual unknown systematic effects in the weak lensing measurements, these nuisance parameters largely absorb the impact of these systematics on the cosmological constraints.

Following the method in~\citet{Hutereretal:06} \citep[also see][]{2021arXiv210100113M}, we model the systematic error in the mean source redshift by shifting the posterior distribution of each source galaxy by the same amount $\Delta z_{\rm ph}$; that is,
\begin{align}
P_{\rm s}(z_{\rm s}) \longrightarrow P_{\rm s}(z_{\rm s}+\Delta z_{\rm ph}).
\end{align}
We then use the shifted distribution to compute the averaged lensing efficiency $\avrg{\Sigma_{\rm cr}^{-1}}_{\rm ls}$ and the weight $w_{\rm ls}$ for the source-lens pairs using the actual HSC-Y1 and SDSS catalogs (Eqs.~\ref{eq:ave_Sigmacr_def} and \ref{eq:wls_def}) and then determine the lensing signal as before using Eq.~(\ref{eq:lens_measurement}). We find that the lensing signal after this shift is well approximated by the following multiplicative form: 
\begin{align}
\widehat{\dSigma}^{(i_{\rm l})}\!(R;\Delta z_{\rm ph}) \simeq f_{\rm ph}^{(i_{\rm l})}(\Delta z_{\rm ph}) \widehat{\dSigma}^{(i_{\rm l})}\!(R;\Delta z_{\rm ph}=0),
\label{eq:dsigmabias}
\end{align}
where $f_{\rm ph}^{(i_{\rm l})}(\Delta z_{\rm ph})$ is the multiplicative factor to model the effect of systematic photo-$z$ error and $i_{\rm l}$ is an index denoting the three lens samples, ``LOWZ'', ``CMASS1'' and ``CMASS2''. Note that we properly take into account the dependence of $f_{\rm ph}^{(i_{\rm l})}(\Delta z_{\rm ph})$ on the assumed cosmology, $\Omega_{\rm m}$ for a flat $\Lambda$CDM model, in parameter inference, using the similar method to that described in Section~\ref{sec:gglensing}. We find that the shift $\Delta z_{\rm ph}$ leads to different changes in the amplitudes of $\dSigma$ for the different lens samples (LOWZ, CMASS1, and CMASS2) depending on the lens redshift. Because we have used a single population of source galaxies, we can use the differences in the $\dSigma$ amplitudes at different lens redshifts to determine $\Delta z_{\rm ph}$, simultaneously with cosmological parameter estimation. That is, we are carrying out a self-calibration of the average photo-$z$ error using the method proposed by \citet{OguriTakada:11}. We will show below that this method indeed enables a self-calibration of the photo-$z$ uncertainty to the level allowed by the current statistical errors.

The nuisance parameter for photo-$z$ systematics we employ is only the mean shift of $P_{\rm s}(z_{\rm s})$. To check if this parametrization is adequate, we perform the following test using the reweighted COSMOS 30-band photo-$z$ (see Section~\ref{sec:gglensing_estimator}). The COSMOS 30-band photo-$z$'s have a much lower outlier rate and higher precision than do our photo-$z$'s because of the wide wavelength coverage and deeper photometry. We compute the possible bias in $\langle \Sigma_{\rm cr}^{-1} \rangle$ due to the use of $P_{\rm s}(z_{\rm s})$ using the method given by Eq.~(11) in \citet{2019ApJ...875...63M} \citep[also see][for the original discussion of this method]{Mandelbaumetal:08}. Specifically, we comput the ratio between $\langle \Sigma_{\rm cr}^{-1} \rangle$ based on the reweighted COSMOS photo-$z$ and that based on our $P_{\rm s}(z_{\rm s})$. We find that, for the entire lens sample and our source galaxy sample, the ratio is 1.005 (that is, the fractional change is only $0.5\%$). As described in Section~\ref{sec:parameter_estimaton_method}, we employ a Gaussian prior for $\Delta z_{\rm ph}$ with $\sigma(\Delta z_{\rm ph})$=0.1 in our baseline setup. We confirmed that $\langle \Sigma_{\rm cr}^{-1} \rangle$ is changed by $\sim-4$\% ($\sim+5$\%) for $\Delta z_{\rm ph}=-0.1$ ($\Delta z_{\rm ph}=0.1$), which is larger than the difference between the reweighted COSMOS photo-$z$ and $P_{\rm s}(z_{\rm s})$ of our source galaxies. We thus conclude that our parametrization of photo-$z$ systematics and its prior effectively absorbs all other photo-$z$ systematics. In what follows, we will also employ an even wider prior of $\sigma(\Delta z_{\rm ph})=0.2$ to study the ability of our method to self-calibrate possible unknown photo-$z$ errors in the current HSC-Y1 data.

In our method, we follow Eq.~\ref{eq:dsigmabias} and divide by the photo-$z$ error factor for an assumed $\Delta z_{\rm ph}$:
\begin{align}
\dSigma^{\rm model}(R)\rightarrow \frac{\dSigma^{\rm model}(R)}{f^{(i_{\rm l})}_{\rm ph}(\Delta z_{\rm ph})},
\label{eq:dSigma_dz_ph}
\end{align}
for each of the LOWZ, CMASS1 and CMASS2 samples. We do this rather than redoing the weak lensing measurement incorporating the photo-$z$ bias.  Because we are not changing the data vector, we can use the same covariance matrix in the cosmology inference. 
\begin{table}
\caption{Differences in the galaxy-galaxy lensing signals computed using the different photo-$z$ catalogs, compared to the lensing signal with the fiducial photo-$z$ catalog ({\tt MLZ}). We use the same method (Eq.~\ref{eq:sourcegalaxy_cut}) to select source galaxies for each of the different photo-$z$ catalogs, and compute $\Delta\Sigma$ with the same binning scheme as the fiducial measurement. We then subtract the fiducial signal, computed the covariance of the difference signal (for details of the covariance calculation, see Appendix~\ref{sec:lens_systematics}), and average over $3\le R/[h^{-1}{\rm Mpc}]\le 30$ with inverse-variance weighting, for each of the lens samples (LOWZ, CMASS1 and CMASS2). The resulting average and standard deviation is shown in each case.}
\label{tab:photo-z_test}
\begin{center}
\begin{tabular}{l|ccc}
\hline\hline
photo-$z$ & LOWZ & CMASS1 & CMASS2 \\ 
method & \\
\hline 
{\tt demp}&$-0.054 \pm 0.037$& $-0.011 \pm 0.029$& $-0.034 \pm 0.029$\\ 
{\tt ephor\_ab}&$-0.051 \pm 0.038$& $0.057 \pm 0.036$& $0.142 \pm 0.060$\\ 
{\tt frankenz}&$-0.004 \pm 0.025$& $0.002 \pm 0.030$& $0.003 \pm 0.033$\\ 
{\tt mizuki}&$-0.047 \pm 0.024$& $-0.055 \pm 0.013$& $-0.045 \pm 0.013$\\ 
{\tt nnpz}&$-0.054 \pm 0.033$& $-0.020 \pm 0.047$& $0.067 \pm 0.055$\\ 
\hline
\end{tabular}
\end{center}
\end{table}
As a final sanity check, we also study the impact of different photo-$z$ methods on the cosmological results. Table~\ref{tab:photo-z_test} gives the weighted average of differences in $\dSigma(R)$ between the measurements with the different photo-$z$ methods and those with the fiducial photo-$z$ method. Note that we repeat the cut of Eq.~(\ref{eq:sourcegalaxy_cut}) to define the source galaxy sample for each catalog, so the source samples are different for different photo-$z$ catalogs. The lensing signals changes between photo-$z$ algorithms are different between the LOWZ, CMASS1, and CMASS2 samples, some of which show 2-3$\sigma$ differences. We will explicitly study to what extent the cosmological results are changed by using the different photo-$z$ methods in Section~\ref{sec:lcdm} and Appendix~\ref{sec:test_analysis_setup}.

\subsubsection{A model for the residual shear calibration factor: $\Delta m_{\gamma}$}
\label{sec:nuisance_dm_gamma}
An accurate weak lensing measurement requires an unbiased measurement of the shapes of an ensemble of galaxies used to measure the shear. This is not straightforward \citep{2018MNRAS.481.3170M}, and an imperfect shape measurement leads to a residual systematic error in the $\dSigma$ measurements. To model the impact of a residual systematic error in the shear calibration, we introduce a nuisance parameter, $\Delta m_\gamma$, and then shift the theoretical template as
\begin{align}
\dSigma^{\rm model}(R) \rightarrow (1+\Delta m_\gamma)\dSigma^{\rm model}(R;\Delta m_\gamma=0). 
\label{eq:dsigma_dm_gamma}
\end{align}
We vary $\Delta m_\gamma$ with a Gaussian prior in the parameter inference (see Section~\ref{sec:parameter_estimaton_method} for details). Since we use a single population of source galaxies, we can use the same parameter $\Delta m_\gamma$ for the lensing signals of all three lens samples (LOWZ, CMASS1, and CMASS2), following \citet{OguriTakada:11}. This is a good approximation as long as the source galaxies are well separated from or physically independent of the lens galaxies. Thus the effect of $\Delta m_\gamma$ does not depend on the lens redshift, allowing us to distinguish  systematic effects of $\Delta z_{\rm ph}$ and $\Delta m_\gamma$ from one another.  That is, in principle, we can make a self-calibration of both $\Delta z_{\rm ph}$ and $\Delta m_\gamma$. One of the most pernicious systematic effects in shape measurements for deep imaging such as HSC-Y1 is blending effects in the source galaxies. However, this systematic can be absorbed by the nuisance parameter $\Delta m_\gamma$ because any blending effect affects the galaxy-galaxy weak lensing for the three lens samples in the same way.

As described above, because we treat the effects of systematic errors ($\Delta z_{\rm ph}$ and $\Delta m_\gamma$) as multiplicative functions, we can include both effects jointly by multiplying  the multiplicative functions in the cosmology analysis.

\subsection{Galaxy-galaxy clustering: $w_{\rm p}(R)$}
\label{sec:gg_clustering}
As another clustering observable, we use the projected correlation function, denoted as $\wgg(R)$, measured for each of the lens galaxy samples: LOWZ, CMASS1 and CMASS2. Here we describe how we measure $\wgg(R)$. 

First, we measure the three-dimensional galaxy-galaxy correlation function using the \citet{Landy:1993} estimator:
\begin{equation}
    \widehat{\xi}_{\rm gg}(R, \Pi) = \frac{DD-2D\mathscr{R}+\mathscr{R}\mathscr{R}}
    {\mathscr{R}\mathscr{R}},
\end{equation}
where $R$ and $\Pi$ are the projected separation and line-of-sight separation between galaxy pairs, respectively, and $DD$, $D\mathscr{R}$, and $\mathscr{R}\mathscr{R}$ are the counts of galaxy pairs, galaxy-random pairs, and random pairs in a given separation bin of ($R, \Pi$). Note that the notation ``$\mathscr{R}$'' is used to denote random points, which should not be confused with the responsivity ${\cal R}$ in Eq.~(\ref{eq:responsibity_def}). Throughout this paper we assume the distant observer approximation to estimate the separations $(R,\Pi)$ from the observed redshifts and angular positions (RA, Dec) of galaxies or randoms for each pair. We then project the three dimensional correlation function to the projected correlation function as 
\begin{equation}
\widehat{w}_{\rm p}(R)=2\int_0^{\Pi_{\rm max}}\mathrm{d}\Pi~ \hat{\xi}_{\rm gg}(R, \Pi),
\label{eq:wp_def}
\end{equation}
where we choose the fiducial value of $\Pi_{\rm max} = 100~\hiMpc$ for the projection length. The projected correlation function minimizes the effect of redshift-space distortions (RSD), which makes the modeling somewhat easier. However the RSD effect may not be negligible for large projected separations, and we include the RSD effect in the theoretical model predictions using the method in \citet{vandenBosch:2013} \citep[also see][]{2021arXiv210100113M}. When calculating the integral we adopt a binning of $\Delta \Pi = 1~\hiMpc$. We employ 30 bins logarithmically evenly spaced over $0.5\le R/[h^{-1}{\rm Mpc}]\le 80$. For the cosmology analysis below, we use 16 bins in the range of $2\le R/[h^{-1}{\rm Mpc}]\le 30$ as our fiducial choice.

As in the $\dSigma$ measurement, the conversion of angular separation and redshift difference between paired galaxies to the three-dimensional separation ($R,\Pi$) requires the use of a reference cosmology, which will in general differ from the true cosmology. We use the method in Ref.~\cite{2013ApJ...777L..26M} to correct for the conversion with varying $\Omega_{\rm m}$ for the flat $\Lambda$CDM cosmology.

We estimate the covariance matrix of $\wgg(R)$ using the jackknife method of the actual SDSS data for each of the LOWZ, CMASS1 and CMASS2 samples. We utilize 192 jackknife regions of the SDSS survey footprint \citep[see][for details]{Miyatakeetal:15}, measure $\wgg(R)$ from each jackknife region and then estimate the covariance matrix from the measured $\wgg(R)$ from all the jackknife realizations.

In the bottom panels of Fig.~\ref{fig:lens_signal} we show the measured signals of $\wgg(R)$ for the LOWZ, CMASS1 and CMASS2 samples. The unshaded region displays the range of separations, $2\le R/[h^{-1}{\rm Mpc}]\le 30$ (16 bins), which we use for our fiducial cosmological analysis.

Fig.~\ref{fig:clustering_clumulative_sn_from_largescales} shows the cumulative $S/N$ in $\wgg(R)$ integrated over $R_{\rm min}\le R\le 30~h^{-1}{\rm Mpc}$, as a function of $R_{\rm min}$, similarly to Fig.~\ref{fig:lens_cumulative_sn_from_largescales}. 
The SDSS samples cover $\sim 8000$~\sqdeg, and $\wgg(R)$ is measured with high significance. The $S/N$ values over the fitting range, $2\le R/[h^{-1}{\rm Mpc}]\le 30$, are consistent with those we found from the mock catalogs for the {\it Planck}-like cosmology \citep[see Table~III in Ref.][]{2021arXiv210100113M}. Nevertheless, as we will show below, analyzing $\dSigma(R)$ and $\wgg(R)$ jointly is essential to break degeneracies between galaxy bias uncertainties and cosmological parameters and therefore to obtain stringent constraints on cosmological parameters \citep{More:2013b,2015ApJ...806....2M,2021arXiv210100113M}. Either $\dSigma(R)$ or $\wgg(R)$ alone suffer from severe parameter degeneracies. 

\section{Analysis method}
\label{sec:analysis}
In this section, we describe theoretical templates to model the clustering observables, $\dSigma(R)$ and $\wgg(R)$, that we use for our cosmological analysis. Details and validation of the theory using mock galaxy catalogs are given in \citet{2021arXiv210100113M}. Both of the clustering observables, $\dSigma(R)$ and $\wgg(R)$, depend only on the clustering properties of the SDSS galaxies and not on those of HSC galaxies used as the source galaxy sample in the $\dSigma$ measurements. Thus, the theoretical templates are designed to model the clustering properties of the SDSS galaxies. 

\subsection{Model}
\label{sec:model}

\subsubsection{Dark Emulator}
\label{sec:dark_emulator}
\begin{table}
\begin{center}
\caption{The set of 6 cosmological parameters used in our analysis, which specify a model within the flat-geometry $\Lambda$CDM framework. For an input $\Lambda$CDM model, 
{\tt Dark Emulator} outputs the halo clustering quantities (see text for details). The column labeled  ``parameters'' lists 6 cosmological parameters. The column labeled ``supported range'' denotes the range of parameters that is supported by {\tt Dark Emulator}.
 \label{tab:cosmological_parameters_supportingrange}}
\begin{tabular}{l|l} \hline\hline
parameters & supported range [min,max]  
\\ \hline
$\Omega_{\rm de}$ &  $[0.54752,0.82128]$\\ 
$\ln (10^{10}A_{\rm s})$ &  $[2.4752,3.7128]$\\ \hline
$\omega_{\rm b}$&  $[0.0211375,0.0233625]$\\
$\omega_{\rm c}$& $[0.10782,0.13178]$\\
$n_{\rm s}$ &  $[0.916275,1.012725]$\\
\hline\hline
\end{tabular}
\end{center}
\end{table}
In this paper we extensively use the publicly-available code, {\tt Dark Emulator} \footnote{\url{https://github.com/DarkQuestCosmology/dark_emulator_public}}, developed in \citet{2018arXiv181109504N}. {\tt Dark Emulator} is a software package enabling fast, accurate computations of halo clustering quantities for an input flat $w$CDM cosmological model. They constructed an ensemble set of cosmological $N$-body simulations, each of which was performed with $2048^3$ particles for a box with length $1$ or $2~{\rm Gpc}/h$ on a side, for 101 flat $w$CDM cosmological models. The $w$CDM cosmology is parametrized by 6 parameters, ${\bf p}=\{\omega_{\rm b},\omega_{\rm c}, \Omega_{\rm de},\ln(10^{10}A_{\rm s}),n_{\rm s}, w_{\rm de}\}$, where $\omega_{\rm b}(\equiv \Omega_{\rm b}h^2)$ and $\omega_{\rm c}(\equiv \Omega_{\rm c}h^2)$ are the physical density parameters of baryons and CDM, respectively, $h$ is the Hubble parameter, $\Omega_{\rm de}\equiv 1-(\omega_{\rm b}+\omega_{\rm c}+\omega_\nu)/h^2$ is the density parameter of dark energy for a flat-geometry universe, $A_{\rm s}$ and $n_{\rm s}$ are the amplitude and tilt parameters of the primordial curvature power spectrum normalized at $k_{\rm pivot}=0.05~{\rm Mpc}^{-1}$, and $w_{\rm de}$ is the equation of state parameter for dark energy. In the following we focus on flat $\Lambda$CDM cosmological models with $w_{\rm de}=-1$.

For the $N$-body simulations, they included the neutrino mass effect fixing the neutrino density parameter $\omega_\nu\equiv \Omega_\nu h^2$ to 0.00064, corresponding to 0.06~eV for the total mass of three neutrino species that is the lower bound of the normal mass hierarchy as in \citet{2019JHEP...01..106E}. They included the effect of massive neutrinos in an approximate manner only through the present-day linear matter transfer function, which was then scaled to the initial redshift of the simulations using the linear growth factor computed \textit{without} neutrinos in setting up the initial conditions. The subsequent nonlinear growth was followed consistently in an $N$-body simulation, ignoring the neutrino effects \citep[see][for details]{2018arXiv181109504N}. Since we focus on the $\sigma_8$ parameter \footnote{$\sigma_8$ is the parameter often used in the literature for the normalization of the linear matter power spectrum, corresponding to the rms linear mass density fluctuations within a top-hat sphere of radius $8\,h^{-1}{\rm Mpc}$.}, i.e., the present-day normalization of the linear matter power spectrum instead of the amplitude of the primordial fluctuations, this approximate treatment has little impact on our primary constraints from the HSC-Y1 and SDSS data. 

The particle mass for the fiducial {\it Planck} cosmology is $m=1.02\times 10^{10}~h^{-1}M_\odot$ for the higher resolution simulations used as the basis for {\tt Dark Emulator}. The emulator uses halos with mass greater than $10^{12}~h^{-1}M_\odot$, corresponding to about 100 simulation particles.

For each $N$-body simulation realization (each redshift output) for a given cosmological model, they constructed a catalog of halos using \texttt{Rockstar} \citep{Behroozi:2013}, which identifies halos and subhalos based on clustering of $N$-body particles in phase space (position and velocity space). Then they constructed the catalog of central halos in each output. In this step, halo mass is defined using the spherical overdensity with respect to the halo center (defined as the position with the maximum mass density): $M\equiv M_{\rm 200 m}=(4\pi/3)R_{\rm 200m}^3\times (200\bar{\rho}_{\rm m0})$, where $R_{\rm 200m}$ is the spherical halo boundary radius within which the mean mass density is 200 times $\bar{\rho}_{\rm m0}$. By combining the outputs of $N$-body simulations and the halo catalogs at multiple redshifts in the range $z=[0,1.48]$, they built an emulator, dubbed {\tt Dark Emulator}, which enables fast and accurate computations of the halo mass function, halo-matter cross-correlation, and halo auto-correlation as a function of halo mass, redshift, spatial separation and cosmological model.  

For host halos of SDSS LOWZ and CMASS galaxies, which have a minimum (typical) mass of $M_{\rm 200m}\sim10^{12}M_\odot$  ($10^{13}M_\odot$), {\tt Dark Emulator} was shown to achieve sufficient accuracy for these observable quantities compared to the statistical measurement errors of $\dSigma$ and $\wgg$ expected from the HSC-Y1 and SDSS data, as shown in Fig.~31 of Ref.~\cite{2018arXiv181109504N}. In summary, {\tt Dark Emulator} outputs the following quantities:
\begin{itemize}
\item $\frac{\mathrm{d}n_{\rm h}}{\mathrm{d}M}(M; z,{\bf p})$: the halo mass function for halos in the mass range $[M,M+\mathrm{d}M]$,
\item $\xi_{\rm hm}(r; M, z, {\bf p})$: the halo-matter cross-correlation function for a sample of halos in the mass range $[M,M+\mathrm{d}M]$, and
\item $\xi_{\rm hh}(r; M, M', z,{\bf p})$: the halo-halo auto-correlation function for two samples of halos with masses $[M,M+\mathrm{d}M]$ 
and $[M',M'+\mathrm{d}M']$
\end{itemize}
for an input set of parameters, halo mass $M$ (and $M'$ for the cross-correlation function between two halo samples), redshift $z$, and cosmological parameters ${\bf p}$. 

In addition, {\tt Dark Emulator} outputs ancillary quantities, such as the linear halo bias (the large-scale limit of the halo bias), the Tinker model of the linear halo bias \citep{2010ApJ...724..878T} (see below), the linear matter power spectrum, the linear rms mass fluctuations of halo mass scale $M$ ($\sigma^L_{\rm m}(M)$), and $\sigma_8$. 

The supported range of each cosmological parameter for {\tt Dark Emulator} is given in Table~\ref{tab:cosmological_parameters_supportingrange}. These ranges are sufficiently broad that they cover the range of cosmological constraints from current state-of-the-art large-scale structure probes such as the Subaru HSC cosmic shear results \cite{2019PASJ...71...43H,2020PASJ...72...16H}. Since $\sigma_8$ and $\Omega_{\rm m}$ are primary parameters to which large-scale structure probes are sensitive, we also quote the supported ranges of these \textit{derived} parameters: $0.55\lesssim \sigma_8\lesssim1.2$ and $0.17\lesssim \Omega_{\rm m}\lesssim0.45$, as shown in Fig.~2 of \citet{2018arXiv181109504N}. In this paper we use {\tt Dark Emulator} to perform cosmological parameter inference in a multi-dimensional parameter space by comparing the model templates of $\dSigma$ and $\wgg$ with the signals measured from the SDSS and HSC-Y1 data.

A Bayesian parameter inference method might sample some models that are outside the supported range of $\Lambda$CDM 
models in {\tt Dark Emulator}. In this case, we make the following, simple extrapolation of the model predictions: 
\begin{align}
\xi_{\rm hm}(r; {\bf p}_{\notin})&\rightarrow 
\frac{b^{\rm Tinker}({\bf p}_{\notin})}{b^{\rm Tinker}({\bf p}_{\rm edge})}\frac{\xi^L_{\rm mm}(r;{\bf p}_{\notin})}{\xi^L_{\rm mm}(r;{\bf p}_{\rm edge})}
\xi^{\rm DE}_{\rm hm}(r; {\bf p}_{\rm edge}),\nonumber\\
\xi_{\rm hh}(r; {\bf p}_{\notin})&\rightarrow \left(
\frac{b^{\rm Tinker}({\bf p}_{\notin})}{b^{\rm Tinker}({\bf p}_{\rm edge})}\right)^2\nonumber\\
&\hspace{5em}\times
\frac{\xi^L_{\rm mm}(r;{\bf p}_{\notin})}{\xi^L_{\rm mm}(r;{\bf p}_{\rm edge})}
\xi^{\rm DE}_{\rm hh}(r; {\bf p}_{\rm edge}),
\end{align}
where ${\bf p}_{\notin}$ is a set of 6 cosmological parameters that are outside the supported range (Table~\ref{tab:cosmological_parameters_supportingrange}), ${\bf p}_{\rm edge}$ is a set of parameters at the edge of the supported range, $b^{\rm Tinker}(\bf{p}_{\notin})$ and $b^{\rm Tinker}({\bf p}_{\rm edge})$ are the linear bias parameters at models of ${\bf p}_{\notin}$ and ${\bf p}_{\rm edge}$ that are computed based on the fitting formula of \citet{2010ApJ...724..878T}, $\xi^L_{\rm mm}$ is the linear-theory prediction for the matter two-point correlation function at the respective model, and $\xi_{\rm hh}^{\rm DE}$ and $\xi^{\rm DE}_{\rm hm}$ are the {\tt Dark Emulator} outputs at the edge model. Here we use {\tt CLASS} \citep{Lesgourgues:2011, Blas2011} to compute the linear-theory matter correlation, $\xi^{L}_{\rm mm}(r)$, for models outside the supported range.  We define ${\bf p}_{\rm edge}$ by replacing only the parameter(s) outside the supported range with their value(s) at the edge of the supported range, while keeping the other parameter(s) at their input value(s). In the above extrapolation, we simply assume that the halo-matter cross-correlation and the halo auto-correlation follow the linear theory predictions ($\xi_{\rm hm}\simeq b\xi_{\rm mm}$ and $\xi_{\rm hh}\simeq b^2\xi_{\rm mm}$), and that the ratio of $\xi_{\rm hh}^{\rm L}({\bf p}_{\notin})$ and $\xi_{\rm hh}({\bf p}_{\notin})$ can be accurately captured by a similar ratio between $\xi_{\rm hh}^{\rm L}({\bf p}_{\rm edge})$ and $\xi_{\rm hh}({\bf p}_{\rm edge})$. Including automated outputs of the model predictions for models outside the supported range is important, because we perform a blinded cosmological analysis of the HSC and SDSS data. If {\tt Dark Emulator} provides an error message indicating that an outside model has been sampled, we could unintentionally and prematurely unblind our analysis. For the extrapolation we can adopt any input value for $A_{\rm s}$, but need to adopt values in the  specific ranges for $\omega_{\rm c}$ and $\Omega_{\rm de}$ as we will explain around Table~\ref{tab:parameters}.

After unblinding our cosmology analysis, we confirmed that all models within the 95\% credible interval of $S_8$ in the chains for our baseline analysis are within the emulator supported range \footnote{For this discussion, we used the chains for the models that have $\omega_{\rm c}$ within a $\pm5\sigma$ range of the {\it Planck} constraint, because $\omega_{\rm c}$ is not well-constrained by the observables used in this paper.}. 

\subsubsection{Galaxy-galaxy weak lensing: $\dSigma(R)$}
\label{sec:dsigma_theory}
Our galaxy-galaxy weak lensing observable $\dSigma(R)$ depends only on the clustering properties of SDSS lens galaxies, and not on the redshifts of HSC source galaxies. The ensemble average of the galaxy-galaxy weak lensing estimator has two contributions: 
\begin{align}
\dSigma^{\rm model}(R;z_{\rm l})\simeq \dSigma(R;z_{\rm l})+\dSigma^{\rm mag}(R;z_{\rm l}). 
\label{eq:dSigma_two_contributions}
\end{align}
The first term on the right-hand side is the standard contribution to the galaxy-galaxy weak lensing signal: the excess surface mass density profile of lens galaxies. The second term is the contribution caused by the lensing magnification effect, which arises from correlations between shapes of source galaxies and the mass distribution in the foreground structures of lens galaxies along the same line-of-sight directions to source galaxies \citep{2020A&A...638A..96U}. Below we describe our models for each contribution within the $\Lambda$CDM model framework. Throughout this paper, we model the clustering observables of each SDSS galaxy sample using the theoretical model prediction at a representative redshift, denoted as $z_{\rm l}$: $\bar{z}_{\rm l}\simeq 0.26, 0.51$ and $0.63$ for the LOWZ, CMASS1 and CMASS2 samples, respectively. That is, we ignore the possible redshift evolution of the clustering observables within the redshift bin for simplicity.

The excess surface mass density profile $\dSigma$ for a given sample of lens galaxies is expressed as \citep[e.g.][]{Mandelbaumetal:13,2013MNRAS.435.2345H}:
\begin{align}
\dSigma(R;z_{\rm l}) &=\bar{\rho}_{\rm m0}\int\!\!\frac{k\mathrm{d}k}{2\pi}~P_{\rm gm}(k;z_{\rm l})J_2(kR),
\label{eq:dSigma_def}
\end{align}
where $J_2(x)$ is the 2nd-order Bessel function and $P_{\rm gm}(k;z_{\rm l})$ is the cross-power spectrum between galaxies and matter at redshift $z_{\rm l}$. Hereafter we omit $z_{\rm l}$ in the argument for notational simplicity. 

As described above, {\tt Dark Emulator} outputs halo clustering properties for an input cosmology. To obtain the model predictions for the observable quantities for SDSS galaxies, we need a model for the galaxy-halo connection. For this, we use the halo occupation distribution \cite[HOD][]{1998ApJ...494....1J,Zhengetal:05}. In Appendix~\ref{sec:hod} we describe the galaxy-halo connection model (for more details see \citet{2021arXiv210100113M}). Our fiducial model for the galaxy-halo connection has five parameters for each galaxy sample (LOWZ, CMASS1 and CMASS2): $\{M_{\rm min},\sigma_{\log M},\kappa, M_1, \alpha\}$. Here $M_{\rm min}$ and $\sigma_{\log M}$ describe the central galaxy HOD, while the other parameters are for the satellite galaxy HOD. The parameter $\alpha$ is the slope of the satellite occupation number, and is distinct from the parameter $\alpha_{\rm mag}$ used for the slope of the number counts of lens galaxies when  modeling magnification bias. 

The mean number density of galaxies is given by
\begin{align}
\bar{n}_{\rm g}=\int\!\mathrm{d}M~ 
\frac{\mathrm{d}n_{\rm h}}{\mathrm{d}M}
\avrg{N_{\rm c}}(M)\left[1+\lambda_{\rm s}\!(M)\right],
\label{eq:ng}
\end{align}
where $\avrg{N_{\rm c}}\!(M)$ is the HOD of central galaxies, and $\avrg{N_{\rm c}}\!(M)\lambda_{\rm s}\!(M)$ is the HOD of satellite galaxies. Here we use {\tt Dark Emulator} to compute the halo mass function $\mathrm{d}n_{\rm h}/\mathrm{d}M$

As shown in Eq.~(\ref{eq:dSigma_def}), we must compute $P_{\rm gm}$ for a given set of model parameters to obtain a model prediction for  $\dSigma(R)$. We use {\tt Dark Emulator} to compute $P_{\rm gm}$ as
\begin{align}
P_{\rm gm}(k)&=\frac{1}{\bar{n}_{\rm g}}
\int\!\mathrm{d}M\frac{\mathrm{d}n_{\rm h}}{\mathrm{d}M}
\avrg{N_{\rm c}}\!(M)\left[1+\lambda_{\rm s}\!(M)\tilde{u}_{\rm s}(k;M,z)\right]\nonumber\\
&\hspace{5em}\times P_{\rm hm}(k; M),
\label{eq:Pgm_def}
\end{align}
where $\tilde{u}_{\rm s}(k;M)$ is the Fourier transform of the average radial profile of satellite galaxies in a host halo with mass $M$.
Here we use {\tt Dark Emulator}  to compute the halo mass function $\mathrm{d}n_{\rm h}/\mathrm{d}M$ and the halo-matter cross power spectrum, 
$P_{\rm hm}(k; M)$, for an input cosmological model. 
Throughout this paper, we assume that satellite galaxies follow a Navarro-Frenk-White (NFW) profile \citep{Navarroetal:97}. To compute the NFW profile as a function of halo mass and redshift for a given cosmological model, we use the halo mass-concentration relation computed using the publicly-available code {\tt Colossus} \footnote{\url{http://www.benediktdiemer.com/code/colossus/}} \citep{2018ApJS..239...35D}. For our fiducial model, we do not consider the effect of off-centered ``central'' galaxies or the ``incompleteness'' of central galaxies \citep{2012MNRAS.419.3457H,2013MNRAS.435.2345H}, where the incompleteness effect models a possibility that some massive halos might not host a central galaxy in the sample due to color and magnitude cuts. For an extended cosmological analysis, we include parameters to model the off-centering and incompleteness effects to study their impact on the inferred cosmological parameters, following \citet{2021arXiv210100113M}. In order to compute $\dSigma$ for each model, we use the publicly-available {\tt FFTLog} \cite{Hamilton00} code to perform the Hankel transforms in Eq.~(\ref{eq:dSigma_def}).

We model the second term in Eq.~(\ref{eq:dSigma_two_contributions}) using the nonlinear matter power spectrum \cite{2020A&A...638A..96U}:
\begin{align}
\dSigma^{\rm mag}(R)&\simeq 2(\alpha_{\rm mag}-1)
\frac{3}{2}H_0\Omega_{\rm m}
\int_0^{z_{\rm l}}\frac{\mathrm{d}z~ H_0}{H(z)}\frac{(1+z)^2}{1+z_{\rm l}}\nonumber\\
&\hspace{1em}\times \int\!\mathrm{d}z_{\rm s}~ P_{\rm s}(z_{\rm s}) \frac{\chi^2(\chi_{\rm l}-\chi)(\chi_{\rm s}-\chi)}{\chi_{\rm l}^2(\chi_{\rm s}-\chi_{\rm l})}\nonumber\\
&\hspace{1em}\times \bar{\rho}_{\rm m0}
\int\!\frac{k\mathrm{d}k}{2\pi}P_{\rm mm}^{\rm NL}\!(k;z)J_2\left(k\frac{\chi}{\chi_{\rm l}}R\right),
\label{eq:magnification_bias}
\end{align}
where $P_{\rm s}(z_{\rm s})$ is the stacked posterior distirbution of source galaxies and $P^{\rm NL}_{\rm mm}(k)$ is the nonlinear matter power spectrum. We use {\tt halofit} \citep{Takahashi_2012} to model $P^{\rm NL}_{\rm mm}$ for a given cosmological model. Note that $\dSigma^{\rm mag}$ does not depend on galaxy bias. The above expression includes the redshift distribution of source galaxies, but we treat the lens galaxies as all being at their mean redshift for simplicity. As we will show below, $\dSigma^{\rm mag}$ leads to about 1\%, 7\% and 10\% contributions to the $\dSigma^{\rm model}$ for the LOWZ, CMASS1 and CMASS2 samples, respectively, for the {\it Planck} cosmology \cite{PlanckCosmology:16}. Including the $\dSigma^{\rm mag}$ contribution in the theoretical template adds some cosmological information. In our analysis we treat the magnitude slope $\alpha_{\rm mag}$ as a nuisance parameter, with a Gaussian prior with width $\sigma(\alpha_{\rm mag})=0.5$ around the central value taken from the measurement value (see Fig.~\ref{fig:magnification}).
On the other hand, using the mock signals, we checked that, if the magnification bias is ignored in the model template, it could cause $\sim 0.1$--$0.2\sigma$ bias in $S_8$ . 

Exactly speaking we have to use the intrinsic redshift distribution of source galaxies to compute the model prediction of Eq.~(\ref{eq:magnification_bias}). We checked that the model prediction is changed only by up to $\sim$5\% in the amplitude even if  
using the intrinsic redshift distribution estimated by the
reweighting method based on the COSMOS photo-$z$ catalog (see Section~\ref{sec:gglensing}) and including the weights of source and lens galaxies as done in Eq.~(\ref{eq:wls_def}). 
This inaccuracy is safely absorbed by the prior range of $\sigma(\alpha_{\rm mag})=0.5$ in the parameter inference, because 
$\pm 1\sigma$ changes 
in $\alpha_{\rm mag}$ from its central value lead to $\pm 20$--40\% fractional chaneges in the magnification bias (Eq.~\ref{eq:magnification_bias})
for the three lens samples. 

{\tt Dark Emulator} allows us to compute the model predictions, $\dSigma^{\rm model}(R)$, for an input model in a few CPU seconds. This is fast enough to enable cosmological parameter inference in a high-dimensional parameter space (25 parameters for our baseline setup). In \citet{2021arXiv210100113M}, they validated that this fiducial model template has sufficient accuracy to recover the input $S_8$ to within $0.5\sigma$ through a suite of tests using mock signals with varying galaxy properties, such as different HOD implementations, different satellite distributions within halos, central galaxies with off-centering effect (except for extreme cases), central galaxies 
with incompleteness effect, and baryonic effects on the matter distribution. 

\subsubsection{Projected auto-correlation function: $\wgg(R)$}
\label{sec:wgg_theory}
As shown in Eq.~(\ref{eq:wp_def}), we must first compute the three-dimensional correlation function of galaxies for a given set of the model parameters to obtain the model templates for $\wgg(R)$. The three-dimensional correlation function $\xi_{\rm gg}$ is given as
\begin{align}
\xi_{\rm gg}(r;z_{\rm l})=\int_0^\infty\!\frac{k^2\mathrm{d}k}{2\pi^2}~P_{\rm gg}(k;z_{\rm l})j_0(kr),
\label{eq:xi_gg}
\end{align}
where $j_0(x)$ is the zero-th order spherical Bessel function, and $P_{\rm gg}(k)$ is the auto-power spectrum of galaxies. Once the power spectrum $P_{\rm gg}(k)$ is given for an input of model parameters, we can compute the model prediction of $\wgg(R)$ according to Eq.~(\ref{eq:wp_def}). 

In the halo model, $P_{\rm gg}$ can be divided into two contributions, i.e., the 1- and 2-halo terms, as
\begin{align}
P_{\rm gg}(k)=P_{\rm gg}^{\rm 1h}(k)+P_{\rm gg}^{\rm 2h}(k),
\end{align}
where the 1-halo term describes correlations between galaxies within the same host halo, and the 2-halo term describes correlations between galaxies 
residing in different halos. In our method, we compute the auto-power spectrum as
\begin{widetext}
\begin{align}
P^{\rm 1h}_{\rm gg}(k)&= \frac{1}{\bar{n}_{\rm g}^2}
\int\!\mathrm{d}M~\frac{\mathrm{d}n_{\rm h}}{\mathrm{d}M}\avrg{N_{\rm c}}\!(M)
\left[
2\lambda_{\rm s}(M)\tilde{u}_{\rm s}(k;M)+\lambda_{\rm s}(M)^2\tilde{u}_{\rm s}(k;M)^2\right], \nonumber\\
P^{\rm 2h}_{\rm gg}(k)&= \frac{1}{\bar{n}_{\rm g}^2}
\left[\int\!\mathrm{d}M~\frac{\mathrm{d}n_{\rm h}}{\mathrm{d}M}
\avrg{N_{\rm c}}\!(M)\left\{1+\lambda_{\rm s}(M)\tilde{u}_{\rm s}(k;M)
\right\}
\right]\nonumber\\
&\hspace{10em}\times
\left[\int\!\mathrm{d}M'~\frac{\mathrm{d}n_{\rm h}}{\mathrm{d}M'}
\avrg{N_{\rm c}}\!(M')\left\{1+\lambda_{\rm s}(M')\tilde{u}_{\rm s}(k;M')
\right\}
\right]P_{\rm hh}(k;M,M').
\label{eq:Pgg_1h2h}
\end{align}
\end{widetext}
Here we use {\tt Dark Emulator} to compute $\mathrm{d}n_{\rm h}/\mathrm{d}M$ and $P_{\rm hh}(k;M,M')$, the power spectrum between halos with masses $M$ and $M'$ for an input cosmological model. Note that in our fiducial model we assume that satellite galaxies reside in halos that host a central galaxy in our sample. In \citet{2021arXiv210100113M}, they confirmed that fitting the model to mock observables computed for the case that satellite galaxies are populated in halos irrespective of whether the halos host central galaxies in the sample resulted in a negligible shift in $S_8$, for our baseline analysis setup (see below). 

Then we project the computed $\xi_{\rm gg}(r)$ over $\Pi=[0,\pi_{\rm max}]$ to obtain $\wgg(R)$ for each input model, where we employ $\pi_{\rm max}=100\,h^{-1}{\rm Mpc}$ as used in the measurement. We include the residual RSD effect in the $\wgg(R)$ prediction using the method in \citet{vandenBosch:2013} for each input cosmological model \citep[also see][for details]{2021arXiv210100113M}.

For each input model, {\tt Dark Emulator} allows us to compute the model prediction $\wgg(R)$ in $\sim30$ CPU seconds. 
\begin{table}
\caption{Model parameters and priors used in our cosmological inference. The label ``flat'' denotes a flat prior with the range given, while ``Gauss($\mu,\sigma$)'' is a Gaussian prior with mean $\mu$ and width $\sigma$. The parameters above the horizontal double lines are the parameters used in our fiducial analysis: 5 cosmological parameters, 5 HOD parameters for each of the LOWZ, CMASS1 and CMASS2 samples, 2 nuisance parameters to model residual photo-$z$ and multiplicative shear biases, and 3 parameters ($\alpha_{\rm mag}$) to model the magnitude slope of the galaxy number counts that characterizes the magnification bias effect on $\dSigma$ for each of the LOWZ, CMASS1 and CMASS2 samples: $25=5+3\times 5+2+3$ in total. The parameters below the double lines are used in the extended models.}
\label{tab:parameters}
\begin{center}
\begin{tabular}{cc}  \hline\hline
Parameter & Prior \\ \hline
\multicolumn{2}{c}{{\bf Cosmological parameters}}\\
$\Omega_{\rm de}$ & flat$(0.4594, 0.9094)$\\
$\ln(10^{10}A_{\rm s})$ & flat$(1, 5)$\\
$\omega_{\rm b}$ & Gauss$(0.02268, 0.00038)$\\
$\omega_{\rm c}$ & flat$(0.0998, 0.1398)$\\ 
$n_{\rm s}$ & Gauss$(0.9649, 0.0126)$\\ \hline
\multicolumn{2}{c}{{\bf HOD parameters}} \\
$\log M_{\rm min}(z_i)$ & flat$(12.0,14.5)$ \\ 
$\sigma_{\log M}^2(z_i)$ & flat$(0.01,1.0)$ \\ 
$\log M_1(z_i)$ & flat$(12.0,16.0)$\\
$\kappa(z_i)$ & flat$(0.01,3.0)$\\ 
$\alpha(z_i)$ & flat$(0.5,3.0)$\\ \hline
\multicolumn{2}{c}{\bf Magnification bias}\\
$\alpha_{\rm mag}$ (LOWZ) & Gauss: (2.26,0.5) \\
$\alpha_{\rm mag}$ (CMASS1) & Gauss: (3.56,0.5) \\
$\alpha_{\rm mag}$ (CMASS2) & Gauss: (3.73,0.5) \\ \hline
\multicolumn{2}{c}{\bf Photo-$z$/Shear errors}\\
$\Delta z_{\rm ph}$ & Gauss: $(0.0,0.1)$  \\
$\Delta m_\gamma$ & Gauss: $(0.0,0.01)$ \\ \hline\hline
\multicolumn{2}{c}{\bf Add. galaxy-halo connection paras} \\ 
\multicolumn{2}{c}{Off-centering parameters} \\
$p_{\rm off}(z_i)$ & flat$(0,1)$ \\
$R_{\rm off}(z_i)$ & flat$(0.01,1)$ \\
\multicolumn{2}{c}{Incompleteness parameters} \\
$\alpha_{\rm incomp}(z_i)$ & flat$(0,5)$\\ 
$\log M_{\rm incomp}(z_i)$ & flat$(12,15.3)$\\ \hline\hline
\end{tabular}
\end{center}
\end{table}

\subsection{Parameter estimation method}
\label{sec:parameter_estimaton_method}
We assume that the likelihood of the data compared to the model predictions follows a multivariate Gaussian distribution: 
\begin{align}
\ln{\cal L}({\bf d}|\boldsymbol{\theta})=-\frac{1}{2}\sum_{i,j} \left[
d_i - t_i(\boldsymbol{\theta})\right]
{\bf C}^{-1}_{ij} 
\left[
d_j - t_j(\boldsymbol{\theta})\right],
\label{eq:likelihood}
\end{align}
where ${\bf d}$ is the data vector, ${\bf t}$ is the model prediction for the data vector given the model parameters ($\boldsymbol{\theta}$), ${\bf C}^{-1}$ is the inverse of the covariance matrix, and the summation runs over indices corresponding to the dimension of the data vector. In our baseline analysis, the data vector consists of $\dSigma\!(R)$ in 9 logarithmically-spaced radial bins  within $3\le R/[h^{-1}{\rm Mpc}]\le 30$, and $\wgg\!(R)$ in 16 radial bins within $2\le R/[h^{-1}{\rm Mpc}]\le 30$, for each galaxy sample. We therefore use $75(=3\times (9+16))$ data points in total. When we use the data vector in a more limited range of separations, we take the submatrix of the full covariance matrix corresponding to that range of separations, and then invert the matrix to obtain the inverse of the covariance matrix, $[{\bf C}_{\rm sub}]^{-1}$. We also note that our analysis does not include the observed galaxy abundance, $\bar{n}_{\rm g}$.

For the model parameters in our baseline analysis, we include 5 cosmological parameters given by ${\boldsymbol{\theta}}_{\rm cosmo}=\{\Omega_{\rm de},\ln(10^{10}A_{\rm s}),\omega_{\rm b}, \omega_{\rm c},n_{\rm s}\}$ for the flat $\Lambda$CDM framework and 5 HOD parameters for each of the LOWZ, CMASS1, and CMASS2 samples. For $\omega_b$, we employ  a Gaussian prior with a mean and width inferred from Big Bang nucleosynthesis (BBN) experiments \cite{2020A&A...641A...6P, Aver:2015iza, Cooke:2017cwo, Schoneberg:2019wmt}. For $n_{\rm s}$, we employ a Gaussian prior inferred from the {\it Planck} 2018 ``TT,EE,TE+lowE'' constraints \cite{2020A&A...641A...6P}: $n_{\rm s}=0.9649\pm (3\times 0.0042)$, where we employ the Gaussian width three times wider than the $1\sigma$ uncertainty ($0.0042$) of the {\it Planck} constraint. We employ these priors since the clustering observables $\dSigma$ and $w_{\rm p}$ are not sensitive to $\omega_b$ and $n_{\rm s}$. For $\Omega_{\rm de}$ and $\omega_{\rm c}$, we adopt broad, flat priors in the ranges that correspond to about $\pm 30\sigma$ and $\pm 15\sigma$, respectively, compared to the $1\sigma$ error of the {\it Planck} constraints for flat $\Lambda$CDM model. These ranges correspond to the supported range of the extrapolation of {\tt Dark Emulator} (for details see Section~\ref{sec:dark_emulator}). Since there is no limitation on $\ln{(10^{10} A_{\rm s})}$ in the extrapolation, we employ a broad and uninformative flat prior. 

In addition we include $\alpha_{\rm mag}(z_i)$ to model a possible uncertainty in the magnitude slope of the number counts in modeling the magnification for each lens sample: we use the measured value of $\alpha_{\rm mag}$ for the central value (see Section~\ref{sec:gglensing_estimator} and Fig.~\ref{fig:magnification}) and employ a Gaussian prior with a width of $\sigma(\alpha_{\rm mag})=0.5$. This is a conservative choice, since the Gaussian width is much wider than the measurement error on $\alpha_{\rm mag}$, but we show that the results remain almost unchanged when fixing $\alpha_{\rm mag}$ to the measured value. 

Furthermore, we include nuisance parameters, $\Delta z_{\rm ph}$ and $\Delta m_\gamma$, to model residual uncertainties in the source photo-$z$ biases and the multiplicative shear bias. Since we use a single population of source galaxies, we need to adopt just one $\Delta z_{\rm ph}$ and one $\Delta m_\gamma$ parameter to model the impact on the galaxy-galaxy weak lensing signals for all three lens galaxy samples. We use a conservative prior range on $\Delta z_{\rm ph}$ that is wider than that used in \citet{2019PASJ...71...43H} and \citet{2020PASJ...72...16H}. Their prior range was estimated from the difference between the means of the stacked photo-$z$ posterior distributions for different photo-$z$ methods and for the reweighted COSMOS redshift distribution. Our broader prior range allows us to marginalize over possible residual photo-$z$ systematics that may not be captured by the prior range employed based on the COSMOS reweighting method. For $\Delta m_\gamma$, we employ a prior range that corresponds to about $1\sigma$ statistical uncertainties in the shape measurement calibration \cite{2018MNRAS.481.3170M} \citep[also see Table~6 in Ref.][]{2019PASJ...71...43H}. We will discuss the case where the prior range of $\Delta m_\gamma$ is broadened in Section~\ref{sec:lcdm}. We have 5 nuisance parameters of the systematic effects in total: $\alpha_{\rm mag}(z_i)$, $\Delta z_{\rm ph}$ and $\Delta m_\gamma$. Hence we have 25($=5+3\times 5+5$) parameters in total, as summarized in Table~\ref{tab:parameters}.

We then obtain the posterior distribution of our parameters given the data by performing Bayesian inference:
\begin{align}
{\cal P}(\boldsymbol{\theta}|{\bf d})&\propto {\cal L}({\bf d}|\boldsymbol{\theta})\Pi(\boldsymbol{\theta}),
\label{eq:bayes}
\end{align}
where ${\cal P}(\boldsymbol{\theta}|{\bf d})$ is the posterior distribution of $\boldsymbol{\theta}$ given the data vector (${\bf d}$) and $\Pi(\boldsymbol{\theta})$ is the prior distribution. Throughout this paper we focus on the marginalized posterior distributions of the derived parameters, $\Omega_{\rm m}$, $\sigma_8$, and $S_8\equiv \sigma_8(\Omega_{\rm m}/0.3)^{0.5}$,
where $\Omega_{\rm m} = 1-\Omega_{\rm de}$ for a flat cosmological model. While $\ln{(10^{10} A_{\rm s})}$ is sampled in logarithmic space with a flat prior, we account for the Jacobian or weight to effectively produce a flat prior in linear space of $\sigma_8$ when obtaining the posterior distribution of $\sigma_8$ as a derived parameter (see Section~IV~A in \citet{Sugiyama:2020} for a detailed discussion). However, the effect is negligible because the Jacobian is nearly constant in the range of the credible interval of $\sigma_8$ in our constraints. 

To estimate the posterior distribution of parameters in a multi-dimensional parameter space, we use the importance nested sampling algorithm implemented in the publicly-available software package {\tt MultiNest} \cite{Feroz:2008,Feroz:2009,Feroz:2019} through the package {\tt Monte Python} \cite{Audren:2012, Brinckmann:2018}. We set the sampling efficiency parameter $\texttt{efr}=0.8$ and the evidence tolerance factor $\texttt{tol}=0.5$ as recommended by the developers. After extensive convergence tests as described in Appendix~\ref{app:convergence_mcmc}, we confirmed that the chains used in our analysis have converged to the desired degree. In this paper, we report the mode of the 1-dimensional or 2-dimensional posterior distributions as the central value(s) of parameter(s), and the highest density interval of the marginalized posterior distribution to infer the credible interval(s) of parameter(s) (see Eq.~\ref{eq:hdi}).
\begin{table*}
\caption{A summary of the analysis setups. The first column identifies each analysis setup. The scale cuts ``$(X,Y)$'' denote the lower scale cuts applied to $\wgg(R)$ and $\dSigma(R)$, meaning that we use $\wgg$ and $\dSigma$ for $X\le R/[h^{-1}{\rm Mpc}]\le 30$ and $Y\le R/[h^{-1}{\rm Mpc}]\le 30$, respectively, in the cosmology analysis. The column ``sample parameters'' lists the model parameters used in each analysis.The setups labeled ``wide shear prior'' and ``wide photo-$z$ prior'' were identified for study {\it after} unblinding our cosmology results, in order to study the robustness of our cosmological parameter constraints to  the adopted prior width for photo-$z$ biases ($\Delta z_{\rm ph}$) or multiplicative shear biases ($\Delta m_\gamma$).
 \label{tab:analysis_setups}}
\begin{center}
\begin{tabular}{l|ccl}\hline\hline
setup & scale cuts & sample parameters \\
& $[h^{-1}{\rm Mpc}]$ &  
\\ \hline 
baseline & $(2,3)$ &
$(\Omega_{\rm de},\ln(10^{10}A_{\rm s}), \omega_{\rm b},\omega_{\rm c}, n_{\rm s}$)+HOD (3$\times$5 paras.)+mag/photo-$z$/shear (5 paras.) & \\ \hline 
scale cuts 
& $(4,6)$ & --  \\
& $(8,12)$ & -- \\
\hline
no LOWZ & (2,3) & $(\Omega_{\rm de},\ln(10^{10}A_{\rm s}), \omega_{\rm b},\omega_{\rm c}, n_{\rm s}$)+HOD (2$\times$5 paras.)+mag/photo-$z$/shear (4 paras.)  \\ 
no CMASS1 & (2,3) & $(\Omega_{\rm de},\ln(10^{10}A_{\rm s}), \omega_{\rm b},\omega_{\rm c}, n_{\rm s}$)+HOD (2$\times$5 paras.)+mag/photo-$z$/shear (4 paras.)  \\ 
no CMASS2 & (2,3) & $(\Omega_{\rm de},\ln(10^{10}A_{\rm s}), \omega_{\rm b},\omega_{\rm c}, n_{\rm s}$)+HOD (2$\times$5 paras.)+mag/photo-$z$/shear (4 paras.)  \\ 
\hline
no shear error & (2,3) & $(\Omega_{\rm de},\ln(10^{10}A_{\rm s}), \omega_{\rm b},\omega_{\rm c}, n_{\rm s}$)+HOD (3$\times$5 paras.)+mag/photo-$z$ (4 paras.)  \\
no photo-$z$ error& (2,3) & $(\Omega_{\rm de},\ln(10^{10}A_{\rm s}), \omega_{\rm b},\omega_{\rm c}, n_{\rm s}$)+HOD (3$\times$5 paras.)+mag/shear (4 paras.) \\ 
fix mag. bias & (2,3) &$(\Omega_{\rm de},\ln(10^{10}A_{\rm s}), \omega_{\rm b},\omega_{\rm c}, n_{\rm s}$)+HOD (3$\times$5 paras.)+photo-$z$/shear (2 paras.) \\ 
\hline
wide shear prior & (2,3) & same as the baseline analysis, with $\sigma(\Delta m_\gamma)=0.1$  \\ 
wide photo-$z$ prior & (2,3) & same as the baseline analysis, with $\sigma(\Delta z_{\rm ph})=0.2$  \\ 
\hline
off-cent. & (2,3) & $(\Omega_{\rm de},\ln(10^{10}A_{\rm s}), \omega_{\rm b},\omega_{\rm c}, n_{\rm s}$)+HOD w/off-centering (3$\times$7 paras.)+photo-$z$/shear (5 paras.) \\ 
incomp. & (2,3) & $(\Omega_{\rm de},\ln(10^{10}A_{\rm s}), \omega_{\rm b},\omega_{\rm c}, n_{\rm s}$)+HOD w/incompleteness (3$\times$7 paras.)+photo-$z$/shear (5 paras.) \\
\hline
diff. photo-$z$ & (2,3) & same sample parameters as the baseline analysis,  but lensing signal computed \\
& &with different photo-$z$ methods ({\tt DEmP}, {\tt Ephor\_ab}, {\tt Franken-Z}, {\tt Mizuki}, and {\tt NNPZ})\\
\hline
2 cosmo & (2,3) & $(\Omega_{\rm de},\ln(10^{10}A_{\rm s}))$+HOD (3$\times$5 paras.)+mag/photo-$z$/shear (5 paras.)\\
\hline\hline
\end{tabular}
\end{center}
\end{table*}
\subsection{Analysis setups}
\label{sec:analysis_setups}

To perform the cosmological parameter inference, we must specify other aspects of the analysis setup, such as the range of separations and  combinations of observables to use. Table~\ref{tab:analysis_setups} summarizes the setups used in this paper. One important choice in the analysis relates to the range of separations in $\dSigma(R)$ and $\wgg(R)$ used in the cosmological analysis, or ``scale cuts''. There are two competing effects. To increase the  statistical constraining power on the cosmological parameters, we want to include information from $\dSigma(R)$ and $\wgg(R)$ down to smaller separations. However, the observables at such small scales may be more strongly affected by physical systematic effects inherent in galaxy formation/physics, which are difficult to accurately model. As we carefully studied in our validation paper \cite{2021arXiv210100113M}, the scale cuts of $(2,3)~h^{-1}{\rm Mpc}$ for $\wgg$ and $\dSigma$ are reasonable choices to obtain unbiased estimates of the cosmological parameters, with reasonably small credible intervals given the statistical power of HSC-Y1 and SDSS. The scale cuts of $(2,3)\,h^{-1}{\rm Mpc}$ are larger than the virial radii of massive halos, so we do not include information from scales that are deeply in the 1-halo term regime in our cosmology analysis. Nevertheless we note that the galaxy-galaxy weak lensing signal around the scale cut is sensitive to the {\it interior} mass inside that radius, which allows us to extract the average mass of halos hosting the SDSS galaxies and in turn helps constrain the large-scale bias of SDSS galaxies via the scaling relation of halo bias with halo mass, encoded in {\tt Dark Emulator}, when combined with the measurement of $\wgg$. To study the impact of the scale cut choice, we also study the results for $(4,6)$ and ($8,12$)$~h^{-1}{\rm Mpc}$. 

If we use either $\dSigma$ or $\wgg$ alone, the parameter inference suffers from severe degeneracies, especially between the galaxy bias (and therefore the HOD model parameters) and the cosmological parameters that encode information about the power spectrum amplitude, as shown in our validation paper \cite{2021arXiv210100113M} (see Fig.~9 in their paper). Hence, in the following we show only the results of the joint analysis of $\dSigma$ and $\wgg$.

As an internal consistency test, we also perform the analyses excluding some information from the baseline setup: excluding one of the LOWZ, CMASS1 or CMASS2 samples, or either of the residual systematic error parameters, $\Delta z_{\rm ph}$ or $\Delta m_\gamma$ (see Table~\ref{tab:parameters}). We also show the results for extended models that include the effects of off-centered central galaxies or the incompleteness effect of central galaxies. For both extended models, we introduce two additional model parameters, as indicated in the rows  ``{off-cent.}'' or ``{incomp.}'' in Table~\ref{tab:parameters}.

To check for possible systematic biases arising from photo-$z$ estimates, we perform cosmological inference using the lensing signals computed with photo-$z$ methods other than the one used in the baseline analysis, referred to as the ``diff. photo-$z$'' setup in Table~\ref{tab:analysis_setups}.

In addition, after unblinding our cosmology results (see next section), we further decided to introduce the setups labeled  ``wide shear prior'' and ``wide photo-$z$ prior'' in Table~\ref{tab:analysis_setups}. For these, we employ significantly wider Gaussian priors, $\sigma(\Delta m_\gamma)=0.1$ or $\sigma(\Delta z_{\rm ph})=0.2$, in the parameter inference. The purpose of these additional setups is to study the impact of the prior width on the cosmological parameters and to explore the possibility of self-calibration of these nuisance parameters.

After unblinding, we also perform an analysis with cosmological parameters other than $(\Omega_{\rm m}, \ln(10^{10}A_{\rm s}))$ fixed to the {\it Planck} 2015 ``TT,TE,EE+lowP'' constraints \cite{PlanckCosmology:16} to check how the parameters that are not well constrained by our data vector affect our cosmological constraints.  This setup is labeled ``2 cosmo'' in Table~\ref{tab:analysis_setups}.

\section{Blinding scheme and Validation}
\label{sec:blinding}
To avoid confirmation bias we perform our cosmological analysis in a blind fashion. 
The details of the blinding scheme can be found in Section~3.2 of \citet{2019PASJ...71...43H}. We employ a two-tier blinding strategy to avoid unintentional
unblinding during the cosmological analysis. The two tiers are as follows:
\begin{itemize}
\item {\it Catalog level}: The analysis team performs the cosmological analysis using three different weak lensing shape catalogs.  Only one is the true catalog and the other two are fake catalogs (see below for details). The analysis team members do not know which is the true catalog. 
\item {\it Analysis level}: The analysis team does not make plots comparing the measurements with  theoretical models. When the analysis team makes plots showing the credible intervals of cosmological parameters (i.e. the posterior distribution), the central value(s) of parameter(s) are shifted to zero, and only the range of the credible interval(s) can be seen. Finally, the analysis team does not compare the posterior for cosmological parameter(s) or the model predictions with external results such as the {\it Planck} CMB cosmology prior to unblinding. 
\end{itemize}

See Section~3.2 in \citet{2019PASJ...71...43H} for details of how the fake catalogs were constructed in a manner that prevents accidental unblinding by the analysis team.  Use of these catalogs means that the analysis group must perform three analyses, but this method avoids the need for reanalysis once the catalogs are unblinded.

Validation of the cosmological analysis method is demonstrated in \citet{2021arXiv210100113M}. The analysis team promised that the results would be published regardless of the outcome, once the results are unblinded. In addition, the analysis method could not be changed or modified after unblinding. In the following we explicitly flag results obtained after unblinding. 

\section{Results: Cosmological Constraints}
\label{sec:results}
In this section we show the main results of this paper, which are the cosmological parameters estimated 
from the joint measurements of $\dSigma(R)$ and $\wgg(R)$ in the HSC-Y1 and SDSS datasets. 
\begin{figure*}
\begin{center}
\includegraphics[width=0.7\textwidth]{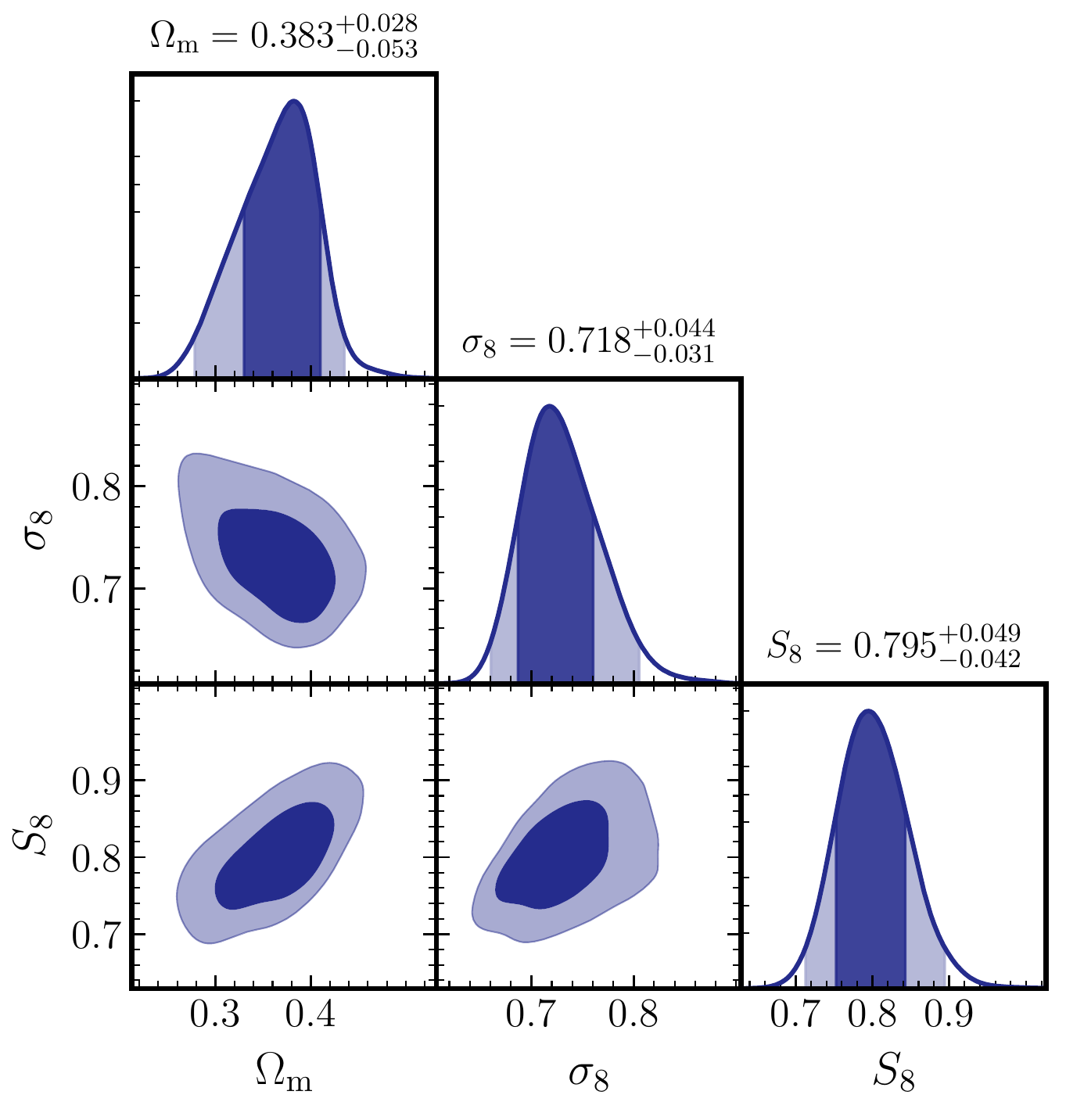}
\end{center}
\caption{The 1-d and 2-d posterior distributions in the sub-space of $S_8$, $\sigma_8$, and $\Omega_{\rm m}$ for the flat $\Lambda$CDM cosmology, obtained from our baseline analysis setup using the lensing signals with $R/[\hiMpc]=[3,30]$ and clustering signals with $R/[\hiMpc]=[2,30]$. The dark (light) shaded regions show the 68\% (95\%) credible intervals, including marginalization over uncertainties in the other parameters. The modes and 68\% credible intervals of each parameter are shown above each panel of the 1-d posterior distributions.}
\label{fig:contour_baseline}
\end{figure*}
\begin{figure*}
\begin{center}
\includegraphics[width=\textwidth]{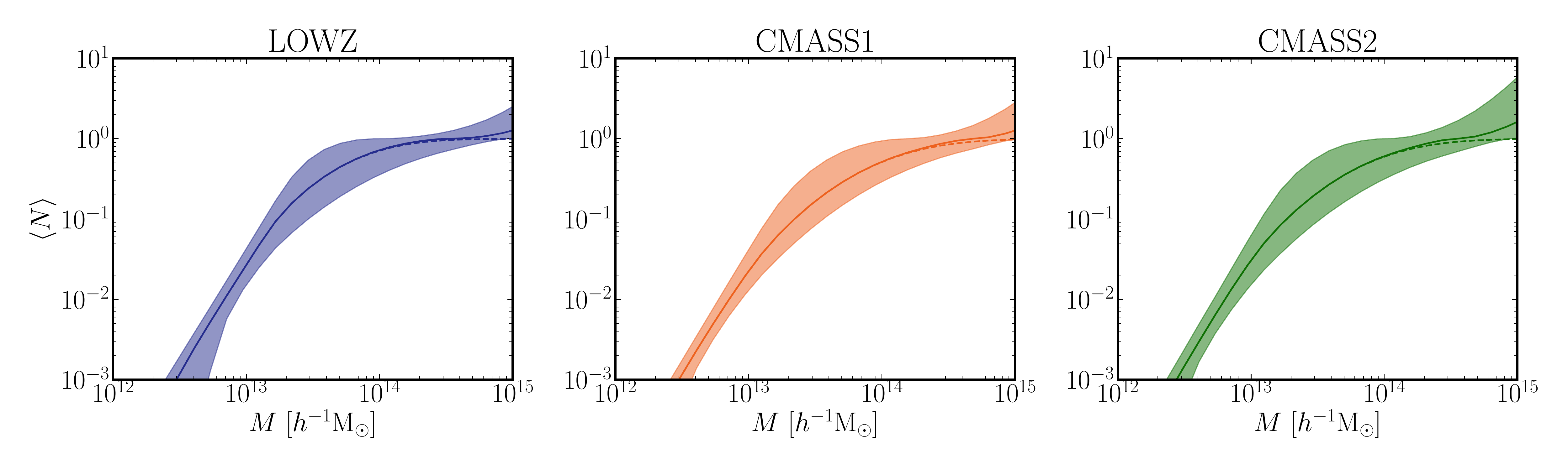}
\caption{The median and the range from the 16th to the 84th percentile of the HOD in each halo mass bin for the LOWZ, CMASS1 and CMASS2 samples. These are computed from the posterior distribution of the model predictions, $\langle N \rangle (M)$, in the chains of the baseline analysis, marginalizing over uncertainties in the cosmological parameters and other model parameters. The solid (dashed) lines show the median for the central+satellite HOD (central HOD), and the shaded region displays the percentile range. Note that, for this figure, Fig.~\ref{fig:cross_correlation_coefficient}, and Table~\ref{tab:halo_properties}, we use the median and percentile to show the range of the model predictions in each bin, because the posterior distribution of a quantity under consideration has a non-Gaussian distribution and its mode and highest density interval are difficult to reliably estimate (while the median and percentile are more stable).}
\label{fig:hod}
\end{center}
\end{figure*}
\subsection{$\Lambda$CDM Constraints}
\label{sec:lcdm}
\begin{figure*}
\begin{center}
\includegraphics[width=\textwidth]{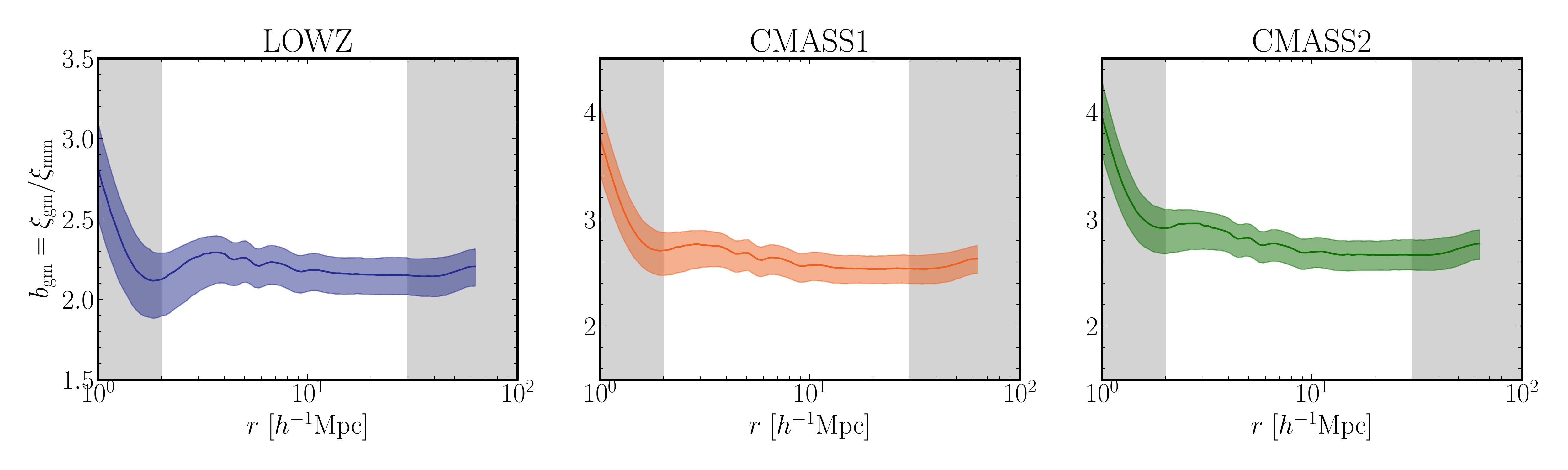}
\includegraphics[width=\textwidth]{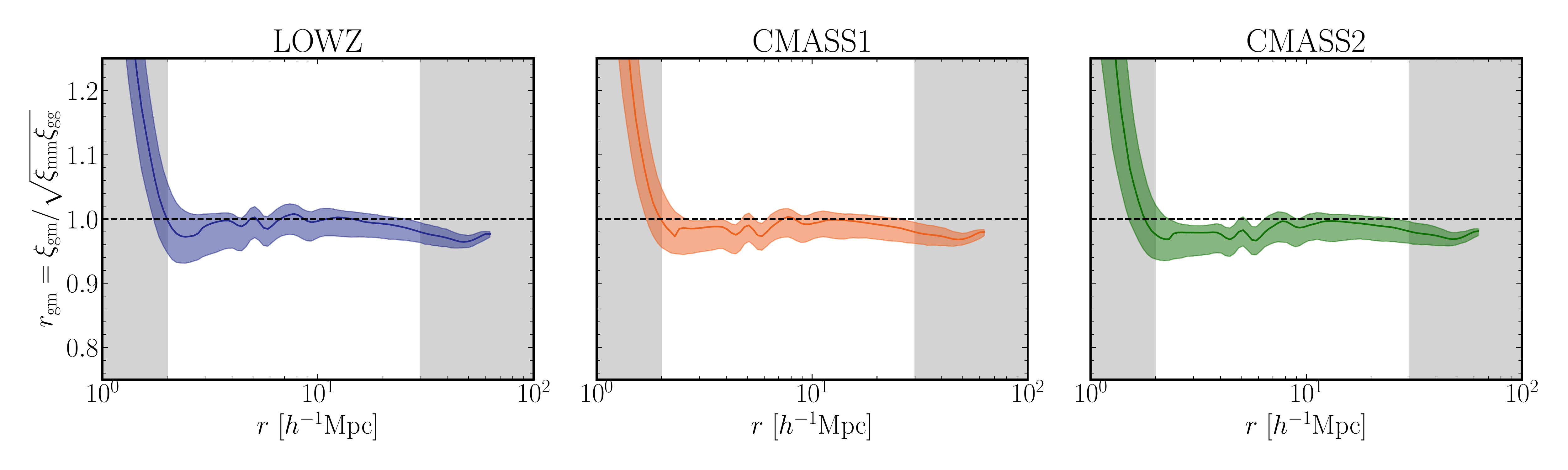}
\caption{{\it Upper panels}: The galaxy bias, defined by $b_{\rm gm}(r)\equiv \xi_{\rm gm}(r)/\xi_{\rm mm}(r)$. The line and shaded region denote the median and the range from the 16th to the 84th percentile of the posterior distribution of the galaxy bias, respectively, for the LOWZ, CMASS1 and CMASS2 samples, as obtained from the chains of the baseline analysis. The unshaded region in each panel shows the range of separations used for the cosmological analysis. {\it Lower panels}: The cross-correlation coefficient, defined by $r_{\rm gm}(r)\equiv \xi_{\rm gm}(r)/[\xi_{\rm gg}(r)\xi_{\rm mm}(r)]^{1/2}$. The horizontal dashed line denotes $r_{\rm gm}(r)=1$ for comparison. \label{fig:cross_correlation_coefficient}}
\end{center}
\end{figure*}
\begin{figure}
\begin{center}
\includegraphics[width=\columnwidth]{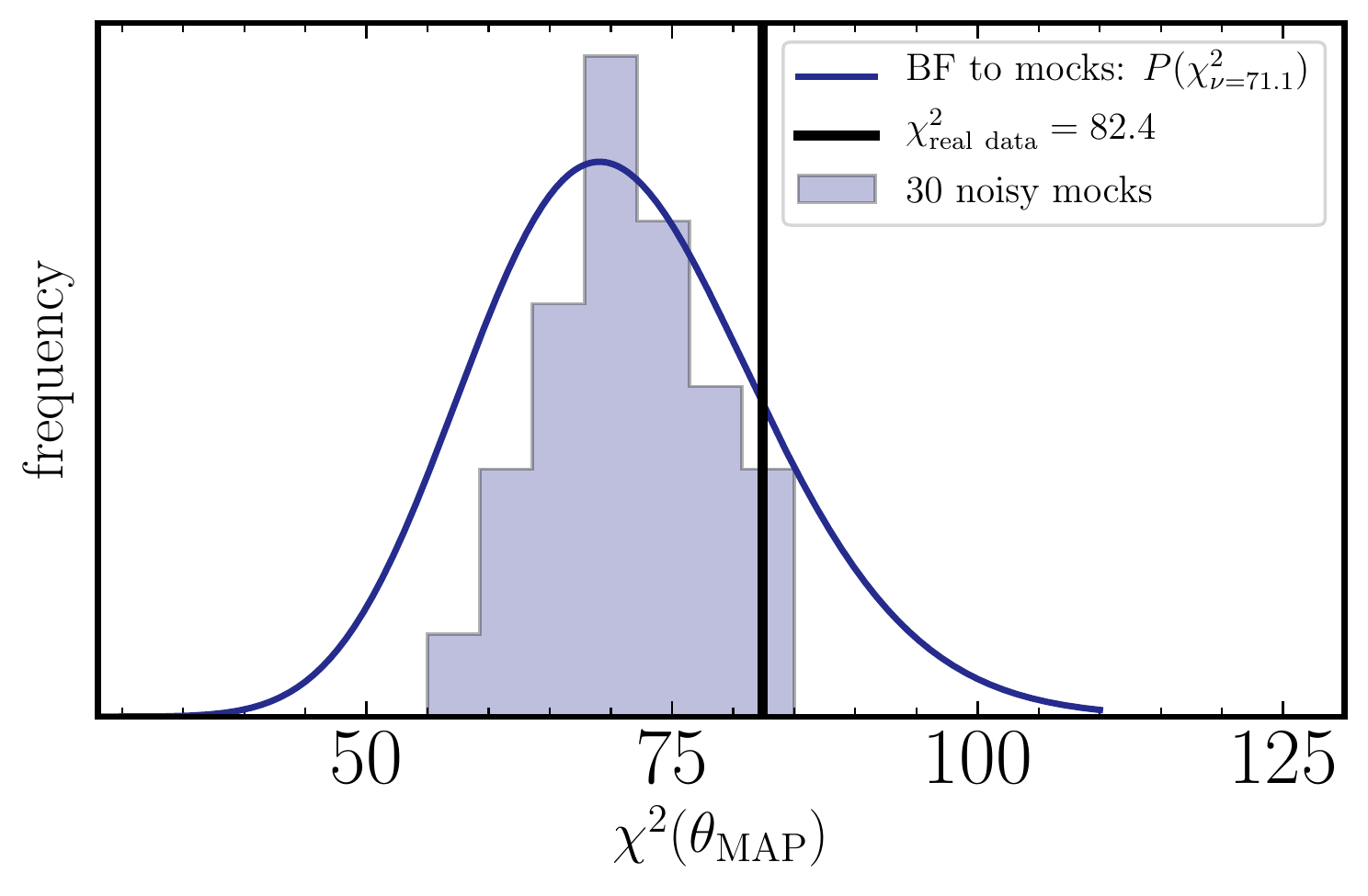}
\end{center}
\caption{An evaluation of the  goodness-of-fit of the best-fit model at {\it maximum a posteriori} (MAP) for the baseline analysis shown in Fig.~\ref{fig:contour_baseline}. The histogram shows the distribution of the $\chi^2$ values of the model at MAP, obtained by applying the same baseline analysis to 30 noisy mock datasets (see text for details). The blue line is the best-fit $\chi^2$ distribution, characterized by the degrees of freedom $\nu=71.1$ estimated from the $\chi^2$ values for 30 noisy mocks data. The vertical black line denotes the $\chi^2$ value ($\chi^2=82.4$) at MAP for the analysis of the HSC-Y1 and SDSS data.}
\label{fig:chi2}
\end{figure}
\begin{figure}
\begin{center}
\includegraphics[width=\columnwidth]{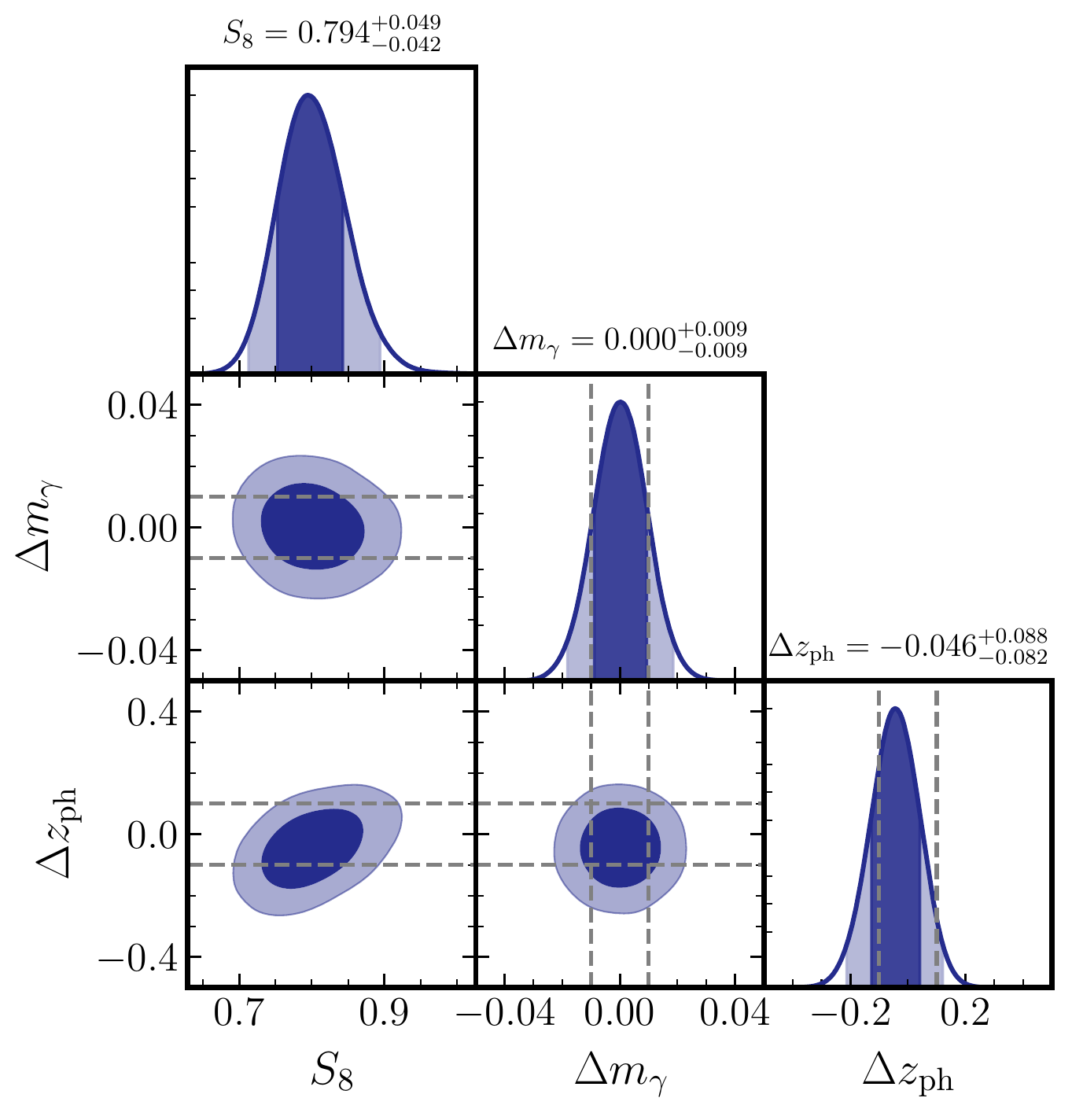}
\end{center}
\caption{Similar to Fig.~\ref{fig:contour_baseline}, but this figure shows the 1-d and 2-d posterior distributions in a subspace of $S_8$ and two nuisance parameters, $\Delta m_\gamma$ and $\Delta z_{\rm ph}$, which model possible residual systematic biases in the multiplicative shear calibration factor and the photometric redshifts for the HSC galaxies used as source galaxies in the galaxy-galaxy weak lensing measurements. The vertical dashed lines in the 1-d posterior distribution of $\Delta m_\gamma$ or $\Delta z_{\rm ph}$ denote the width of the Gaussian prior on the parameter.}
\label{fig:contour_nuisance}
\end{figure}
\begin{figure*}
\begin{center}
\includegraphics[width=2.3\columnwidth]{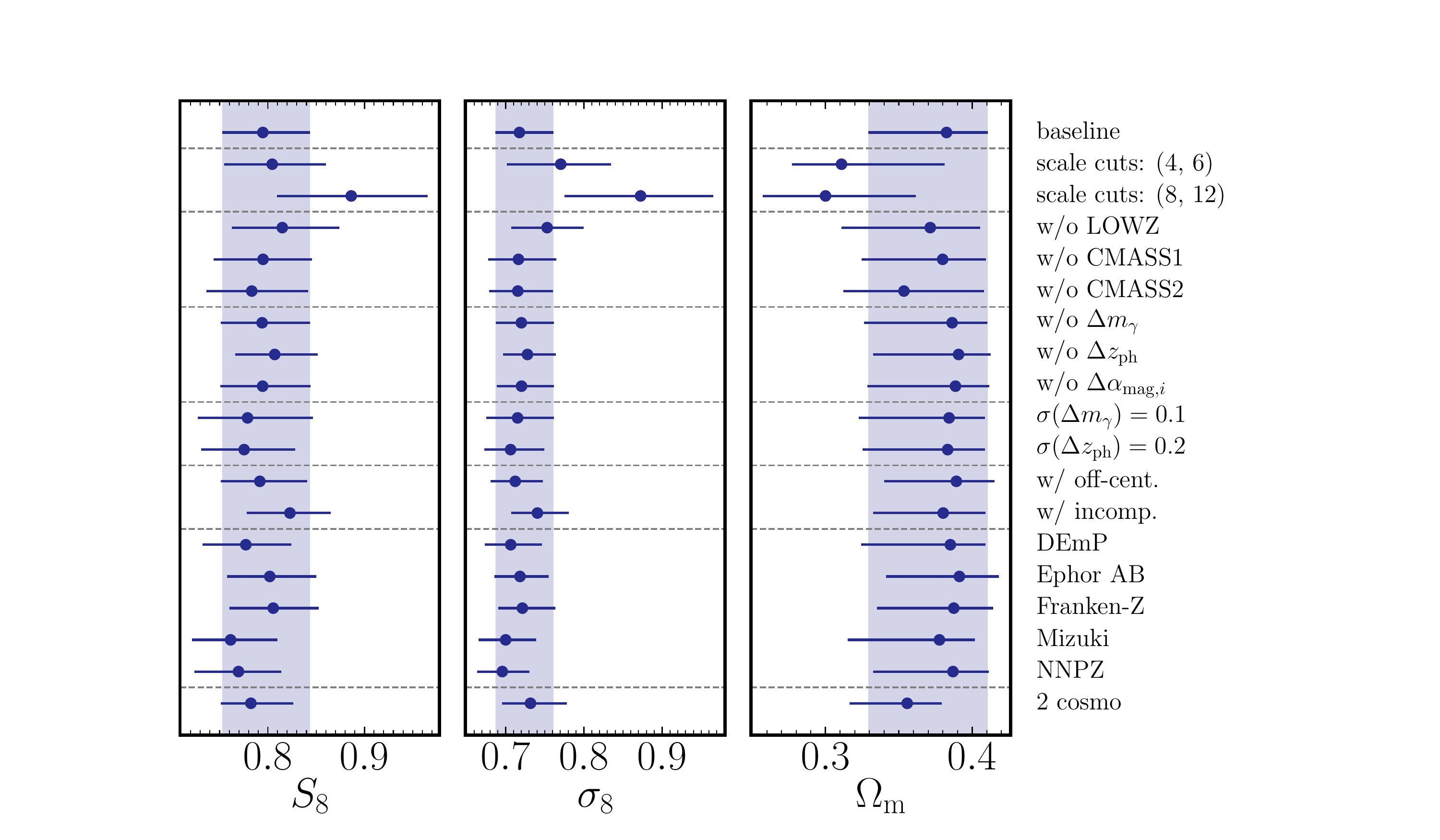}
\end{center}
\caption{A summary of the cosmological parameters, $S_8$, $\sigma_8$ and $\Omega_{\rm m}$, obtained from each analysis setup in Table~\ref{tab:analysis_setups}. Each circle and error bar denotes the mode and 68\% credible interval for one setup, marginalized over the other parameter for that setup.}
\label{fig:summary}
\end{figure*}
{\renewcommand{\arraystretch}{1.4} 
\begin{table*}
\caption{Summary of cosmological constraints with each analysis setup in Table~\ref{tab:analysis_setups}.}
\label{tab:summary}
\begin{center}
\begin{tabular}{l|ccc}
\hline\hline
setup & $S_8 = \sigma_8(\Omega_{\rm m}/0.3)^{0.5}$ & $\sigma_8$ & $\Omega_{\rm m}$\\
\hline
baseline & $0.795^{+0.049}_{-0.042}$ & $0.718^{+0.044}_{-0.031}$ & $0.383^{+0.028}_{-0.053}$\\
\hline
scale cuts: (4, 6) & $0.805^{+0.056}_{-0.050}$ & $0.770^{+0.065}_{-0.069}$ & $0.311^{+0.070}_{-0.034}$\\
scale cuts: (8, 12) & $0.886^{+0.079}_{-0.077}$ & $0.873^{+0.093}_{-0.097}$ & $0.300^{+0.062}_{-0.043}$\\
\hline
w/o LOWZ & $0.815^{+0.059}_{-0.052}$ & $0.753^{+0.047}_{-0.046}$ & $0.371^{+0.034}_{-0.061}$\\
w/o CMASS1 & $0.795^{+0.051}_{-0.051}$ & $0.716^{+0.049}_{-0.039}$ & $0.380^{+0.030}_{-0.055}$\\
w/o CMASS2 & $0.783^{+0.059}_{-0.047}$ & $0.716^{+0.045}_{-0.037}$ & $0.354^{+0.055}_{-0.041}$\\
\hline
w/o $\Delta m_\gamma$ & $0.794^{+0.050}_{-0.043}$ & $0.720^{+0.041}_{-0.033}$ & $0.386^{+0.024}_{-0.060}$\\
w/o $\Delta z_{\rm ph}$ & $0.807^{+0.044}_{-0.041}$ & $0.728^{+0.036}_{-0.032}$ & $0.391^{+0.022}_{-0.058}$\\
w/o $\Delta \alpha_{{\rm mag},i}$ & $0.795^{+0.049}_{-0.044}$ & $0.720^{+0.041}_{-0.032}$ & $0.389^{+0.023}_{-0.060}$\\
\hline
$\sigma(\Delta m_\gamma)=0.1$ & $0.779^{+0.068}_{-0.052}$ & $0.715^{+0.046}_{-0.040}$ & $0.384^{+0.025}_{-0.062}$\\
$\sigma(\Delta z_{\rm ph})=0.2$ & $0.775^{+0.053}_{-0.045}$ & $0.706^{+0.043}_{-0.034}$ & $0.383^{+0.026}_{-0.058}$\\
\hline
w/ off-cent. & $0.792^{+0.049}_{-0.040}$ & $0.712^{+0.035}_{-0.031}$ & $0.389^{+0.026}_{-0.049}$\\
w/ incomp. & $0.823^{+0.042}_{-0.045}$ & $0.741^{+0.040}_{-0.034}$ & $0.380^{+0.029}_{-0.048}$\\
\hline
DEmP & $0.777^{+0.047}_{-0.045}$ & $0.707^{+0.040}_{-0.033}$ & $0.385^{+0.024}_{-0.061}$\\
Ephor AB & $0.802^{+0.048}_{-0.044}$ & $0.718^{+0.036}_{-0.033}$ & $0.391^{+0.027}_{-0.050}$\\
Franken-Z & $0.806^{+0.047}_{-0.045}$ & $0.721^{+0.042}_{-0.031}$ & $0.388^{+0.027}_{-0.052}$\\
Mizuki & $0.761^{+0.049}_{-0.040}$ & $0.700^{+0.039}_{-0.035}$ & $0.378^{+0.024}_{-0.063}$\\
NNPZ & $0.770^{+0.044}_{-0.046}$ & $0.696^{+0.034}_{-0.032}$ & $0.387^{+0.024}_{-0.054}$\\
\hline
2 cosmo & $0.782^{+0.044}_{-0.031}$ & $0.732^{+0.047}_{-0.036}$ & $0.356^{+0.024}_{-0.039}$\\
\hline
\end{tabular}
\end{center}
\end{table*}
}
We show the cosmological parameter constraints for the flat $\Lambda$CDM model, which is the minimum theoretical framework that fairly well reproduces a broad range of cosmological observations. In particular, we will focus on the cosmological parameters $\Omega_{\rm m}$, $\sigma_8$ and $S_8\equiv \sigma_8(\Omega_{\rm m}/0.3)^{0.5}$, which are well-constrained by our measurements. Fig.~\ref{fig:contour_baseline} shows the posterior distributions for $(\Omega_{\rm m},\sigma_8,S_8)$, obtained from the baseline setup in Table~\ref{tab:analysis_setups}. Our results for the cosmological constraints are
\begin{eqnarray}
\Omega_{\rm m}&=&\Omegam \nonumber\\
\sigma_8&=&\sigmaeight \nonumber\\
S_8&=&\Seight . 
\end{eqnarray}
Thus the HSC-Y1 data, combined with the SDSS dataset, can constrain $S_8$ to about 6\% fractional precision. Note that we show the posterior distribution in the full parameter space in Fig.~\ref{fig:full_corner_basline} in Appendix~\ref{app:full_posterior}. 

We check the convergence of our nested sampling results as described in detail in Appendix~\ref{app:convergence_mcmc}. We have confirmed that our run terminates at the point where the relative posterior mass is sufficiently small, and the standard deviation of the mode values of the 1-d projected $S_8$ posterior distributions is $0.0008$ estimated from 4 independent chains. This is about $2$\% of the statistical error in $S_8$.

The best-fit model predictions at {\it maximum a posteriori} (MAP) are shown in Fig.~\ref{fig:lens_signal}, together with the measured signals for each galaxy sample. It is clear that the best-fit model fairly well reproduces the measured signals. 

Fig.~\ref{fig:hod} shows the HODs estimated for each sample. Since we imposed a luminosity cut, the HODs reach unity at a relatively high mass, around $M_{\rm h}\sim10^{14}\ h^{-1}{M}_\odot$. Such high-mass halos host satellite galaxies. This trend is reflected by the high mean halo mass and low satellite fraction given in Table~\ref{tab:halo_properties}, where the mean halo mass and satellite fraction are defined as
\begin{equation}
\langle M_h \rangle=\frac{\int\!\mathrm{d}M\frac{\mathrm{d}n_{\rm h}}{\mathrm{d}M}\langle N_{\rm c}\rangle(M) M}
{\int\!\mathrm{d}M\frac{\mathrm{d}n_{\rm h}}{\mathrm{d}M}\langle N_{\rm c}\rangle(M)}.
\end{equation}
and
\begin{equation}
f_{\rm sat} = \frac{1}{\bar{n}_{\rm g}}\int\!\mathrm{d}M\frac{\mathrm{d}n_{\rm h}}{\mathrm{d}M}\langle N_{\rm c}\rangle(M)\lambda_{\rm s}(M),
\end{equation}
respectively. From the HODs in the chains, we compute predictions for the abundance of each sample, which are consistent with the measured abundances shown in Fig.~\ref{fig:boss_selection}. However, Table~\ref{tab:halo_properties} shows that the number density of each sample is poorly constrained, only by within a factor of 2. This reflects the fact that we did not use abundance information and we employed broad priors for each HOD parameter in our parameter inference. The abundance information could add  significant constraining power in principle if it is reliably used. In other words, our cosmological constraints are purely from the clustering information, and our constraints are considered conservative in this sense. On the other hand, the mean halo mass for each sample is constrained to a fractional precision of $\sim 10\%$, reflecting the fact that the galaxy-galaxy weak lensing can  constrain the mean halo mass, as expected.

The mean halo mass and satellite fraction are generally higher and lower than other similar studies, respectively. This is because our luminosity cut preferentially selects galaxies residing in more massive halos, compared to previous studies. For example, \citet{Miyatakeetal:15} and \citet{2015ApJ...806....2M} used a subsample of CMASS galaxies in the redshift range $z\in[0.43,0.59]$ with stellar mass cuts applied, which resulted in the abundance $\bar{n}_{\rm g}\sim3\times10^{-4} ~(\hiMpc)^{-3}$. They obtained a mean halo mass $\langle M_{\rm h} \rangle \sim 3\times10^{13}~h^{-1}{\rm M}_\odot$ and satellite fraction $f_{\rm sat}\sim8\%$. In \citet{Whiteetal:11}, they used the full CMASS sample from the first semester of BOSS data, and obtained $\langle M_{\rm h} \rangle \sim 2.7\times10^{13}~h^{-1}{\rm M}_\odot$ \footnote{In \citet{Whiteetal:11}, they defined the halo mass such that the enclosed mass is 180 times the background mass density. We converted their halo mass to the definition used in this paper, i.e., $M_{\rm 200m}$.} and $f_{\rm sat}\sim10\%$ from their projected clustering measurement. Note that our validation tests using mock catalogs in \citet{2021arXiv210100113M} indicated that the input HOD parameters are not necessarily well recovered (see Fig.~22 of \citet{2021arXiv210100113M}), partly because our baseline analysis does not use clustering information deeply inside the 1-halo term, which is sensitive to the abundance and spatial distribution of satellite galaxies in host halos. Hence these predictions for the properties of our SDSS galaxy samples should be interpreted with caution.
{\renewcommand{\arraystretch}{1.4} 
\begin{table}
\caption{The predictions for the properties of the SDSS galaxy samples, obtained from the chains of the baseline analysis, similarly to Fig.~\ref{fig:hod}. Here we give the median and the range from the 16th to 84th percentile for the number density, the mean halo mass and the satellite fraction for each of the LOWZ, CMASS1 and CMASS2 samples.}
\begin{center}
\label{tab:halo_properties}
\begin{tabular}{l|ccc}
\hline\hline
sample & $\bar{n}_{\rm g}$ 
& $\langle M_{\rm h} \rangle$ 
& $f_{\rm sat}$ 
\\
&$[10^{-4}(\hiMpc)^{-3}]$ & [$10^{13}h^{-1}M_\odot$] & $[\%]$\\
\hline
LOWZ & $1.11^{+0.70}_{-0.56}$ & $6.59^{+0.85}_{-0.82}$ &
$0.33^{+2.35}_{-0.31}$\\
CMASS1 & $0.59^{+0.65}_{-0.27}$ & $5.76^{+0.55}_{-0.76}$ & 
$0.17^{+1.53}_{-0.16}$\\
CMASS2 & $0.66^{+0.78}_{-0.34}$ & $4.85^{+0.54}_{-0.69}$ &
$0.28^{+2.74}_{-0.27}$\\
\hline
\end{tabular}
\end{center}
\end{table}
}

The ``effective'' bias function for a given galaxy sample can be defined in terms of the clustering correlation functions as
\begin{align}
b_{\rm gm}(r)\equiv \frac{\xi_{\rm gm}(r)}{\xi_{\rm mm}(r)}. 
\end{align}
Another useful quantity is the cross-correlation coefficient function, defined as
\begin{align}
r_{\rm gm}(r)\equiv \frac{\xi_{\rm gm}(r)}{\left[\xi_{\rm gg}(r)\xi_{\rm mm}(r)\right]^{1/2}}. 
\end{align}
We expect $r_{\rm gm}(r)\simeq 1$ on scales where gravitational effects dominate, or equivalently, on scales greater than those affected by nonlinear physics including baryonic physics \citep[][]{2021arXiv210100113M,Hadzhiyska:2021}. Using the chains in the cosmological analysis, we can compute the marginalized posterior distributions of $b_{\rm gm}(r)$ and $r_{\rm gm}(r)$, for each galaxy sample, as shown in Fig.~\ref{fig:cross_correlation_coefficient} \citep[also see Fig.~5 in Ref.~][for a similar approach]{2015ApJ...806....2M}. The figure shows that the large-scale bias $b_{\rm gm}\simeq 2.2$, 2.6 and 2.7 for the luminosity-limited samples of LOWZ, CMASS1 and CMASS2, respectively. These are greater than those of the flux-limited samples, $b\simeq 2.15$, e.g., as shown in Ref.~\cite{2013MNRAS.433.1146C}, but are in good agreement with the bias value $b\simeq 2.5$ for the stellar-mass limited samples of CMASS galaxies at $z\sim 0.5$ in Ref.~\cite{2015ApJ...806....2M}. The figure also shows $r_{\rm gm}\simeq 1$ on scales greater than a few Mpc for all samples, indicating that nonlinear effects are confined to scales smaller than a few Mpc for these SDSS galaxies \citep[e.g.][]{Hadzhiyska:2021}. 

In Fig.~\ref{fig:chi2}, we evaluate the goodness-of-fit of the best-fit model to the measured signals. To do so, we generate noisy data vectors using the ``full'' covariance matrix. The full covariance matrix includes the elements in radial bins outside those used in our cosmology analysis and the cross-covariance terms that describe correlated scatter between the clustering observables, e.g., the galaxy-galaxy weak lensing signals for the different lens samples. Then we apply the same cosmology analysis to each of the mock signals. The histogram in Fig.~\ref{fig:chi2} shows the distribution of the $\chi^2$ values of the model prediction at MAP. We found that the $\chi^2$ values tend to exceed that inferred from the degrees of freedom, $50=75-25$. We ascribe this excess to severe parameter degeneracies; some of our parameters, especially the HOD parameters, are not well constrained by the observables. The histogram can be compared to the $\chi^2$ value of the actual HSC-Y1 and BOSS analysis, denoted by the vertical solid line, showing that the $\chi^2$ value of the real data would occur with a reasonable chance. Hence, we conclude that our model fairly well describes the observables, given the current statistical errors. Note that the degrees of freedom estimated from the analysis of mock signals is much larger than the one estimated using the Gaussian linear model (GLM) \cite{2020PhRvD.101j3527R}, which is $\nu=57.3$. We have confirmed that this discrepancy is due to the strong non-Gaussianity in the posterior distributions of HOD parameters. We rerun the analyses of mock signals with the HOD parameters fixed, and re-estimated the effective degrees of freedom; in this case, the result matches well with the effective degrees of freedom using the GLM with the HOD parameters fixed. 

We emphasize that our cosmological parameter constraints are obtained after marginalizing over the galaxy-halo connection parameters and the nuisance parameters $\Delta z_{\rm ph}$ and $\Delta m_\gamma$. The method outlined in \citet{OguriTakada:11} using a single population of source galaxies allows for an effective marginalization over residual uncertainty in photo-$z$ biases and multiplicative shear biases. Fig.~\ref{fig:contour_nuisance} shows the posterior distributions in a sub-space of $S_8$, $\Delta m_\gamma$ and $\Delta z_{\rm ph}$. The vertical dashed lines in the 1-d posterior distributions of $\Delta m_\gamma$ and $\Delta z_{\rm ph}$ show the width of the Gaussian prior on these parameters. Note that the prior width of $\Delta z_{\rm ph}$, $\sigma(\Delta z_{\rm ph})=0.1$, is much wider than the error inferred from the photo-$z$ method \citep[$\sigma_{\bar{z}_{\rm s}}\simeq 0.04$ as shown in Table~6 of Ref.~][]{2019PASJ...71...43H}. We find that the peak of the 1-d posterior distribution of $\Delta z_{\rm ph}$ is slightly shifted to $-0.046$ from the prior mean ($\Delta z_{\rm ph}=0$), while its width slightly shrinks to $\sigma(\Delta z_{\rm ph})\simeq 0.085$ from the prior width ($0.1$). This implies that the nuisance parameter $\Delta z_{\rm ph}$ is calibrated to some degree, but the constraint is still prior-dominated. The mode $\Delta z_{\rm ph}=-0.046$ corresponds to a decrease in $\langle \Sigma_{\rm cr}^{-1}\rangle$ by $\sim-2$\%, which is the opposite direction compared to the shift estimated using reweighted COSMOS photo-$z$ ($+0.6\%$), as described in Section~\ref{sec:systeamtics_nuisance}. However, the small size of the latter shift and of the COSMOS field ($\sim 2$~\sqdeg) means we can draw no conclusion regarding the difference between these shifts. As a post-unblinding analysis, we performed an additional cosmology analysis employing an even wider prior of $\sigma(\Delta z_{\rm ph})=0.2$ to study the impact of the prior width on our results. We find $S_8=0.775^{+0.053}_{-0.045}$ and $\Delta z_{\rm ph}=-0.113^{+0.135}_{-0.152}$ (see Fig.~\ref{fig:contour_nuisance_ldpz}), which indicates we are entering the self-calibration regime, because the obtained uncertainty on $\Delta z_{\rm ph}$ is smaller than the prior width. The central value of $S_8$ is shifted to a lower value by $\sim 0.5\sigma$ compared to the baseline result ($S_8=\Seight$). In summary these results provide some reassurance that our results are robust against residual photo-$z$ biases, even if they exist.

Similarly, the posterior for $\Delta m_\gamma$ in Fig.~\ref{fig:contour_nuisance} is also prior dominated. As a post-unblinding analysis, we employed a wider prior width of $\sigma(\Delta m_\gamma)=0.1$, which is much wider than the shear calibration uncertainty estimated from image simulations \citep{2018MNRAS.481.3170M}. We find that the posterior width of $\Delta m_\gamma$ slightly shrinks by $\sim$15\%, implying that the data provides some contribution to the posterior due to self-calibration. In this case, we find a slightly lower value of $S_8$ than our baseline result in Table~\ref{tab:summary}: $S_8=0.778^{+0.066}_{-0.053}$. It is interesting to note that the results with wider priors on $\Delta z_{\rm ph}$ and $\Delta m_\gamma$ both prefer a slightly lower value of $S_8$ than the baseline result. 

We also note that $S_8$ is largely unchanged even if we fix $\Delta z_{\rm ph}=0$ or $\Delta m_\gamma=0$ in the parameter inference. When we fix $\Delta m_\gamma=0$, the statistical uncertainty on $S_8$ does not change. However, when we fix $\Delta z_{\rm ph}=0$, the $S_8$ uncertainty decreases by $\sim7$\%. This disparity in behavior is due to the difference in the prior width for these nuisance parameters. 
\begin{figure*}
\begin{center}
\includegraphics[width=1.8\columnwidth]{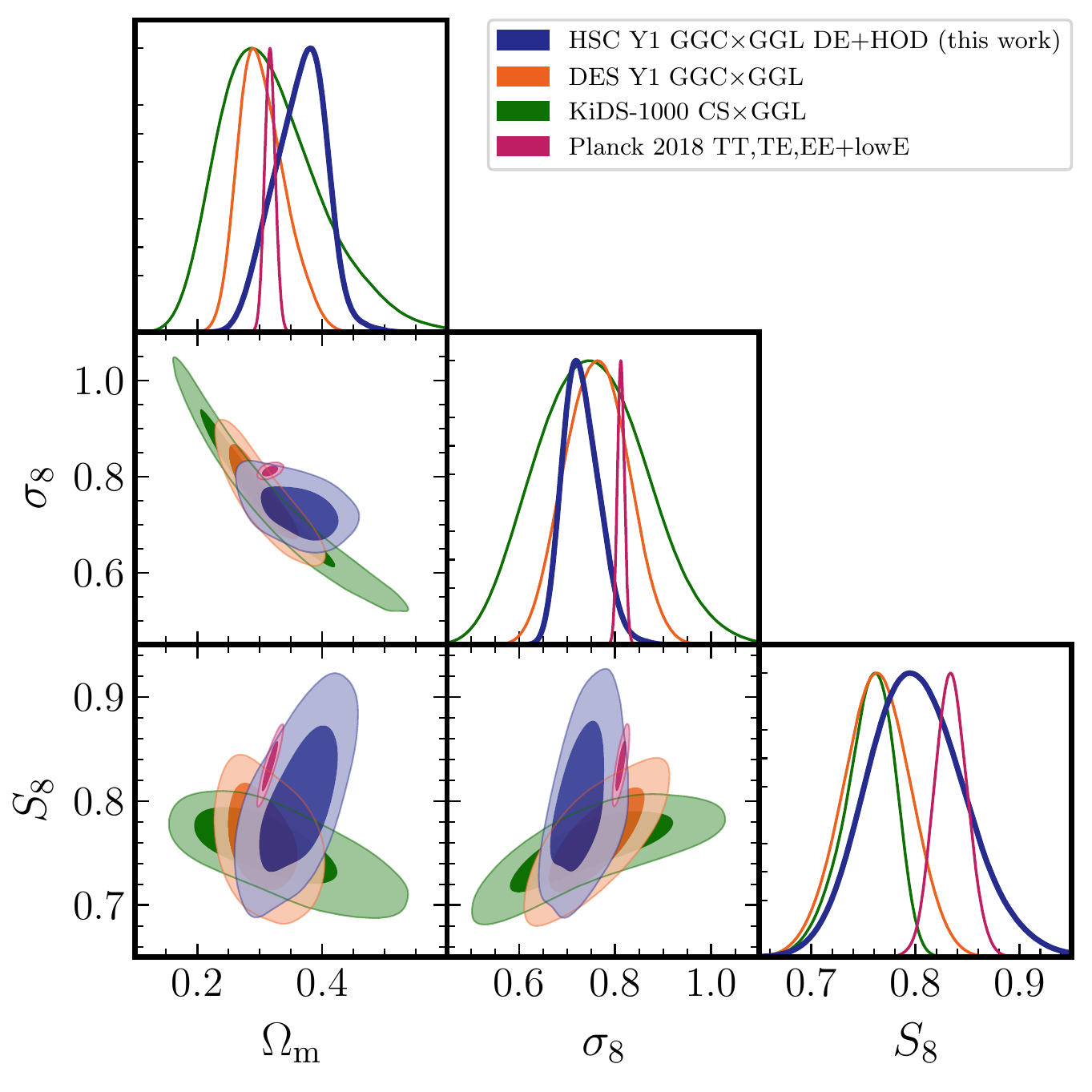}
\end{center}
\caption{Similar to Fig.~\ref{fig:contour_baseline}, but this figure compares our results with those from other cosmological experiments assuming a flat $\Lambda$CDM cosmology. The blue curves (contours) are the main results of this paper, the same as those in Fig.~\ref{fig:contour_baseline}. The orange curves are from the DES-Y1 analysis using galaxy-galaxy lensing (``GGL'') and clustering signals (``GGC'') \cite{2018PhRvD..98d3526A}. The green curves are constraints from the KiDS1000 analysis with cosmic shear (``CS'') and GGL \cite{Heymansetal:2021}. The red curves are the {\it Planck} 2018 results using the primary CMB anisotropy information (``TT,TE,EE+lowE'')
\citep{2020A&A...641A...6P}.}
\label{fig:contour_ext}
\end{figure*}
\begin{figure}
\begin{center}
\includegraphics[width=\columnwidth]{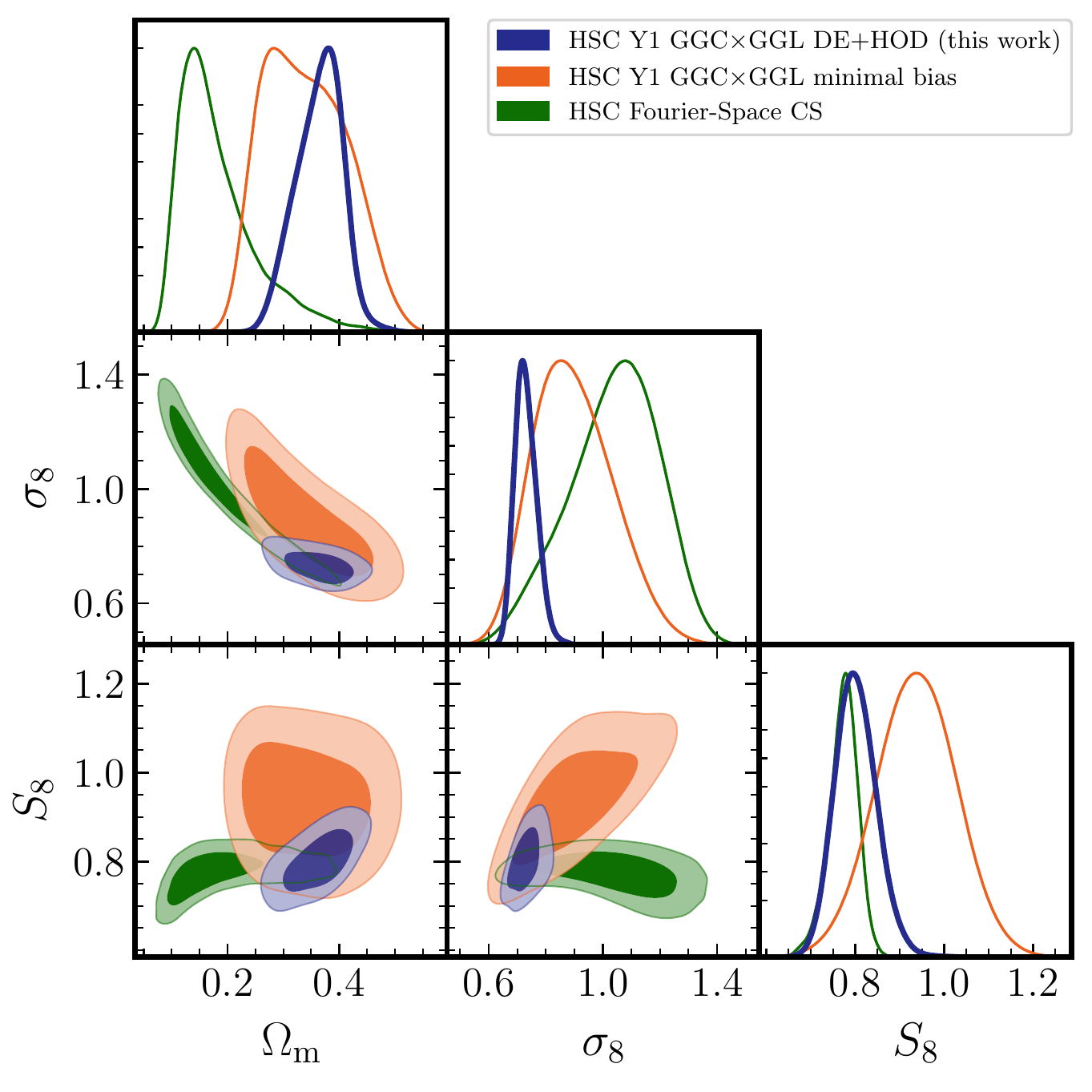}
\end{center}
\caption{Similar to Fig.~\ref{fig:contour_baseline}, but compared with the results from different observables or analysis methods, assuming a flat $\Lambda$CDM cosmology. The blue curves (contours) are the main result of this paper. The orange curves are the constraints from our companion paper \cite{Sugiyama:2021} where the ``minimal'' bias model motivated by the perturbation theory of structure formation is used, as a theoretical template, to interpret the large-scale information of the same signals 
$\dSigma$ and $\wgg$ used in this paper. The green curves are from the Fourier-space cosmic shear measurements of the HSC-Y1 data \citep{2019PASJ...71...43H}.}
\label{fig:contour_comp_hsc}
\end{figure}

Table~\ref{tab:summary} and Fig.~\ref{fig:summary} summarize cosmological parameter constraints for different analysis setups in the flat $\Lambda$CDM model (see Table~\ref{tab:analysis_setups} for the definition of the analysis setups). All the results are consistent with the baseline result, except for the result using stricter scale cuts, $(8,12)\,h^{-1}{\rm Mpc}$. We do not identify any signature of failure or inconsistency in our cosmology analysis, to within the statistical uncertainties. In Appendix~\ref{app:scale_cuts}, we provide a detailed discussion of the results for each analysis setup. The results for the larger scale cuts $(8,12)$ indicate a sizable shift in parameter values compared to the credible intervals. Using the noisy mock observables generated using the covariance matrix, we find that this scatter can happen due to sample variance, as detailed in Appendix~\ref{app:scale_cuts}.

\subsection{Post-unblinding cosmological results}
\label{sec:post-unblinded}

\subsubsection{Comparison with other weak lensing surveys}
\label{sec:comparison_weak_lensing}

In Fig.~\ref{fig:contour_ext} we compare our cosmological constraints on $S_8$, $\sigma_8$, and $\Omega_{\rm m}$ for the flat $\Lambda$CDM cosmology to those from the {\it Planck} 2018 cosmology analysis \citep{2020A&A...641A...6P} and from other weak lensing surveys. For a fair comparison, we infer the other experiment results under similar assumptions/setups. 

For the {\it Planck} 2018 constraints, we employed the fixed neutrino mass $m_{\nu,{\rm tot}}=0.06\,$eV as in our analysis, and ran the {\it Planck} likelihood code \footnote{\url{https://wiki.cosmos.esa.int/planck-legacy-archive/index.php/Cosmological_Parameters}} to estimate cosmological parameters assuming the data vector of the primary CMB information, more specifically ``TT,EE,TE+lowE'' according to the notation used in \citep{2020A&A...641A...6P}. For the DES and KiDS1000 results, we use the public chains available from \footnote{\url{http://desdr-server.ncsa.illinois.edu/despublic/y1a1_files/chains/wg_l3.txt}} and \footnote{\url{http://kids.strw.leidenuniv.nl/DR4/data_files/KiDS1000_3x2pt_fiducial_chains.tar.gz}} to infer the constraints primarily from the similar observables. For DES-Y1, we use the results from joint analysis of the angular galaxy clustering (``GGC'') and the galaxy-galaxy weak lensing (``GGL''), i.e. 2$\times$2pt, from \citet{2018PhRvD..98d3526A}. For KiDS1000, we use the result from joint analysis of cosmic shear (``CS'') and GGL, where a spectroscopic (flux-limited) SDSS galaxy sample is used for the GGC analysis. Note that the 3$\times$2pt result of KiDS1000 includes BAO information that can give a tighter constraint on  $\Omega_{\rm m}$, so we instead use the above result of CS$\times$GGL. For both the DES-Y1 and KiDS1000 analyses, the observables were angular correlation functions -- $\gamma_T(\theta)$ and/or $w(\theta)$ -- rather than  $\dSigma(R)$ and $\wgg(R)$.

Fig.~\ref{fig:contour_ext} shows that our results are generally consistent with other results to within the credible intervals. To be more quantitative, using a tension metric proposed by \citet{Park:2020}, we find that the tension between our result and the {\it Planck} constraint is $0.9\sigma$, or 0.36 in terms of probability-to-exceed. However, a closer look reveals some interesting differences. First, if we take $\Omega_{\rm m}\simeq 0.3$, as inferred from the BAO measurements using SDSS galaxies \citep{Alametal:16,Kobayashi:2021}, the intersection of the $(\Omega_{\rm m},S_8)$-posterior distributions at $\Omega_{\rm m}=0.3$ indicates a possible tension of $S_8$ between our result and the {\it Planck} 2018 result, similar to that reported by previous weak lensing constraints (green, red, and purple contours). However, the significance is weak, so we need more HSC data to reach a definite conclusion.

Our cosmological constraints may appear weaker than those from other surveys because of the broader $S_8$ constraint. However, as shown in Fig.~\ref{fig:contour_ext}, the degeneracy direction in the $\sigma_8$-$\Omega_{\rm m}$ plane is different for this analysis compared to other surveys. The $S_8$ parameter was originally introduced to describe the combination of $\sigma_8$ and $\Omega_{\rm m}$ that cosmic shear can best constrain. However, this combination is not generally optimal for other observables. Thus, we extract the ``optimal'' constraints for each survey by defining the parameter $S_8(\alpha)=\sigma_8(\Omega_{\rm m}/0.3)^\alpha$ and varying $\alpha$ to find the tightest constraint. For each survey, we find the tightest 68\% credible interval: $S_8(\alpha=0.17)=\Seightopt$ for HSC-Y1, $S_8(\alpha=0.61)=0.764^{+0.030}_{-0.030}$ for DES-Y1, and $S_8(\alpha=0.58)=0.758^{+0.017}_{-0.019}$ for KiDS-1000. Therefore, our HSC-Y1 analysis has comparable constraining power to DES-Y1. We expect the cosmological parameter constraints to tighten significantly due to the larger area coverage of subsequent HSC datasets \citep{Li2021} and by further combining the cosmic shear information with the joint-probe measurements in this paper, i.e., 3$\times$2pt cosmology (see below for further discussion).

\subsubsection{Comparison with different analysis methods of HSC-Y1}
\label{sec:comparison_hsc-y1}

In Fig.~\ref{fig:contour_comp_hsc} we compare the cosmological parameters obtained from the different analysis methods and/or observables using the same HSC-Y1 dataset. The orange contours, denoted by the ``minimal bias'' model, are from our companion paper \cite{Sugiyama:2021}, obtained using the same signals as used in this paper, but with even more restrictive scale cuts, because they used a perturbation theory-inspired model as the theoretical template to interpret the large-scale information in the $\dSigma$ and $\wgg$ signals. Hence, the results of the minimal bias model can be considered a conservative estimate of cosmological parameters. From the comparison, it is clear that interpreting the small-scale information using {\tt Dark Emulator} improves our ability to constrain cosmological parameters, even after marginalizing over nuisance parameters. However, there is a shift in the central values of cosmological parameters between our results and the minimal bias method, despite using the same dataset. As discussed above, we used noisy mock data vectors to test whether such shifts in cosmological parameter constraints for these two methods can occur due to sample variance. We found that 3 out of 10 noisy mock realizations exhibit shifts in cosmological parameters similar to those found in the real data. Hence we conclude that the difference between the central parameter values for our method and \citet{Sugiyama:2021} are likely due to sample variance. 

The HSC-Y1 cosmic shear result \citep{2019PASJ...71...43H}, which does not use any information from the SDSS galaxies, is consistent with our results to within the credible intervals, but the central values display a offset, especially in the $\Omega_{\rm m}$ direction. While our constraints are dominated by the SDSS clustering information, we note that \citet{2020PASJ...72...16H}, in their Section~6.7, showed that apparently large scatters in the central values of the cosmological parameters between different analysis methods could occur due to the sample variance and the use of different ranges of scales even for cosmology inference using the same HSC-Y1 dataset. It is interesting to note that the degeneracy directions, e.g., in the ($\Omega_{\rm m},S_8$) space, are different, and thus a combination of these observables may yield even tighter constraints.

\section{Summary and Discussion}
\label{sec:summary}

In this paper, we have reported cosmological constraints from the blinded joint analysis of galaxy-galaxy weak lensing ($\dSigma$) and projected clustering correlation function ($\wgg$) (a $2\times$2pt joint-probe analysis), measured  from the first year imaging galaxy catalog of the Subaru HSC SSP survey (HSC-Y1) and the spectroscopic galaxy catalog of SDSS-III/BOSS DR11. To perform a robust analysis, we have defined the luminosity-limited, rather than the flux-limited, samples from the SDSS galaxies to serve as the tracers of $\wgg$ in the three redshift bins in the range $0.15<z<0.7$, and as the lens samples of $\dSigma$. For the $\dSigma$ measurements, we have opted to use a single sample of background galaxies selected from the weak lensing HSC source catalog with photo-$z$ information greater than 0.75. The HSC-Y1 dataset, despite the relatively small area (about $140$\,\sqdeg),  allows for a significant detection of the galaxy-galaxy weak lensing signals thanks to both the depth and the high imaging-quality of the data (see Fig.~\ref{fig:lens_signal}). 
On the theory side, we have employed the public code {\tt Dark Emulator} to accurately model the clustering observables down to small scales. The validation of our cosmological analysis was demonstrated in previous work \citep{2021arXiv210100113M}. 

With the above joint-probe cosmology, we are able to obtain stringent constraints on the cosmological parameters in flat $\Lambda$CDM model, represented by $S_8\equiv \sigma_8(\Omega_{\rm m}/0.3)^{0.5}=\Seight$ (Fig.~\ref{fig:contour_baseline} and Table~\ref{tab:summary}). An important feature of our results is that our constraints are robust against the possibility of residual photo-$z$ biases in the HSC source sample -- one of the main systematic effects in weak lensing cosmology. By adopting the single sample of source galaxies, we are able to calibrate out the nuisance parameter related to a residual photo-$z$ bias. This is achieved  by comparing the galaxy-galaxy weak lensing amplitudes for the SDSS lens-galaxy samples in the three spectroscopic redshift bins, following the method described in \citet{OguriTakada:11}. Figs.~\ref{fig:contour_nuisance} and \ref{fig:contour_nuisance_ldpz} show that this method enables constraints on the photo-$z$ bias parameter. 

Our results are generally consistent with both {\it Planck} and other weak lensing constraints (DES and KiDS) to within the statistical errors (Fig.~\ref{fig:contour_ext}). However, if we take the intersection of the $(\Omega_{\rm m},S_8)$-posterior distributions at $\Omega_{\rm m}=0.3$, our result indicates a possible tension for $\sigma_8$ compared to the {\it Planck} 2018 result, similarly to those indicated by other weak lensing constraints \citep[e.g.][]{2019PASJ...71...43H}. We performed various tests using the different scale cuts, the extended theoretical template, and the different combinations of datasets, but did not identify any significant signature of residual systematics in our results (Fig.~\ref{fig:summary}). 

There are several ways in which the constraints from this paper could be improved. As can be seen from Fig.~\ref{fig:contour_ext}, the degeneracy direction in the $\Omega_{\rm m}$ and $\sigma_8$ space for our method is different compared to that in the cosmic shear constraint, which is usually characterized by the direction of $\sigma_8(\Omega_{\rm m}/0.3)^{0.5}={\rm const}$ that motivates the definition of $S_8$. The degeneracy direction of our constraint is characterized by the direction of $\sigma_8(\Omega_{\rm m}/0.3)^{0.17}={\rm const}$. This means that the combination of our method ($2\times$2pt) together with cosmic shear would yield improved cosmological constraints. However, this would also require to include additional nuisance parameters to model contaminating effects on the cosmic shear signals such as intrinsic alignments and baryonic effects. In this regard, it is worth considering a $3\times$2pt analysis in future work.

Another promising method would be to combine the HSC galaxy-galaxy lensing information of SDSS galaxies with redshift-space clustering signals of the same SDSS galaxies. In this paper, by intention, we used the ``projected'' clustering information of SDSS galaxies and did not include the geometrical BAO information on large scales. The BAO information, the Alcock-Paczy\'nski effect, and redshift-space distortions due to peculiar velocities of galaxies are powerful probes of $\Omega_{\rm m} $, $\sigma_8$ and the growth rage of large-scale structure. Recently, we developed an emulator based method to model the redshift-space power spectrum of galaxies using an HOD model \cite{Kobayashi:2020b,Kobayashi:2021}. \citet[][]{Kobayashi:2021} applied this method to the redshift-space power spectrum of SDSS galaxies and obtained stringent constraints on $S_8=\sigma_8(\Omega_{\rm m}/0.3)^{0.5}=0.742^{+0.035}_{-0.036}$. This represents a precision of $\sigma(S_8)\simeq 0.04$ compared to our precision of $\sigma(S_8)\simeq 0.05$. Galaxy-galaxy weak lensing probes the Fourier modes that are perpendicular to the line-of-sight direction and are almost independent to those probed by the redshift-space power spectrum \citep{2013arXiv1308.6070D}. Furthermore, the overlap between the HSC and SDSS survey footprints is small and therefore the constraints from HSC are almost independent from those from SDSS. The combination of HSC galaxy-galaxy weak lensing together with the redshift-space power spectrum of SDSS galaxies would be a promising way to improve constraints on both cosmological parameters and the galaxy-halo connection parameters. We have developed the self-consistent emulator-based halo model pipeline needed to perform this joint cosmology analysis, 
and this is our future work.

The HSC survey is ongoing \citep{HSCoverview:17} and currently has the Year 3 shape catalog of galaxies \citep{Li2021} that covers an area of about 430\,\sqdeg~ which is three times larger than the HSC-Y1 data. Hence the HSC Year 3 will enable improved measurements in all weak lensing observables. In future work, we will apply the methods developed in this paper to the HSC Year 3 data, together with other clustering observables, as described above. In addition, HSC already covers more than $1,000$\,\sqdeg\ and will be completed by the end of 2021 or 2022. The full HSC survey will enable us to place one of the tightest cosmological constraints, comparable to other Stage-III surveys such as DES and KiDS.

\acknowledgments
We would like to thank Ryoma~Murata for his contribution during the early phase of this work. This work was supported in part by World Premier International Research Center Initiative (WPI Initiative), MEXT, Japan, and JSPS KAKENHI Grant Numbers JP15H03654, JP15H05887, JP15H05893, JP15H05896, JP15K21733, JP17H01131, JP17K14273, JP18H04350, JP18H04358, JP19H00677, JP19K14767, JP20H00181, JP20H01932, JP20H04723, JP20H05850, JP20H05855, JP20H05856, JP20H05861, JP21J00011, JP21J10314, JP21H01081, and JP21H05456 by Japan Science and Technology Agency (JST) CREST JPMHCR1414, by JST AIP Acceleration Research Grant Number JP20317829, Japan, and by Basic Research Grant (Super AI) of Institute for AI and Beyond of the University of Tokyo. HM and MSi were supported by the Jet Propulsion Laboratory, California Institute of Technology, under a contract with the National Aeronautics and Space Administration. SS is supported by International Graduate Program for Excellence in Earth-Space Science (IGPEES), World-leading Innovative Graduate Study (WINGS) Program, the University of Tokyo. YK was supported by the Advanced Leading Graduate Course for Photon Science at the University of Tokyo. KO is supported by JSPS Research Fellowships for Young Scientists.  RM is supported by a grant from the Simons Foundation (Simons Investigator in Astrophysics, Award ID 620789). AL is supported by the U.D Department of Energy, Office of Science, Office of High Energy Physics under Award Number DE-SC0019301

The Hyper Suprime-Cam (HSC) collaboration includes the astronomical
communities of Japan and Taiwan, and Princeton University. The HSC
instrumentation and software were developed by the National Astronomical
Observatory of Japan (NAOJ), the Kavli Institute for the Physics and
Mathematics of the Universe (Kavli IPMU), the University of Tokyo, the
High Energy Accelerator Research Organization (KEK), the Academia Sinica
Institute for Astronomy and Astrophysics in Taiwan (ASIAA), and
Princeton University. Funding was contributed by the FIRST program from
Japanese Cabinet Office, the Ministry of Education, Culture, Sports,
Science and Technology (MEXT), the Japan Society for the Promotion of
Science (JSPS), Japan Science and Technology Agency (JST), the Toray
Science Foundation, NAOJ, Kavli IPMU, KEK, ASIAA, and Princeton
University.
This paper makes use of software developed for the Large Synoptic Survey
Telescope. We thank the LSST Project for making their code available as
free software at \url{http://dm.lsst.org}

The Pan-STARRS1 Surveys (PS1) have been made possible through
contributions of the Institute for Astronomy, the University of Hawaii,
the Pan-STARRS Project Office, the Max-Planck Society and its
participating institutes, the Max Planck Institute for Astronomy,
Heidelberg and the Max Planck Institute for Extraterrestrial Physics,
Garching, The Johns Hopkins University, Durham University, the
University of Edinburgh, Queen's University Belfast, the
Harvard-Smithsonian Center for Astrophysics, the Las Cumbres Observatory
Global Telescope Network Incorporated, the National Central University
of Taiwan, the Space Telescope Science Institute, the National
Aeronautics and Space Administration under Grant No. NNX08AR22G issued
through the Planetary Science Division of the NASA Science Mission
Directorate, the National Science Foundation under Grant
No. AST-1238877, the University of Maryland, and Eotvos Lorand
University (ELTE) and the Los Alamos National Laboratory.

Based in part on data collected at the Subaru Telescope and retrieved
from the HSC data archive system, which is operated by Subaru Telescope
and Astronomy Data Center at National Astronomical Observatory of Japan.

\bibliography{refs}

\appendix

\begin{figure*}
\includegraphics[width=\columnwidth]{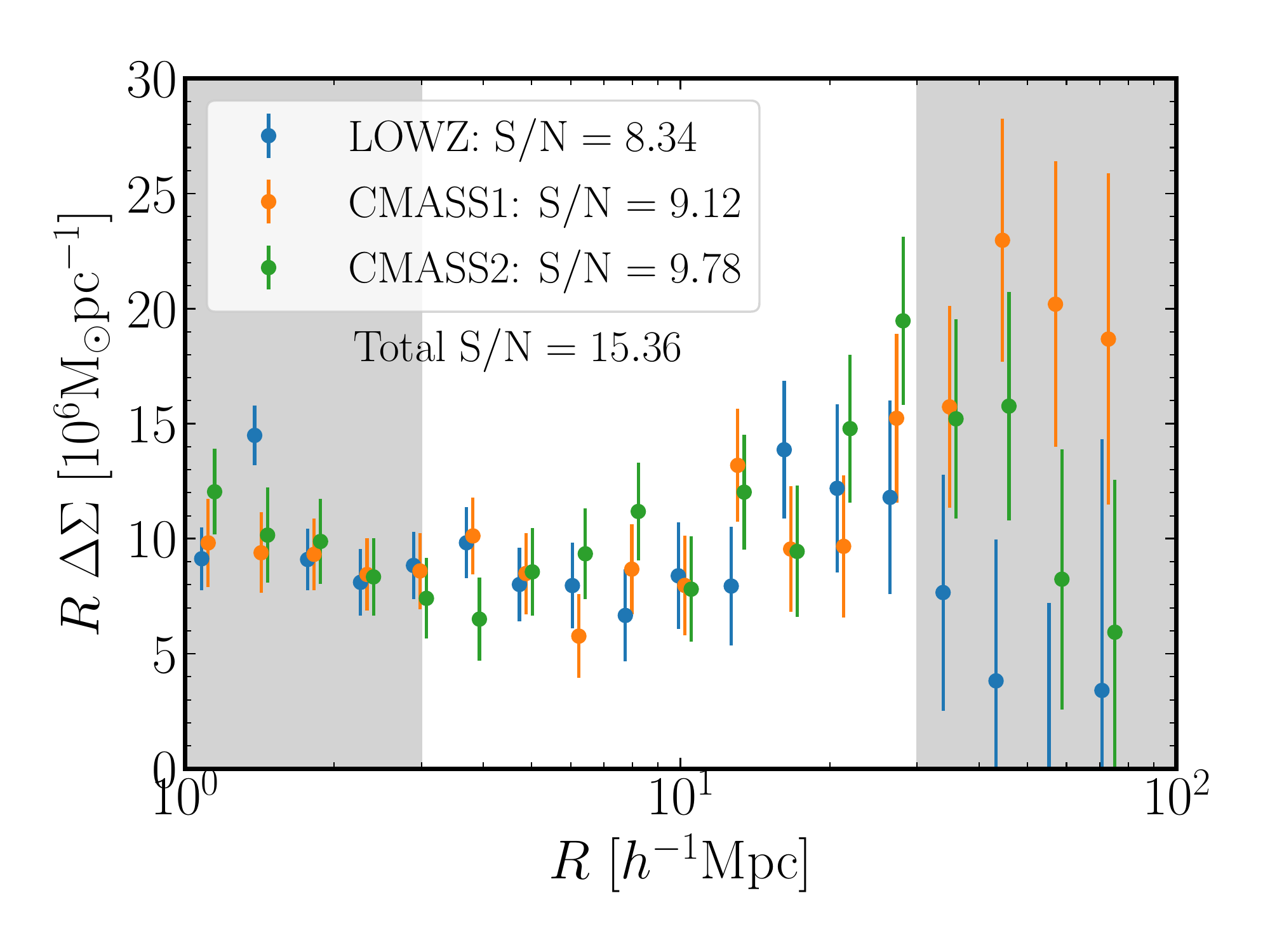}
\includegraphics[width=\columnwidth]{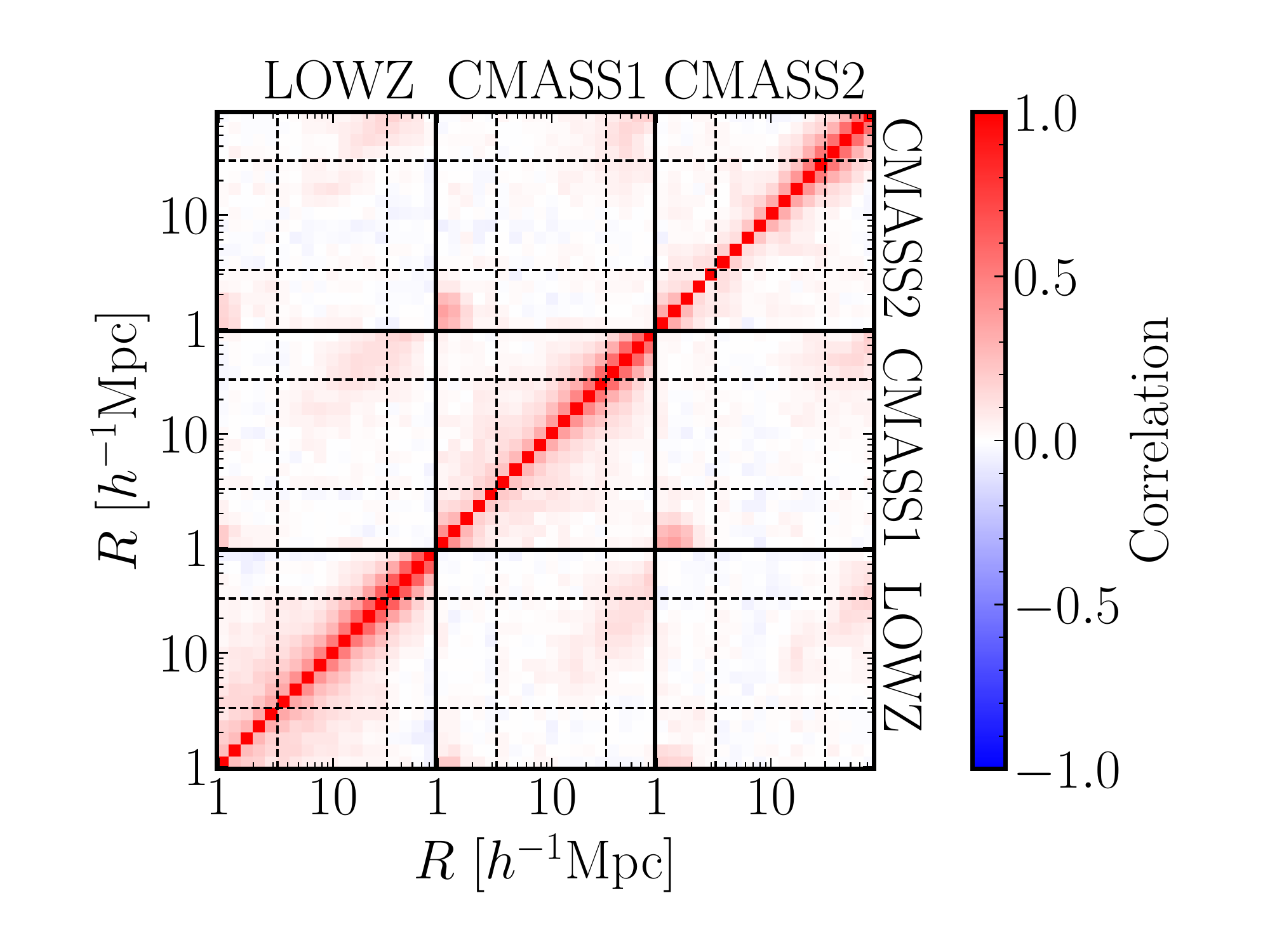}
    \caption{{\it Left panel:} Weak lensing signal around SDSS galaxies measured with the HSC-Y1 source galaxies. The shaded region is excluded from our fiducial cosmology analysis. The signal-to-noise ratios are computed within the scales used for our analysis. The lensing signals for the CMASS1 and CMASS2 samples  have offsets along the $x$-axis for illustrative purposes. {\it Right panel:} Correlation coefficient matrix of the lensing profile ($\dSigma$), defined as $r_{ij}\equiv {\rm Cov}_{ij}/[{\rm Cov}_{ii}{\rm Cov}_{jj}]^{1/2}$. The bold solid lines indicate the different lens samples (LOWZ, CMASS1 and CMASS2). In each sub-matrix, elements inside the dashed lines denote the scales used for our fiducial cosmology analysis: $3\le R/[h^{-1}{\rm Mpc}]\le 30$.}
    \label{fig:lens_signal_covariance}
\end{figure*}
\begin{figure}
\includegraphics[width=\columnwidth]{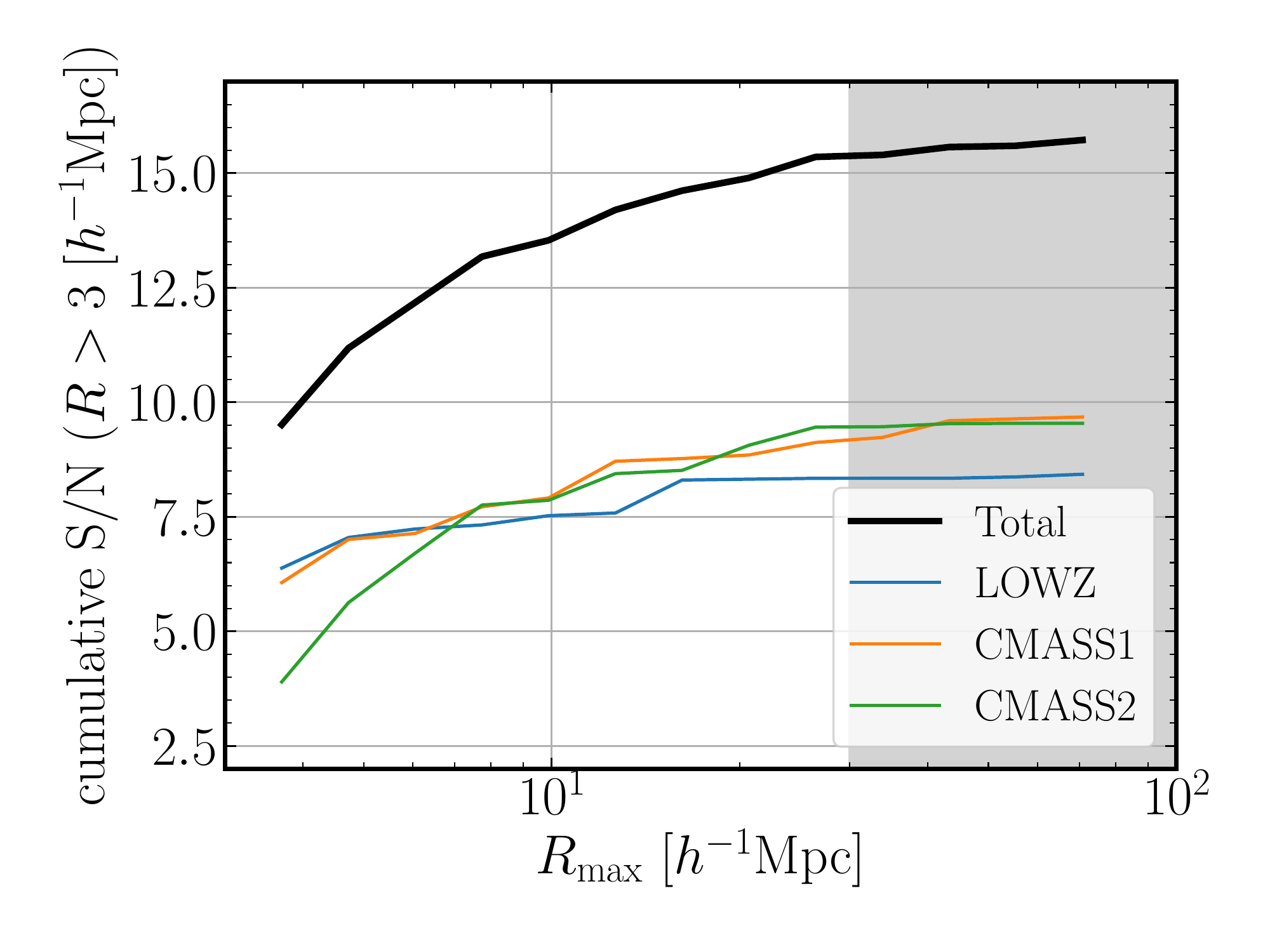}
    \caption{Cumulative signal-to-noise ratio of the weak lensing signal, integrated over $3\le R/[h^{-1}{\rm Mpc}]\le R_{\rm max}$. The cumulative signal-to-noise ratio saturates above $R_{\rm max}>30$~$\hiMpc$ which validates the upper limit of our scale cut.}
    \label{fig:lens_cumulative_sn}
\end{figure}
\begin{figure*}
    \includegraphics[width=2.2\columnwidth]{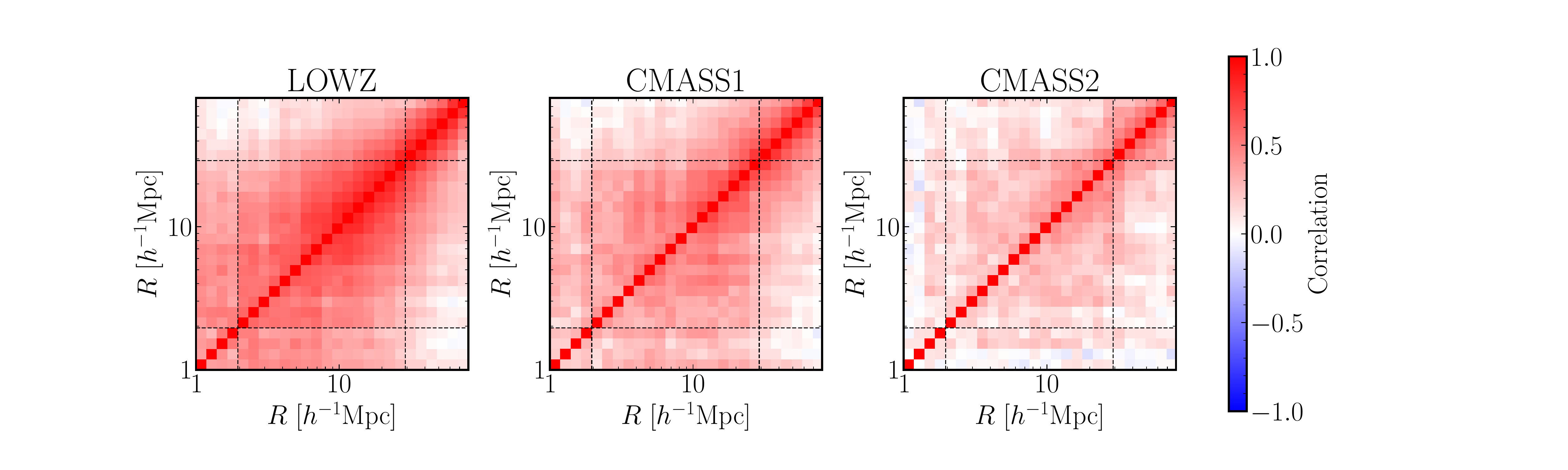}
    \caption{Correlation coefficient matrix of $\wgg(R)$ for each of the LOWZ, CMASS1, and CMASS2 samples, estimated from 192 jackknife samples. In this case, the cross-covariance between the different samples vanishes, in contrast to the covariance matrix for the lensing signals (Fig.~\ref{fig:lens_signal_covariance}).}
    \label{fig:clustering_covariance}
\end{figure*}
\begin{figure}
\includegraphics[width=\columnwidth]{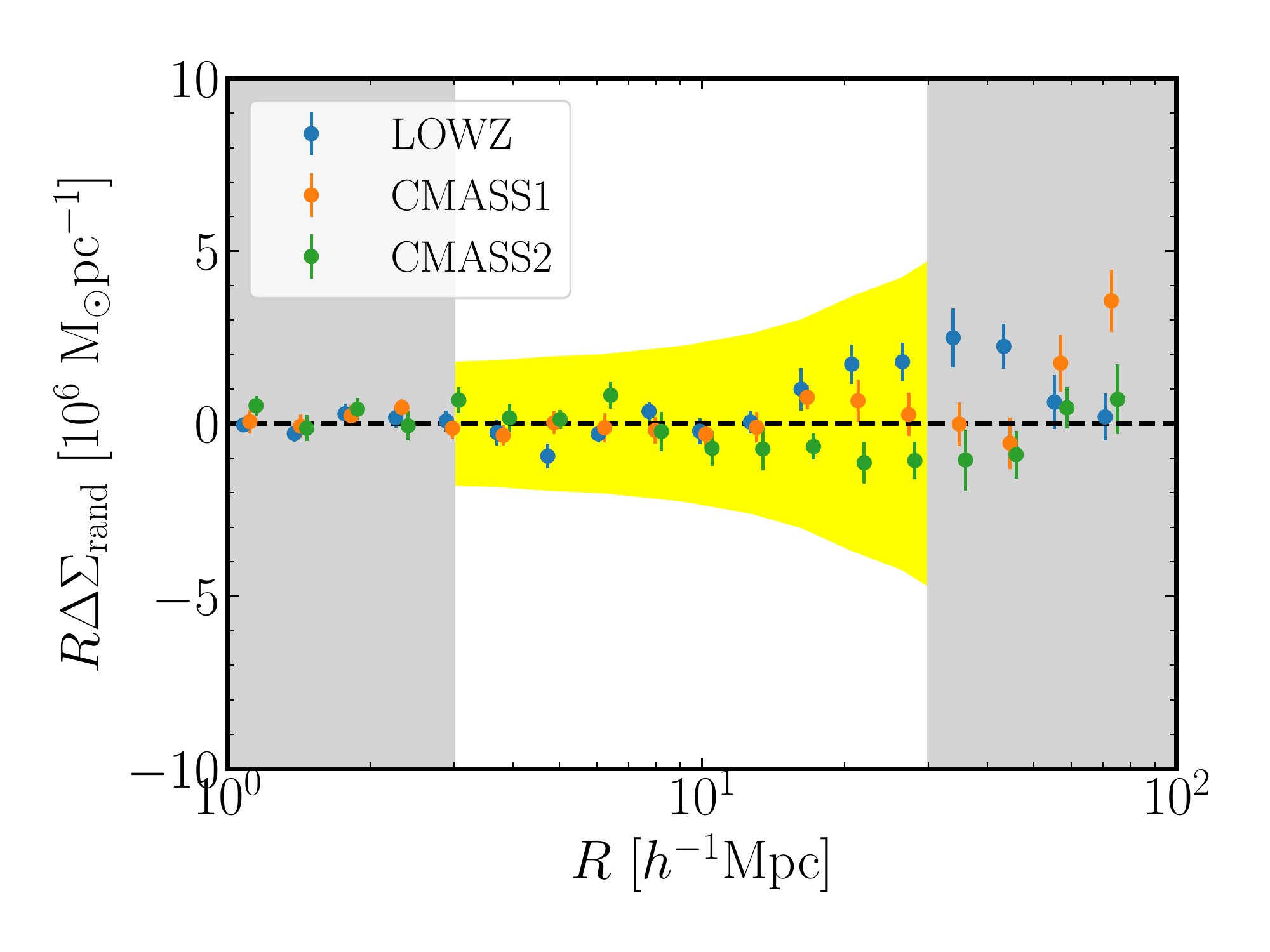}
\caption{The lensing signals measured around random points measured in the same way as the lensing signals around the lens galaxies (the second term in Eq.~\ref{eq:lens_measurement}). Error bars denote the standard deviation of the results for 20 realizations of random points, where each realization contains the same number of random points as that of lens galaxies for each sample. The shaded yellow region denotes the statistical errors of the lensing measurements, computed from the diagonal components of the covariance matrix in Fig.~\ref{fig:lens_signal_covariance}; the statistical errors in each $R$ bin are connected to compute the shaded region for illustrative purpose. The shaded gray region denotes the range excluded from our baseline cosmology analysis. The unshaded region denotes the range used in the cosmology analysis.}
    \label{fig:random_signal}
\end{figure}

\begin{figure*}
\includegraphics[width=\columnwidth]{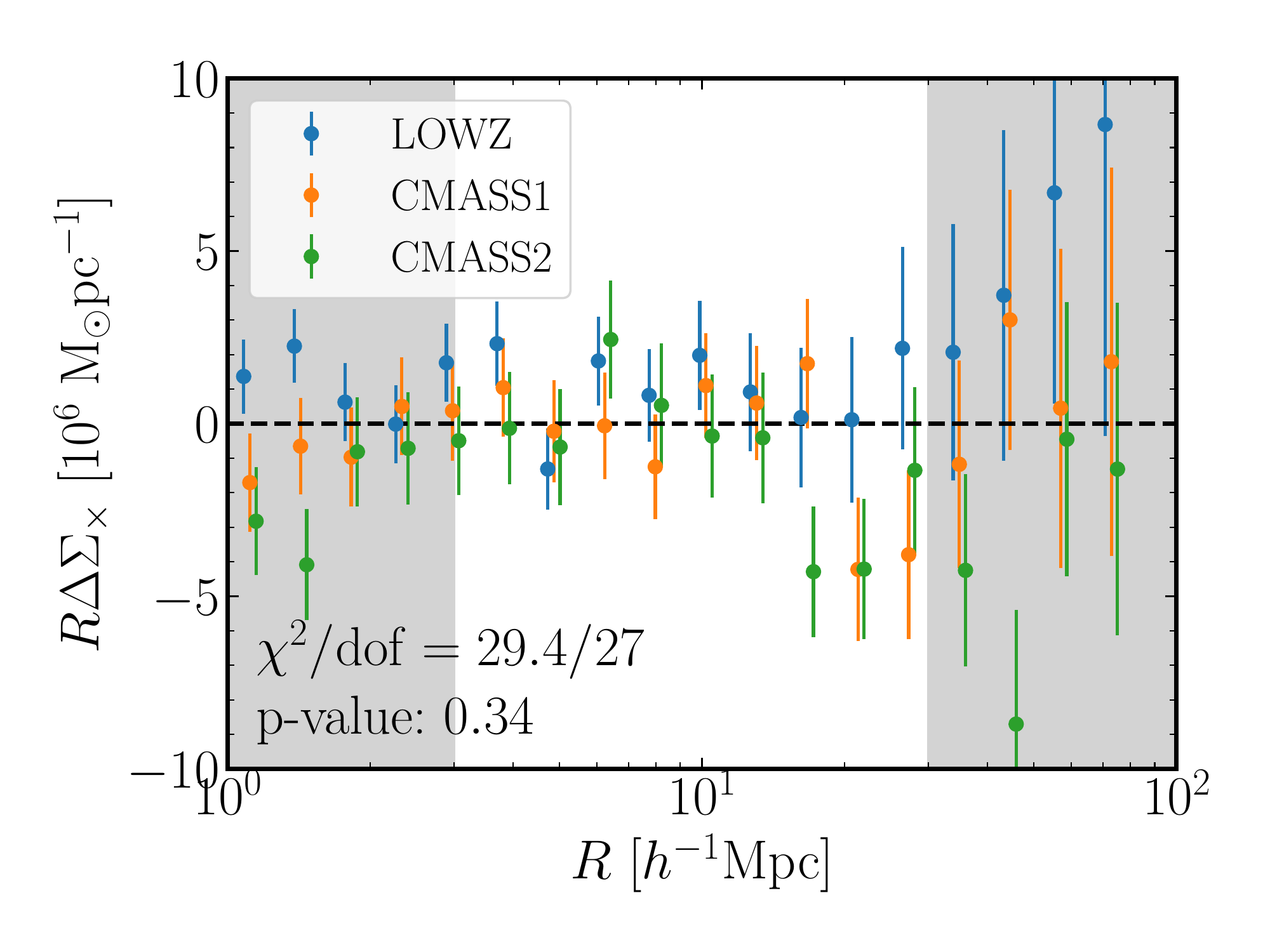}
    \includegraphics[width=\columnwidth]{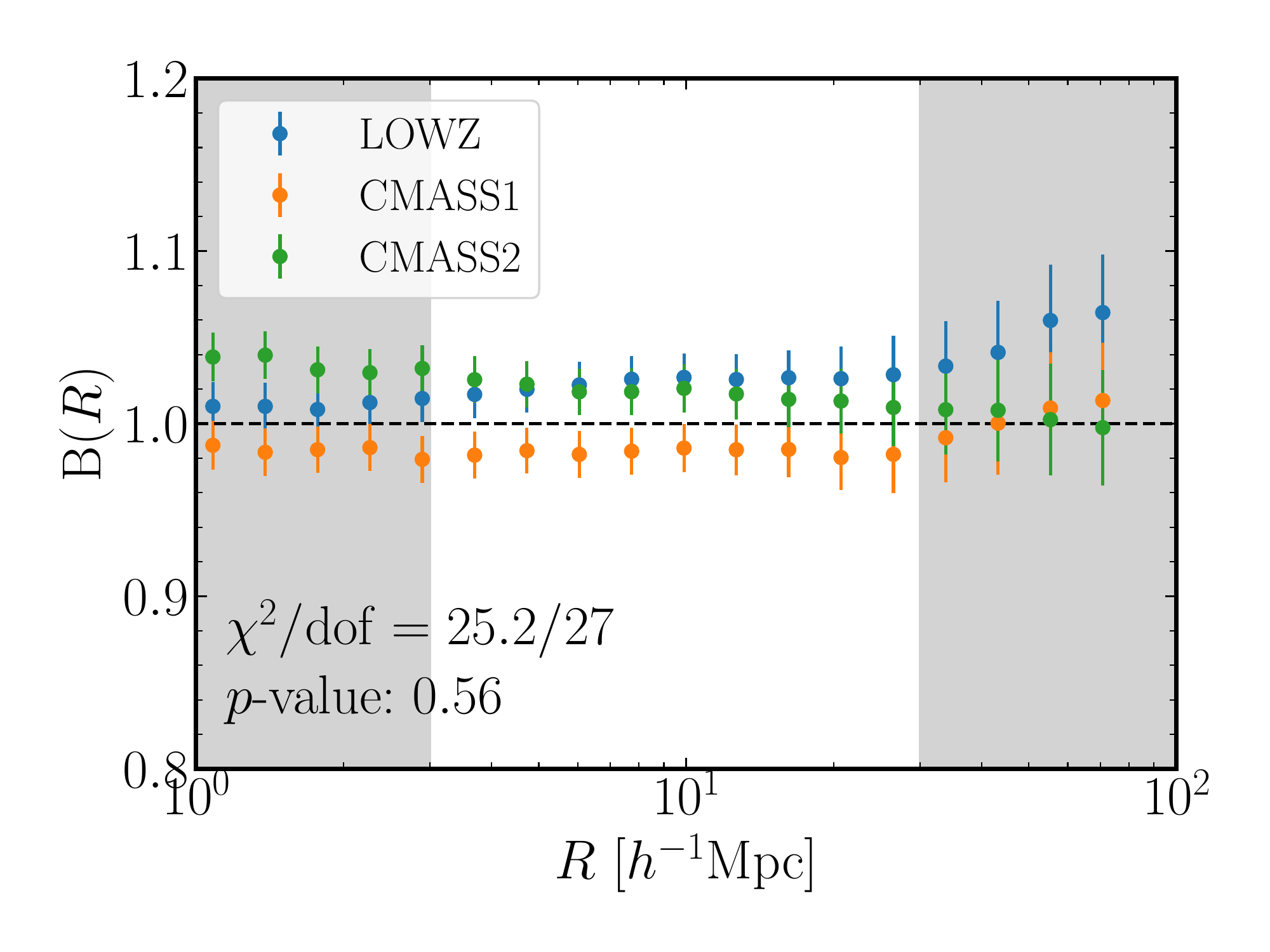}
    \caption{Systematic tests of galaxy-galaxy weak lensing measurements. {\it Left panel}: the $B$-mode signals around galaxies in each of the LOWZ, CMASS1 and CMASS2 samples, measured using the 45-degree component (non-lensing components) of source ellipticities. Error bars denote the diagonal components of the covariance matrix of the lensing profile in each $R$ bin. {\it Right panel}: A ``boost'' factor  $B(R)$, which measures an excess or deficiency in the number of lens/source pairs compared to random/source pairs (see text for details). The error bars are estimated using the same mock catalogs used in the covariance estimation. We performed the same boost factor measurements for each of the mock realizations, and estimated the error bars from the standard deviations. The reduced chi-square values and $p$-values are computed using the data points and the covariance over the range indicated by the unshaded region}
    \label{fig:lens_signal_systematics}
\end{figure*}
\begin{figure*}
\includegraphics[width=\columnwidth]{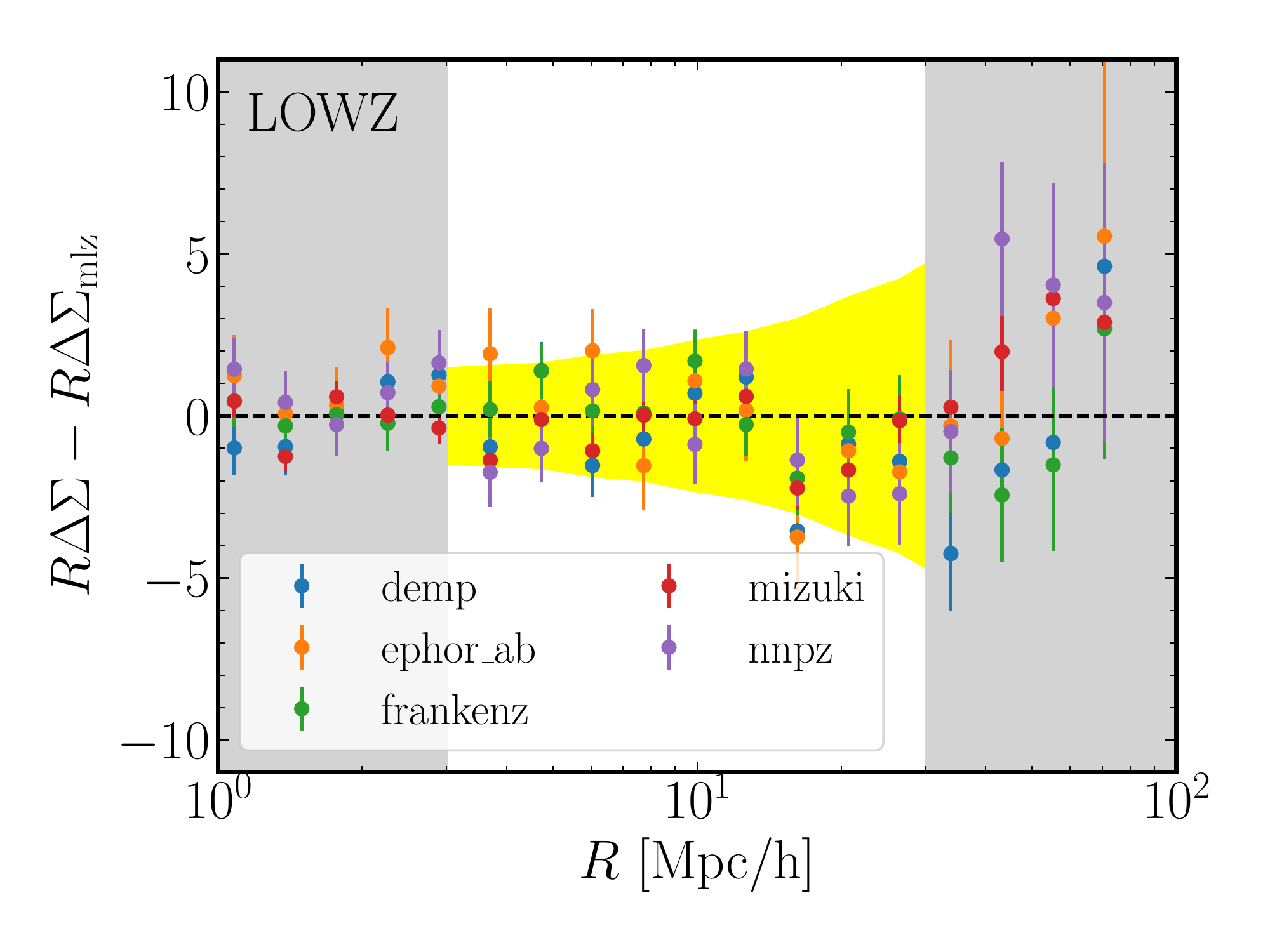}
\includegraphics[width=\columnwidth]{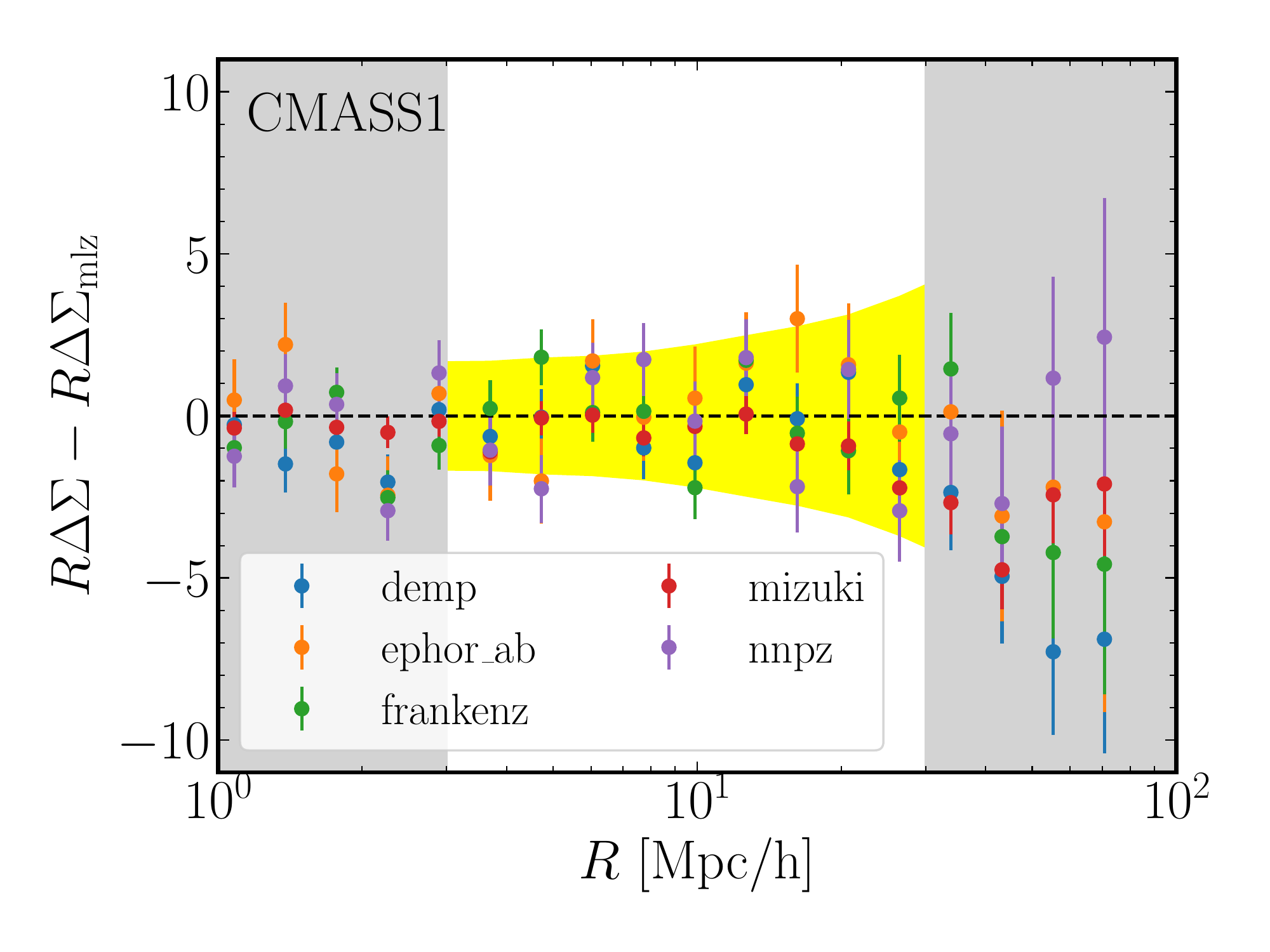}
\includegraphics[width=\columnwidth]{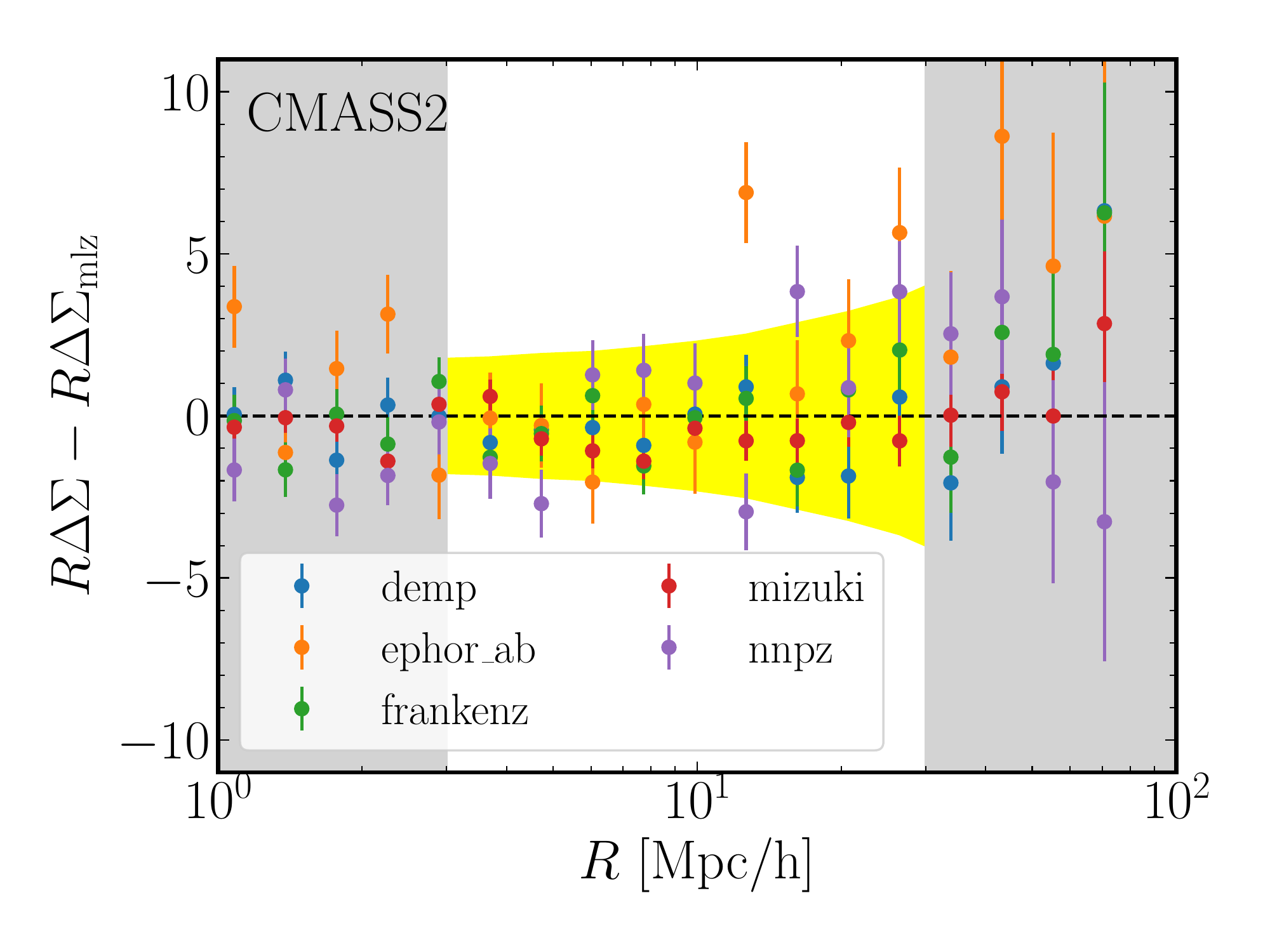}
\includegraphics[width=\columnwidth]{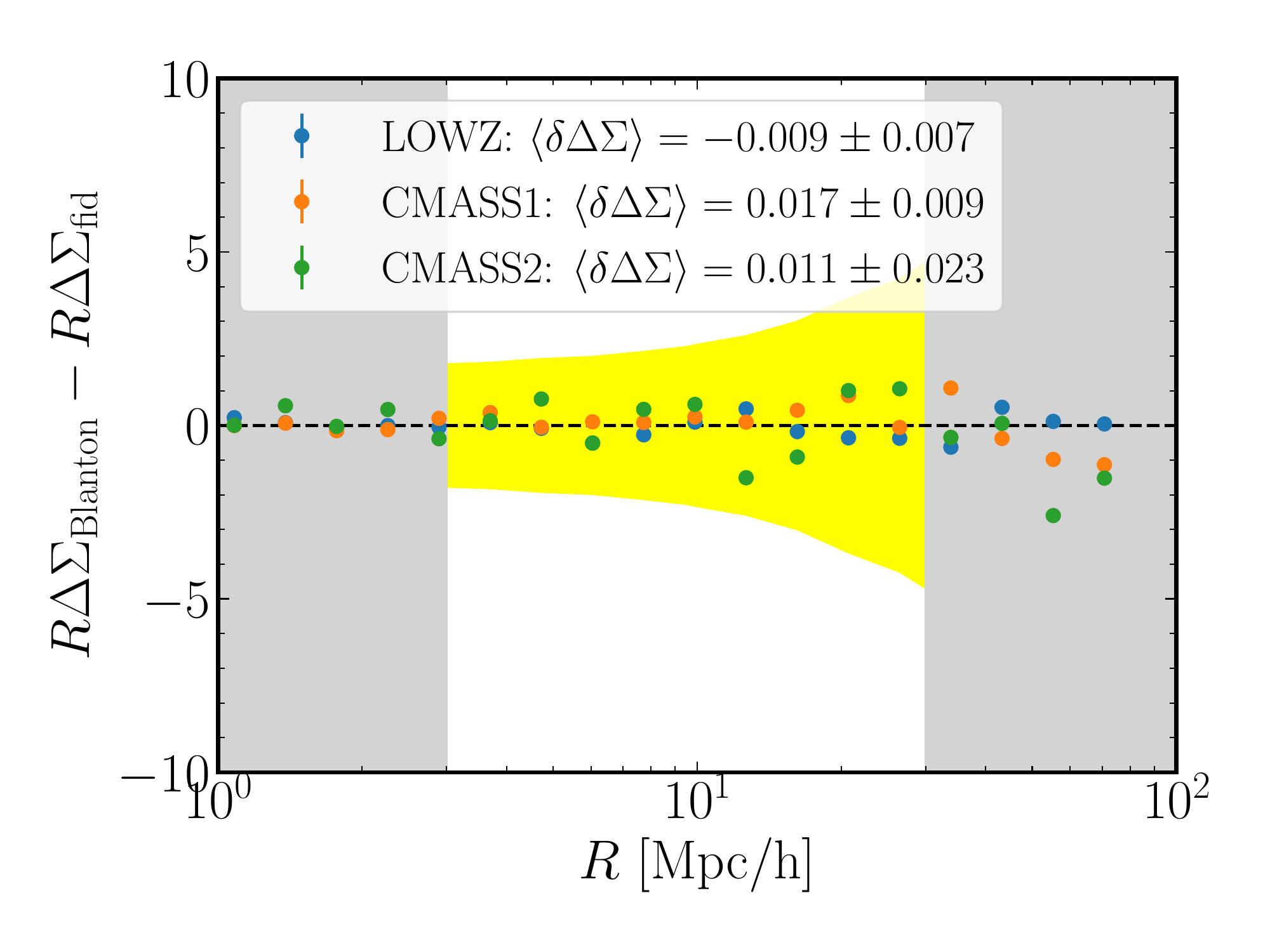}
    \caption{Difference between the fiducial lensing signals and those computed with different analysis choices. The top-left, top-right, bottom-left panels shows the difference when using other photo-$z$ methods. The bottom right panel shows the difference when considering a different k-correction method. The yellow region show the range of statistical uncertainties for our lensing signal. The unshaded region show the range  used in the cosmology analysis}
    \label{fig:lens_signal_photoz_kcorr}
\end{figure*}
\section{Covariance matrix}
\label{sec:covariance}
As described in Appendix~B of \citet{2021arXiv210100113M} \citep[also see][]{Shirasakietal:17,2019MNRAS.486...52S}, we construct the covariance matrix that describes the statistical errors of the $\dSigma$ measurement using light-cone mock catalogs of both SDSS and HSC galaxies.  

To construct these mock catalogs, we use the full-sky, light-cone simulations from \citet{2017ApJ...850...24T}. Each of the 108 light-cone realizations consists of multiple spherical shells at different radii with an observer located at the center of the sphere. Each spherical shell contains the lensing fields and the halo distribution. The lensing fields at the representative redshift of each shell can be used to simulate the lensing distortion effects on a galaxy due to foreground structures if the galaxy is located in the shell. The halo distribution in each shell reflects a realization of halos  at the redshift corresponding to the shell. 

We populate each of the light-cone simulation realizations with SDSS- and HSC-like galaxies. For HSC, we use the HSC-Y1 source catalog and populate each HSC galaxy into the corresponding shell in the light-cone simulation according to its angular position (RA and Dec) and photometric redshift (best-fit photo-$z$). After randomly rotating each galaxy shape to erase the real weak lensing effect, we simulate the lensing effect on each galaxy using the lensing information of the light-cone simulation. Thus, the mock HSC catalog reproduces the angular positions, the distributions of ellipticities, and the photo-$z$'s of HSC galaxies. Because the HSC-Y1 survey footprint covers a small area, we can extract 21 HSC-Y1 realizations in each of the 108 all-sky simulations. Hence we generate 2268 HSC mock catalogs in total. 

For SDSS, we populate each of the light-cone mocks with SDSS-like galaxies based on the HOD method. We built mock catalogs for the LOWZ, CMASS1 and CMASS2 samples in their corresponding redshift ranges in each full-sky simulation. We imprint the SDSS survey footprint onto each full-sky simulation realization.

Thus our light-cone mock catalogs contain both HSC- and SDSS-like galaxies. We apply the same measurement pipeline as used in the actual analysis to each of the mock realizations and compute the covariance matrices of $\dSigma(R)$. The covariance matrices of $\dSigma$ include the cross-covariance between the different lens samples. This arises from the shape noise of source galaxies and from cosmic shear contamination. However, we ignore the cross-covariance between $\dSigma$ and $\wgg$, because the overlap between the HSC-Y1 and SDSS survey footprints is small (only 140~\sqdeg compared to 8000~\sqdeg). The covariance matrix estimated in this method properly includes the super sample covariance contribution \cite{TakadaHu:13}.

We treat shape noise in this covariance calibration as follows. Shapes in the mock source catalog are the sum of the intrinsic shapes of galaxies (taken from the actual shape catalog after applying a random rotation) and lensing shear from the ray-tracing. In the mock source catalog, the multiplicative bias is not incorporated, and thus if we naively compute the covariance from the mock data, there would be inconsistency between the signal and covariance. To properly incorporate the multiplicative bias, we first need to separate out the intrinsic shape component to obtain the lensing shear component alone, and then add the intrinsic shape component scaled by the multiplicative bias. In practice, we compute the covariance as 
\begin{eqnarray}
\cov^{\dSigma}_{i,j} &=& \cov^{\dSigma^{\rm mock}}_{i,j} - \cov^{\dSigma^{\rm mock}_{\rm rand.~shape}}_{i,j} \nonumber\\
&& + \frac{\cov^{\dSigma^{\rm data}_{\rm rand.~shape}}_{i,j}}{(1+K(R_i))(1+K(R_j))},
\end{eqnarray}
where $i$ and $j$ are the indices of radial bins running over all of the lens samples. The first term on the right-hand side is the covariance from the mock data, where a mock signal of each realization is computed following Eq.~\eqref{eq:lens_measurement} but without applying any multiplicative bias correction. The second term is the covariance from the mock data with randomly-rotated shapes. The third term is the covariance from the real data with randomly-rotated shapes scaled by the multiplicative bias factors in the $i$-th and $j$-th radial bins. Note that, when computing the second and third terms, we follow Eq.~\eqref{eq:lens_measurement} (without the multiplication bias correction) but did not subtract the random signal. This is because the random signal subtraction is primarily for subtracting the lensing signal that arises from large-scale structure but this signal no longer exits in the randomized shapes \citep{Shirasakietal:17}. Practically, ignoring the random signal significantly reduces computing time. Fig~\ref{fig:lens_signal_covariance} shows the covariance matrix estimated based on the above method. 

Fig.~\ref{fig:lens_cumulative_sn} shows the cumulative signal-to-noise ratio integrated over $3\le R/[h^{-1}{\rm Mpc}]\le R_{\rm max}$, where we vary the maximum separation $R_{\rm max}$ as denoted in the $x$-axis. The cumulative signal-to-noise ratio does not increase above $R_{\rm max}>30$~$\hiMpc$, which is due to the increase in the sample covariance at large scales due to large-scale structure. Since the cumulative signal-to-noise ratio is a proxy for the information content in the lensing amplitudes over the range of separation, our fiducial choice of $R_{\rm max}=30\,h^{-1}{\rm Mpc}$ in the cosmology analysis is nearly optimal, because we focus on the cosmological parameters $S_8$, $\Omega_{\rm m}$ and $\sigma_8$ which primarily determine the lensing amplitude. 

As described in Section~\ref{sec:gg_clustering}, we use the jackknife method to estimate the covariance matrix for the projected correlation function, $\wgg(R)$, for LOWZ, CMASS1 and CMASS2. The jackknife method has the advantage that it captures all contributions to the covariances (survey geometry, inhomogeneities. etc.). Fig.~\ref{fig:clustering_covariance} shows the results for the covariance matrix. It is clear that there is significant cross-covariance between the $\wgg(R)$ signals in different separation bins, and therefore it is important to properly take into account the cross-covariance.  

\section{Tests of lensing systematics}
\label{sec:lens_systematics}
In this section, we describe our lensing systematic tests.

As described in Section~\ref{sec:gglensing_estimator}, we subtract the signal around random points from the signal around lens galaxies to obtain an unbiased estimate of galaxy-galaxy weak lensing \citep{Mandelbaum:05b}. We use 20 times more random points than that of lens galaxies. Fig.~\ref{fig:random_signal} shows the signal around random points measured in the same way as described in Section~\ref{sec:gglensing}. The error bars are estimated from 20 realizations of random points. The random signal starts to deviate from zero at $R\simgt15\hiMpc$, but is still smaller than the statistical uncertainties. This result shows that the random point correction is not significant. Nonetheless, we subtract the random signal from the signals around lens galaxies to obtain an unbiased estimate of $\dSigma$.

Galaxy-galaxy weak lensing, which arises from the scalar gravitational potential, induces only $E$-mode signal or the tangential shear pattern around lensing galaxies on average in the weak lensing regime. On the other hand, the 45-degree rotated component from the tangential shear, or the $B$-mode, should vanish if the lensing measurement is perfect or if systematic effects are negligible in the data. We compute the $B$-mode signal by replacing the tangential component of individual background galaxy shape in Eq.~\eqref{eq:lens_measurement} with the 45-degree rotated component. This is shown in the left panel of Fig.~\ref{fig:lens_signal_systematics}. The measured $B$-mode signals are consistent with the null signal to within the error bars over the range of separation bins that we use for the cosmological analysis (the unshaded region). The $p$-value for a null B-mode signal (calculated for the three lens samples within the fiducial scale cuts) is 0.34. Hence we do not find evidence for a residual $B$-mode signal for all the three measurements of galaxy-galaxy weak lensing. 

Another important systematic effect is the ``boost'' factor which quantifies an excess or deficiency in the number of lens and source galaxy pairs compared to that of random point and source galaxy pairs \cite[for details, see][]{Mandelbaum:05b, Mandelbaumetal:13}. A non-zero boost factor arises from systematic effects. For example, if some source galaxies are actually in the lens redshift range and are therefore physically associated with lens galaxies, due to the imperfect photo-$z$ estimates, then the number of lens and source pairs appears to be in excess. Or if some source galaxies are difficult to detect in the vicinity of lens galaxies on the sky due to imperfect photometry such as a flux contamination of bright lens galaxies to background HSC galaxies, the number of source galaxies near lenses could appear to be in deficiency. Following \cite{Miyatakeetal:15}, we define the boost factor as
\begin{equation}
B(R_i) = \frac{\sum_{{\rm ls}\in R_i} w_{\rm ls}\left/\sum_{\rm l} w_{\rm l} \right.}{\sum_{{\rm rs}\in R_i} w_{\rm rs}\left/\sum_{\rm r} w_{\rm r} \right.},
\end{equation}
where ``r'' in the summation runs over random points. The numerator and denominator essentially count the averaged number of source galaxies around each lens galaxy and random points, respectively. We estimate the covariance of the boost factor using the mock catalogs of source galaxies. When estimating the boost factor and the covariance, we limit the lens and random points within the FDFC region; otherwise the covariance becomes too large due to the partial use of annulus bins around galaxies and randoms outside of the FDFC region. We show the measured boost factor in the right panel of Fig.~\ref{fig:lens_signal_systematics}. The measured boost displays an offset from unity, but does not show any strong $R$ dependence, which would be observed if it was due to contaminating galaxies clustered with the lenses. The $p$-value of the boost factor being non unity (calculated for the three lens samples within the scale cuts) is 0.56. Fig.~\ref{fig:lens_signal_systematics} gives a misleading impression because the radial bins are highly correlated. We do not make any correction for the boost factor in the weak lensing signals. However, note that we will introduce nuisance parameters of the residual photo-$z$ errors and the multiplicative shear bias, which cause an overall (almost $R$-independent) offset in the weak lensing signals. This nuisance parameters can capture possible residual effect in the boost factor if it is present. 

We also quantify how the lensing signal changes with different photo-$z$ estimates. Fig.~\ref{fig:lens_signal_photoz_kcorr} shows the difference between the lensing signals computed with the fiducial method and one of the non-fiducial photo-$z$ methods. The covariance of the difference signal is estimated using difference signals measured from 256 realizations of source catalogs with randomly-rotated shapes, as detailed in \citet{2019ApJ...875...63M}. The weighted mean and the standard deviation of the difference signals are shown in Table~\ref{tab:photo-z_test}. The resulting impact on the cosmological parameters is summarized in Table~\ref{tab:summary} and Fig.~\ref{fig:summary}. These show that all results are consistent with those of the fiducial analysis to within the error bars. 

Our fiducial sample is defined using the k-correction method described in \citet{2006MNRAS.372..537W}, where we k-corrected the magnitudes of LOWZ galaxies to a redshift of $0.20$ and those of CMASS galaxies to a redshift of $0.55$. We also created a sample using the k-correction method with {\tt kcorrect\_v4.3} \cite{Blanton:2007} to quantify how the lensing signals change with the different k-correction methods. The difference in the signals is shown in Fig.~\ref{fig:lens_signal_photoz_kcorr}, which does not show any significant change in the signals.

In summary we do not find any strong evidence of the residual systematic effects in our weak lensing measurements. This reflects the high-quality of the HSC-Y1 shape catalog, at least compared to the statistical errors of the HSC-Y1 survey volume.

\section{Halo occupation distribution (HOD)}
\label{sec:hod}
In the halo model we assume that all matter is associated with halos, and that the correlation function of matter is given by contributions from pairs of matter in the same halo and those in two different halos. These are referred to as the 1- and 2-halo terms, respectively. We employ the halo occupation distribution \citep[HOD][]{1998ApJ...494....1J,PeacockSmith:00,Scoccimarroetal:01}. The HOD model gives the mean number of central and satellite galaxies in halos of mass $M$ as
\begin{align}
\avrg{N}\!(M)=\avrg{N_{\rm c}}\!(M)+\avrg{N_{\rm s}}\!(M),
\label{eq:N_HOD}
\end{align}
where $\avrg{\hspace{1em}}\!(M)$ denotes the average of a quantity for halos of mass $M$. 

We employ the mean HOD for central galaxies, given as
\begin{align}
\avrg{N_{\rm c}}\!(M)=\frac{1}{2}\left[1+{\rm erf}\left(\frac{\log M-\log M_{\rm min}}{\sigma_{\log M}}\right)\right],
\label{eq:Nc}
\end{align}
where ${\rm erf}(x)$ is the error function and $M_{\rm min}$ and $\sigma_{\log M}$ are model parameters. 

For the mean HOD of satellite galaxies, we employ the following form: 
\begin{align}
\avrg{N_{\rm s}}\!(M)\equiv \avrg{N_{\rm c}}\!(M)\lambda_{\rm s}(M)=\avrg{N_{\rm c}}\!(M)\left(\frac{M-\kappa M_{\rm min}}{M_1}\right)^{\alpha},
\end{align}
where $\kappa,M_1$ and $\alpha$ are model parameters, and we have introduced the notation $\lambda_{\rm s}(M)=[(M-\kappa M_{\rm min})/M_1]^{\alpha}$. For our fiducial prescription we assume that satellite galaxies reside only in a halo that already hosts a central galaxy. 

We have 5 model parameters, $\{M_{\rm min},\sigma_{\log M},\kappa, M_1, \alpha\}$, to characterize the central and satellite HODs for each galaxy sample 
for a given cosmological model.

\section{Convergence test of our nested sampling results}
\label{app:convergence_mcmc}
\begin{figure}
	\includegraphics[width=\columnwidth]{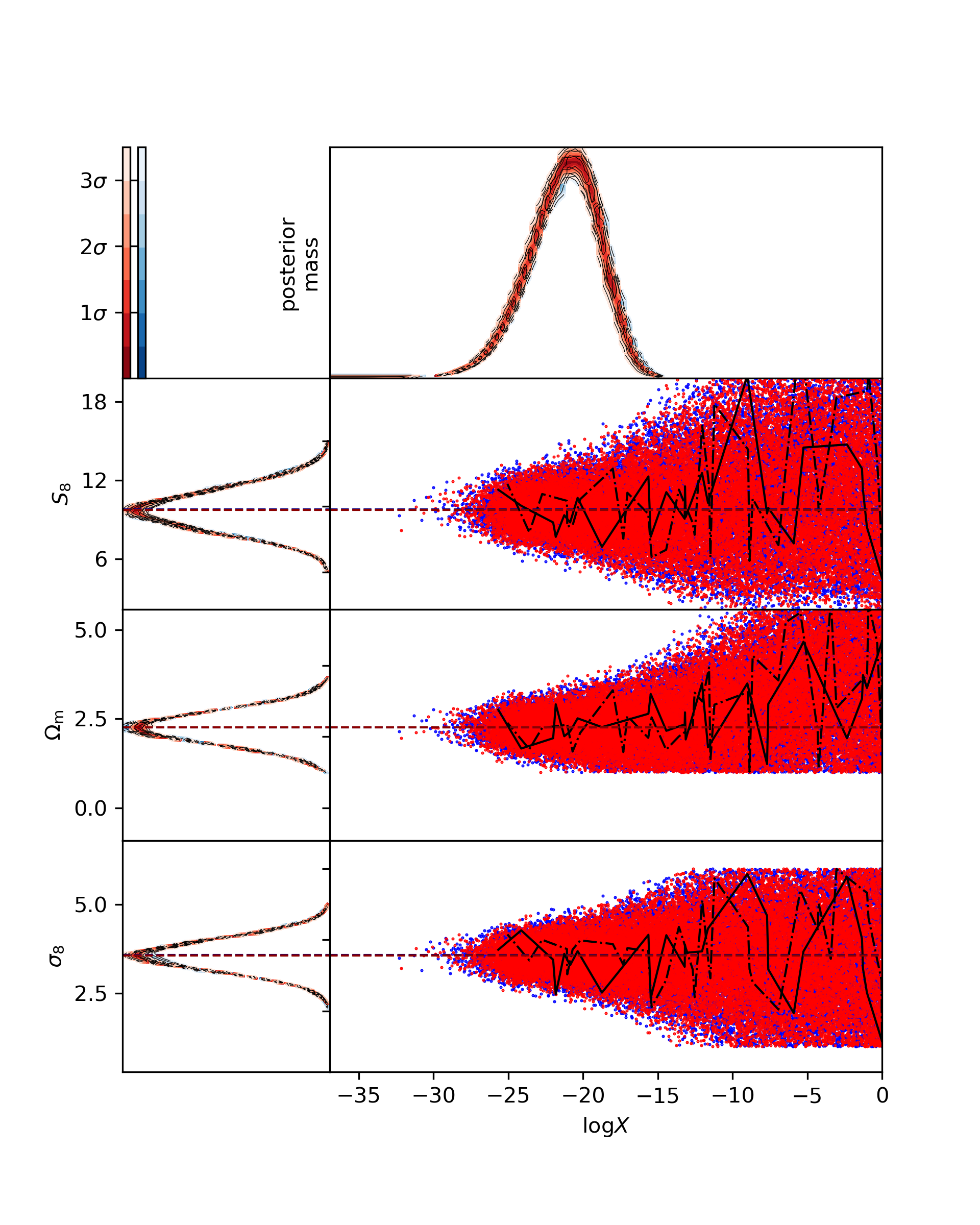}
	\caption{Distributions of $\log X$, the remaining volume of a prior after replacing a live point with the lowest likelihood at each step, for two nested sampling runs. The upper right panel shows the relative posterior mass at each $\log X$ value.}
    \label{fig:nestcheck}
\end{figure}

In this paper, we use the multimodal nested sampling algorithm {\tt MultiNest} \cite{Feroz:2008, Feroz:2009, Feroz:2019} for parameter inference. We test the convergence of our nested sampling results following Appendix~3 in \citet{2019PASJ...71...43H}. Fig.~\ref{fig:nestcheck} shows a diagnostic plot made by the publicly-available software {\tt nestcheck} \cite{higson2018nestcheck, higson2018sampling, higson2019diagnostic}. When our nested sampling runs terminate, the remaining posterior mass is sufficiently small.

The standard deviations of $S_8$, $\sigma_8$, and $\Omega_{\rm m}$ derived from four chains of the baseline analysis setup with different seeds are less than $\sim0.1$\% of the central values, which corresponds to less than $\sim2$\% of the statistical uncertainties, for $S_8$, $\Omega_{\rm m}$, and $\sigma_8$.

\section{Full posterior distributions of the baseline analysis}
\label{app:full_posterior}
\begin{figure*}
\begin{center}
\includegraphics[width=2\columnwidth]{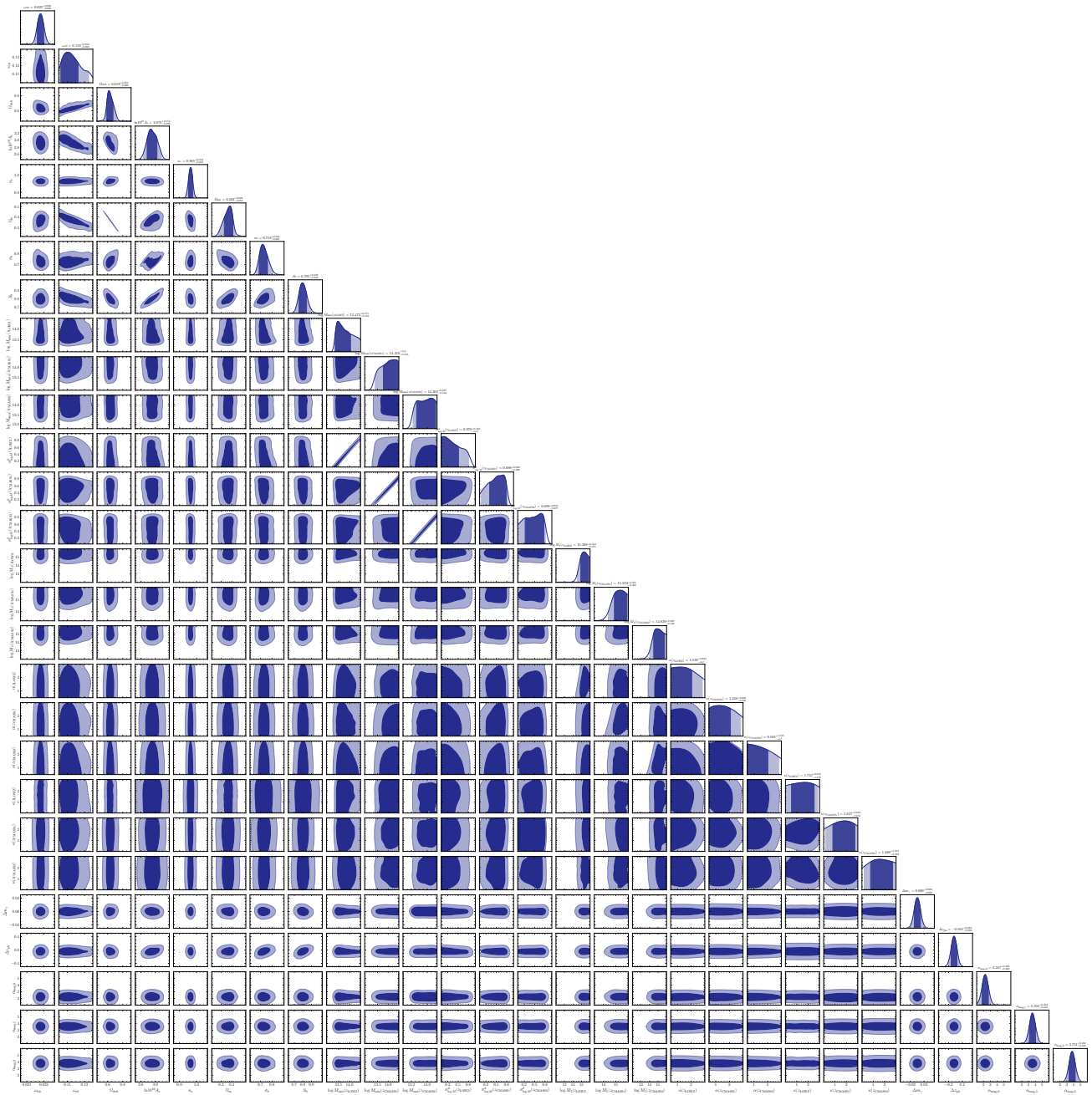}
\end{center}
\caption{The 1-d and 2-d posterior distribution in full parameter space for the baseline analysis.}
\label{fig:full_corner_basline}
\end{figure*}

Fig.~\ref{fig:full_corner_basline} shows the 1-d and 2-d posterior distributions in full parameter space, for the baseline analysis as shown in Fig.~\ref{fig:contour_baseline}. Some of the HOD parameters are not well constrained by the clustering observables, as also found in the validation tests using the mock signals in \citet{2021arXiv210100113M}.
\begin{figure}
\begin{center}
\includegraphics[width=\columnwidth]{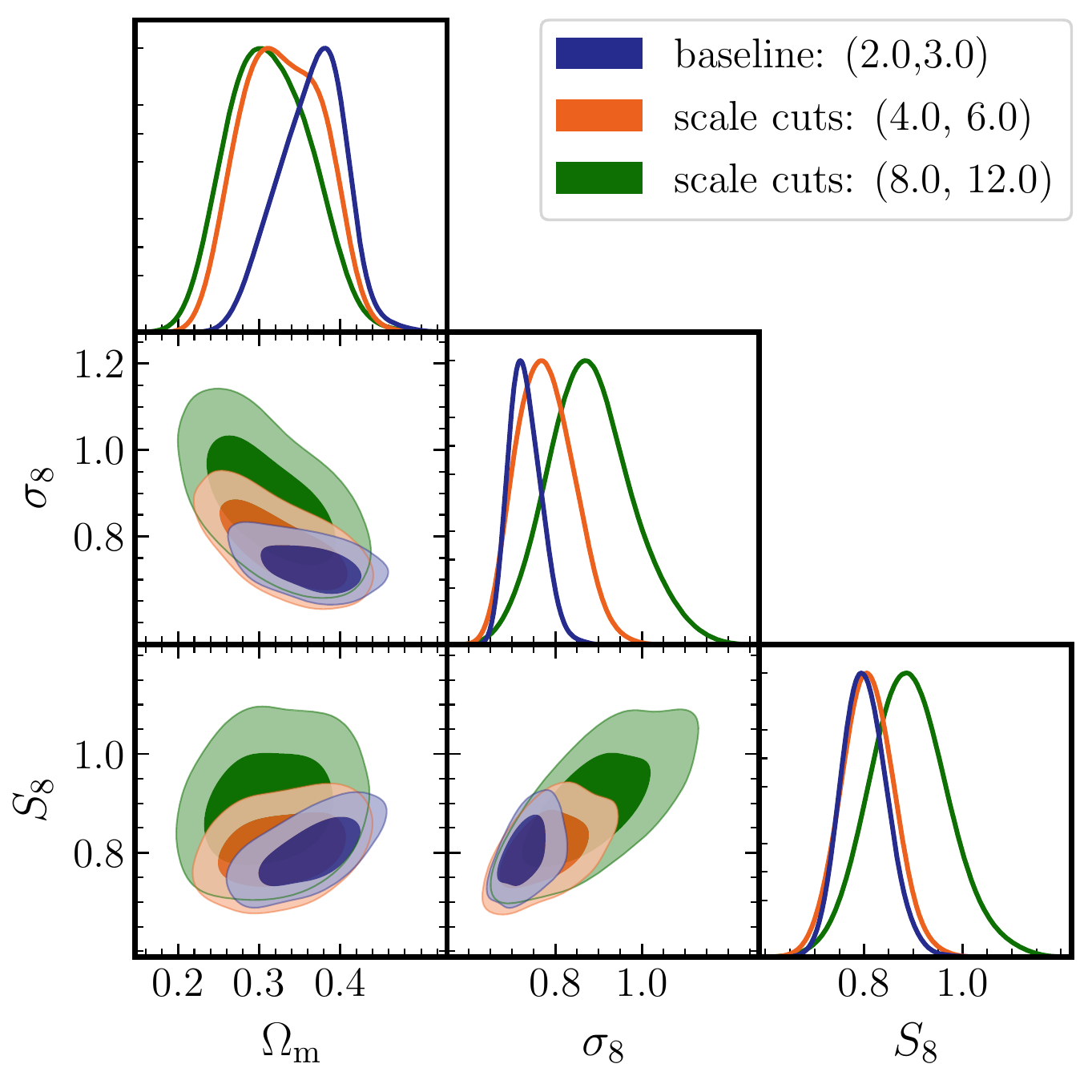}
\end{center}
\caption{Similar to Fig.~\ref{fig:contour_baseline}, but this figure shows the posterior distributions of cosmological parameters obtained by using different scale cuts of $(4,6)~\hiMpc$ and $(8,12)~\hiMpc$ for $\wgg$ and $\dSigma$ instead of our fiducial choice of $(2,3)~\hiMpc$.}
\label{fig:contour_scale_cuts}
\end{figure}
\begin{figure*}
\begin{center}
\includegraphics[width=2\columnwidth]{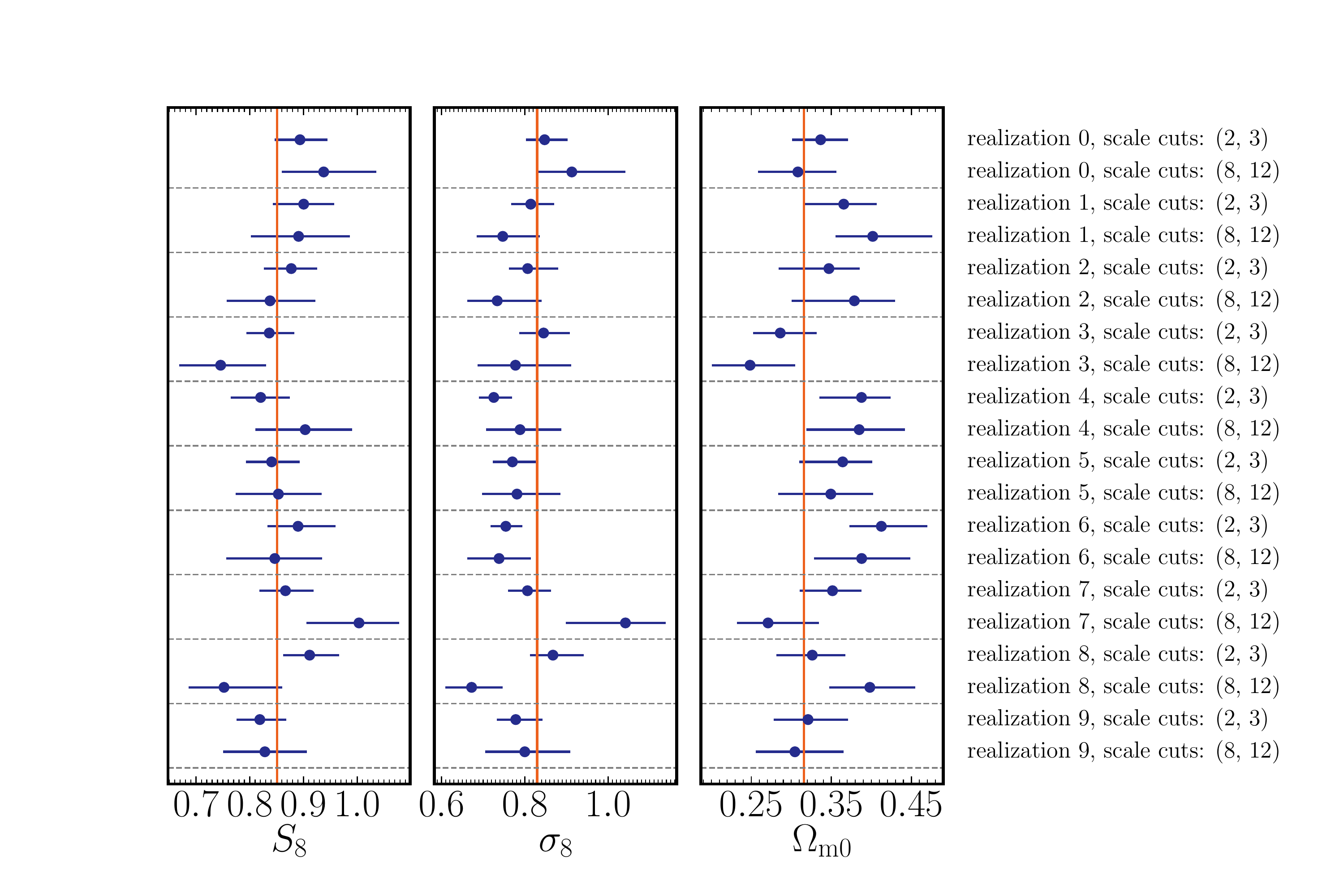}
\end{center}
\caption{The modes and 68\% credible intervals from the cosmology analyses with different scale cuts on 10 realizations of noisy mocks. The orange line denotes input cosmological parameters of mock signals.}
\label{fig:summary_noisy_mocks}
\end{figure*}
\section{The impact of different scale cuts on cosmological parameters}
\label{app:scale_cuts}
We now discuss the results for different scale cuts of  $(4,6)\ \hiMpc$ or $(8,12)\ \hiMpc$ for $\wgg$ and $\dSigma$, instead of our fiducial choice of $(2,3)\ \hiMpc$, as shown in Table~\ref{tab:summary}. The maximum scale cut is kept fixed at 30~$\hiMpc$. The 2-d posterior distribution in each sub-space of $(\Omega_{\rm m},\sigma_8, S_8)$ can be found from Fig.~\ref{fig:contour_scale_cuts}. First of all, the constraining power on $S_8$ is reduced by a factor of 1.2 and 1.9 for the scale cuts of $(4,6)$ and $(8,12)\ \hiMpc$, respectively. A systematic shift in the central value of $S_8$ with the increase of the scale cuts, rather than a random scatter, raises some concern. In the validation paper \citep{2021arXiv210100113M}, we found such a systematic trend in $S_8$ when applying the halo model method to the mock catalogs including a possible assembly bias effect if SDSS galaxies are affected by the effect \citep[see Fig.~17 in][]{2021arXiv210100113M}, where mock SDSS galaxies are preferentially populated into halos with lower concentrations. If the assembly bias effect exists, the large-scale clustering signals cannot be fully characterized by the average mass of halos hosting SDSS galaxies, which is constrained by the small-scale signal of galaxy-galaxy weak lensing, $\dSigma$. On the other hand, the larger scale cuts of $(8,12)\ \hiMpc$ avoid the impact of the assembly bias effect because the cut does not include the small-scale lensing signal. In turn, if assembly bias exists and the halo model is fitted to the signals down to small scales, it would cause a bias in cosmological parameters, which is basically what \citet{2021arXiv210100113M} found. Hence the systematic trend for the different scale cuts might indicate a hint in the assembly bias effect for the SDSS samples \citep[][]{jl2021}. However, note that \citet{2021arXiv210100113M} employed mock catalogs with unexpectedly large assembly bias effects even though there has not been any clear detection of assembly bias from real SDSS data \cite{2016ApJ...819..119L}. On the other hand, we showed that the minimal bias method described in our companion paper \cite{Sugiyama:2021}, is robust against assembly bias effect as long as sufficiently large scale cuts are employed. The scale cuts of $(8,12)\,h^{-1}{\rm Mpc}$ used in this paper (Table~\ref{tab:analysis_setups}) were validated for the perturbation theory-based method using the same mock SDSS catalogs including the assembly bias effect. The result of $S_8$ for the scale cuts of $(8,12)\,h^{-1}{\rm Mpc}$ is, to within the error bar, consistent with that of $S_8$ when using the minimal bias method.

We also study variations in the estimated $S_8$ by applying the halo model method to different realizations of the {\it noisy} mock signals where the statistical errors in each radial bin are added to the {\it noiseless} mock signals using the covariance matrices of $\dSigma$ and $\wgg$. Here we generated the noisy mock signals for all three samples. The generated statistical scatters in each signal include the cross-covariances between the different samples. As shown in Fig.~\ref{fig:summary_noisy_mocks}, we found that 2 among 10 realizations display a similar shift in $S_8$, which is neither a significant nor a negligible fraction. Therefore, we cannot arrive at a definite conclusion: the shift in $S_8$ that we found from the real data might be due to the assembly bias or from the statistical scatters (sample variance). We need more data for a more concrete conclusion. A cosmological analysis of the  redshift-space clustering of SDSS galaxies might also help discriminate the origin of the shift because redshift-space distortions are less affected by the assembly bias effect \citep{Kobayashi:2020a,Kobayashi:2021}.

\begin{figure}
\begin{center}
\includegraphics[width=\columnwidth]{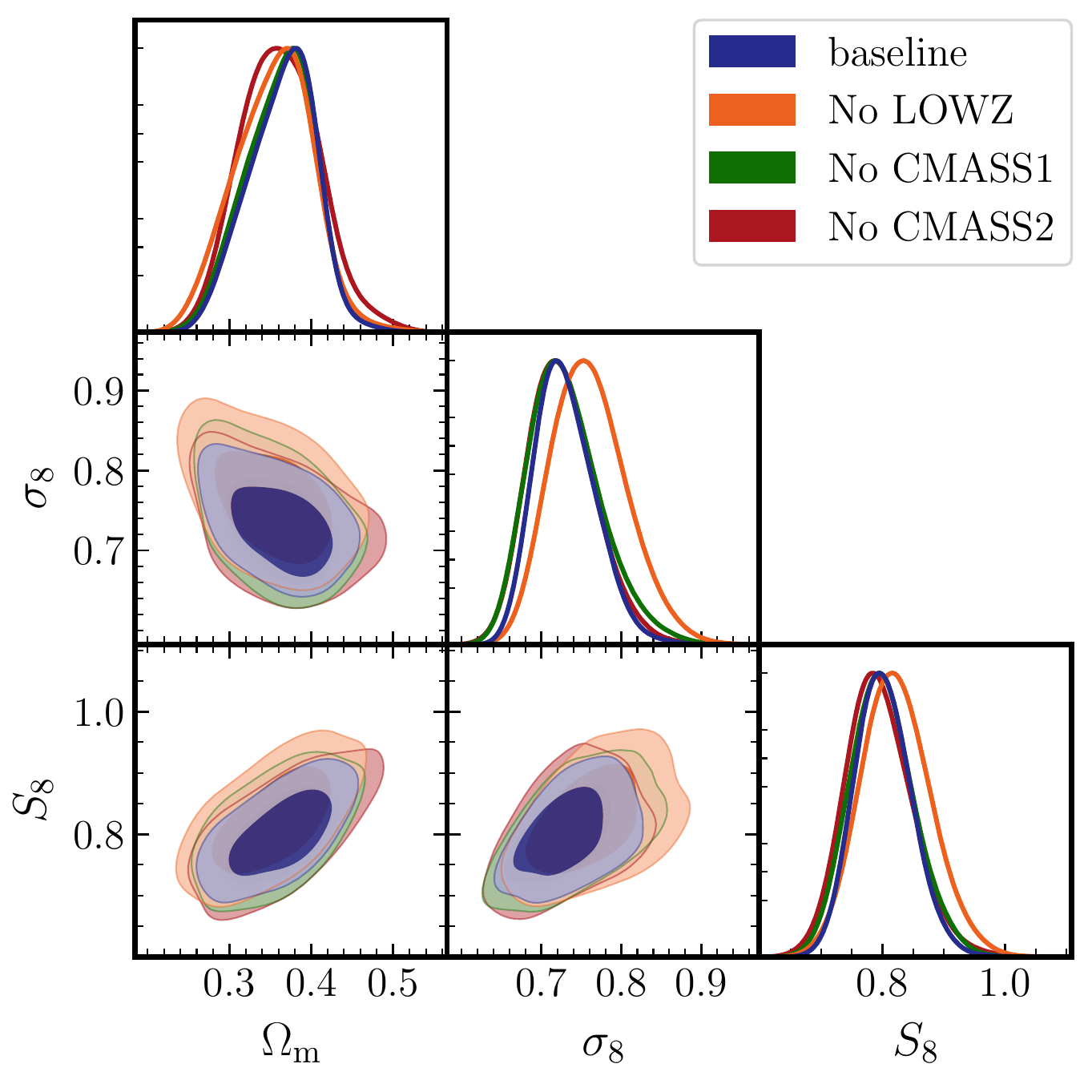}
\end{center}
\caption{Similar to Fig.~\ref{fig:contour_baseline}, but this figure shows the posterior distributions if one of the
LOWZ, CMASS1 or CMASS2 sample is not used in the parameter inference. 
}
\label{fig:contour_minus_lens}
\end{figure}
\begin{figure}
\begin{center}
\includegraphics[width=\columnwidth]{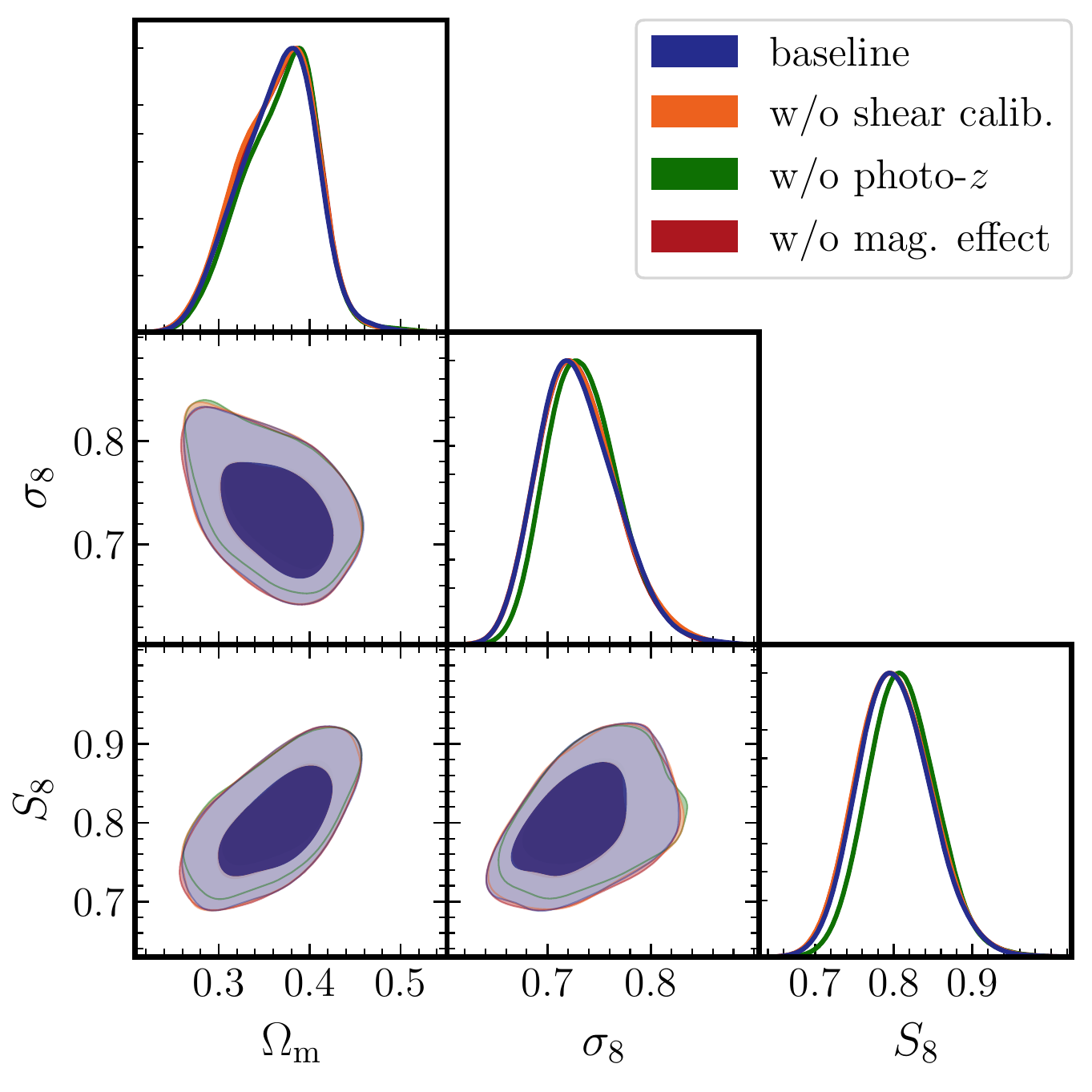}
\end{center}
\caption{Similar to Fig.~\ref{fig:contour_baseline}, but showing the posterior distributions if we fix any of the nuisance parameters ($\Delta z_{\rm ph}$, $\Delta m_\gamma$ or $\alpha_{\rm mag}(z_i)$) to their fiducial value(s) rather than varying them in the parameter inference.}
\label{fig:contour_nuisance_cosmo}
\end{figure}
\begin{figure}
\begin{center}
\includegraphics[width=\columnwidth]{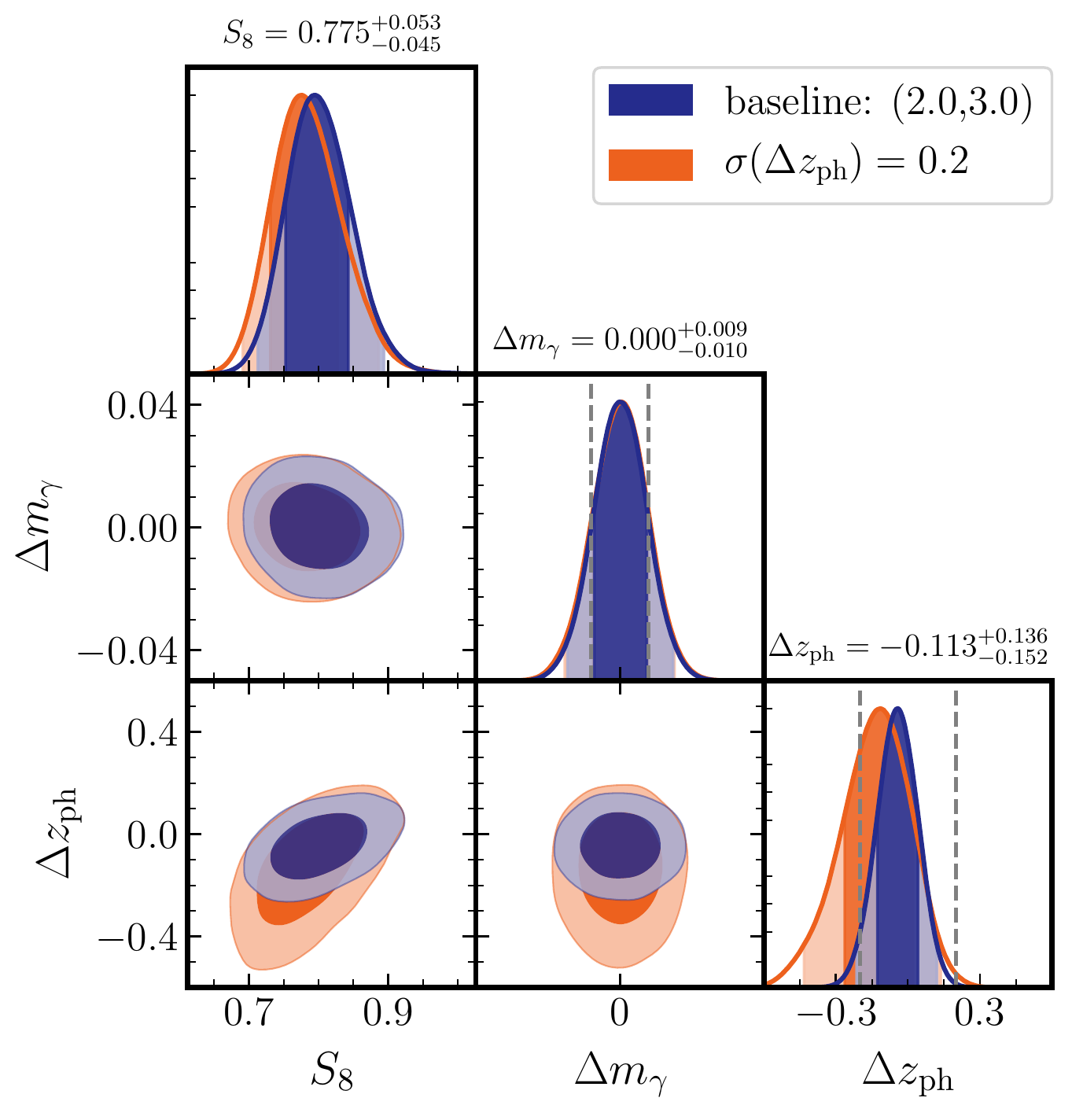}
\end{center}
\caption{Similar to Fig.~\ref{fig:contour_nuisance}, but with the baseline analysis and the $\Delta z_{\rm zp}=0.2$ setup. The numbers on the top diagonal panels are constraints with the $\sigma(\Delta z_{\rm zp})=0.2$ setup. The vertical dashed lines in the 1-d posterior distribution of $\Delta m_\gamma$ or $\Delta z_{\rm ph}$ denote the width of Gaussian prior on the parameter for the $\sigma(\Delta z_{\rm ph})=0.2$ setup.}
\label{fig:contour_nuisance_ldpz}
\end{figure}
\begin{figure}
\begin{center}
\includegraphics[width=\columnwidth]{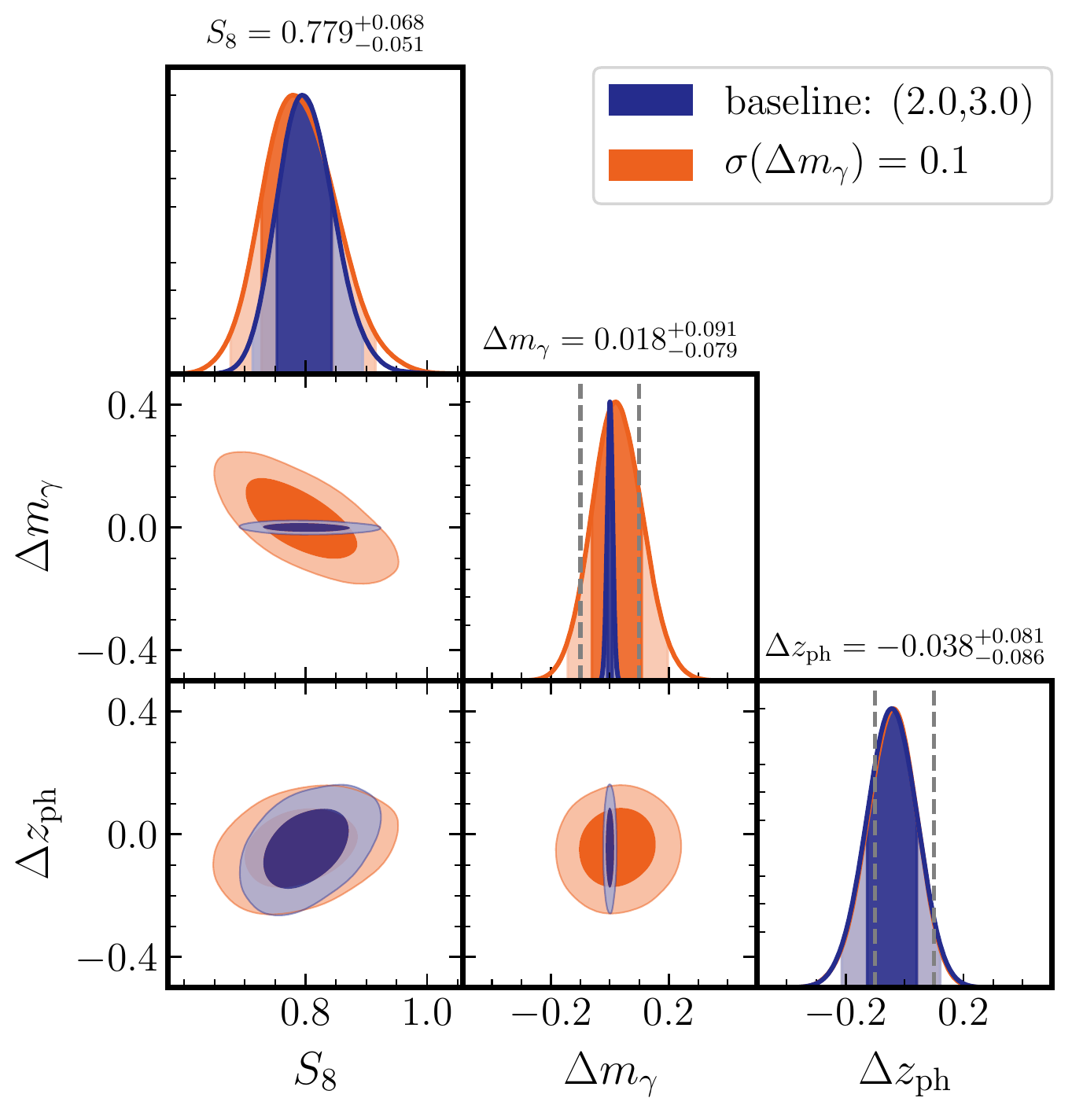}
\end{center}
\caption{Similar to Fig.~\ref{fig:contour_nuisance}, but with the baseline analysis and the $\Delta m_{\gamma}=0.1$ setup. The numbers on the top of the diagonal panels are constraints with the $\sigma(\Delta m_\gamma)=0.1$ setup. The vertical dashed lines in the 1-d posterior distribution of $\Delta m_\gamma$ or $\Delta z_{\rm ph}$ denote the width of Gaussian prior on the parameter for the $\sigma(\Delta m_\gamma)=0.1$ setup.}
\label{fig:contour_nuisance_ldm}
\end{figure}
\begin{figure}
\begin{center}
\includegraphics[width=\columnwidth]{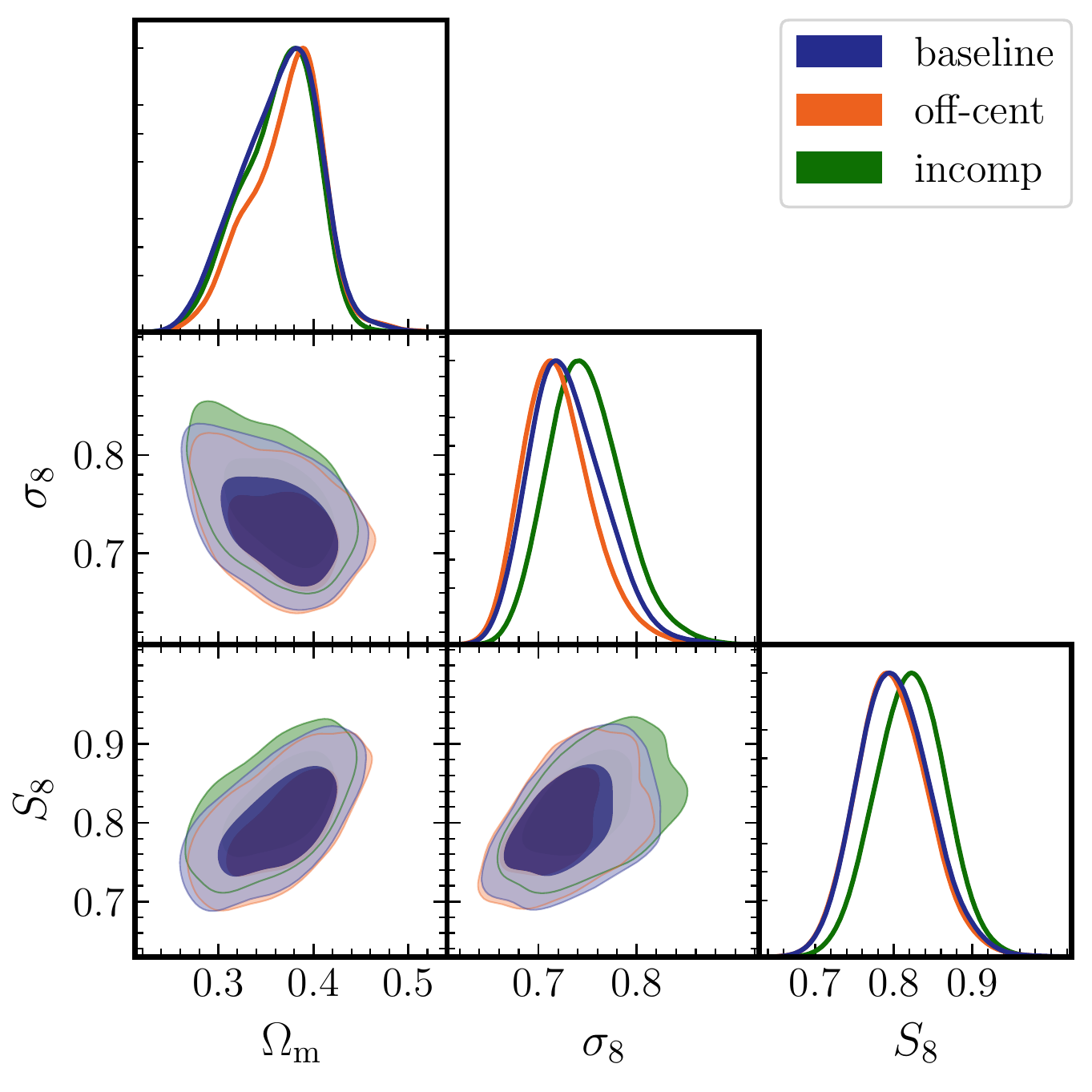}
\end{center}
\caption{Similar to Fig.~\ref{fig:contour_baseline}, but the posterior distributions obtained by using the extended halo model where the off-centering effect or the incompleteness selection for central galaxies are further included (also see Table~\ref{tab:analysis_setups}). For each case, the two additional model parameters for each of the LOWZ, CMASS1 and CMASS2 samples are included. For comparison we also show the results for the baseline setup, which are the same in Fig.~\ref{fig:contour_baseline}.
}
\label{fig:contour_model_variants}
\end{figure}
\begin{figure}
\begin{center}
\includegraphics[width=\columnwidth]{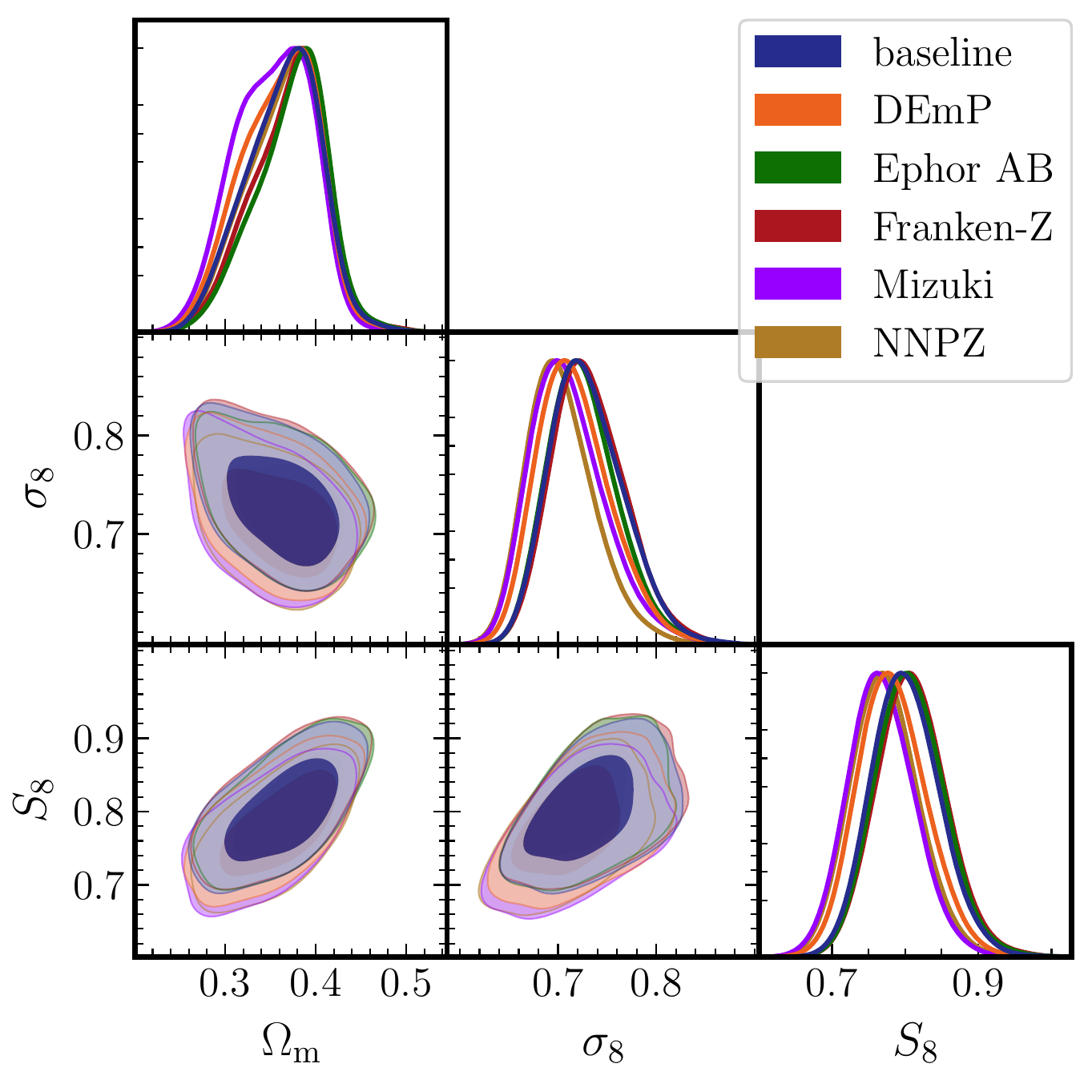}
\end{center}
\caption{Similar to Fig.~\ref{fig:contour_baseline}, but this figure shows the posterior distributions if the different photo-$z$ catalogs, as denoted by legend, are used to define the source galaxy sample for the $\dSigma$ measurement. 
}
\label{fig:contour_different_photoz}
\end{figure}
\begin{figure}
\begin{center}
\includegraphics[width=\columnwidth]{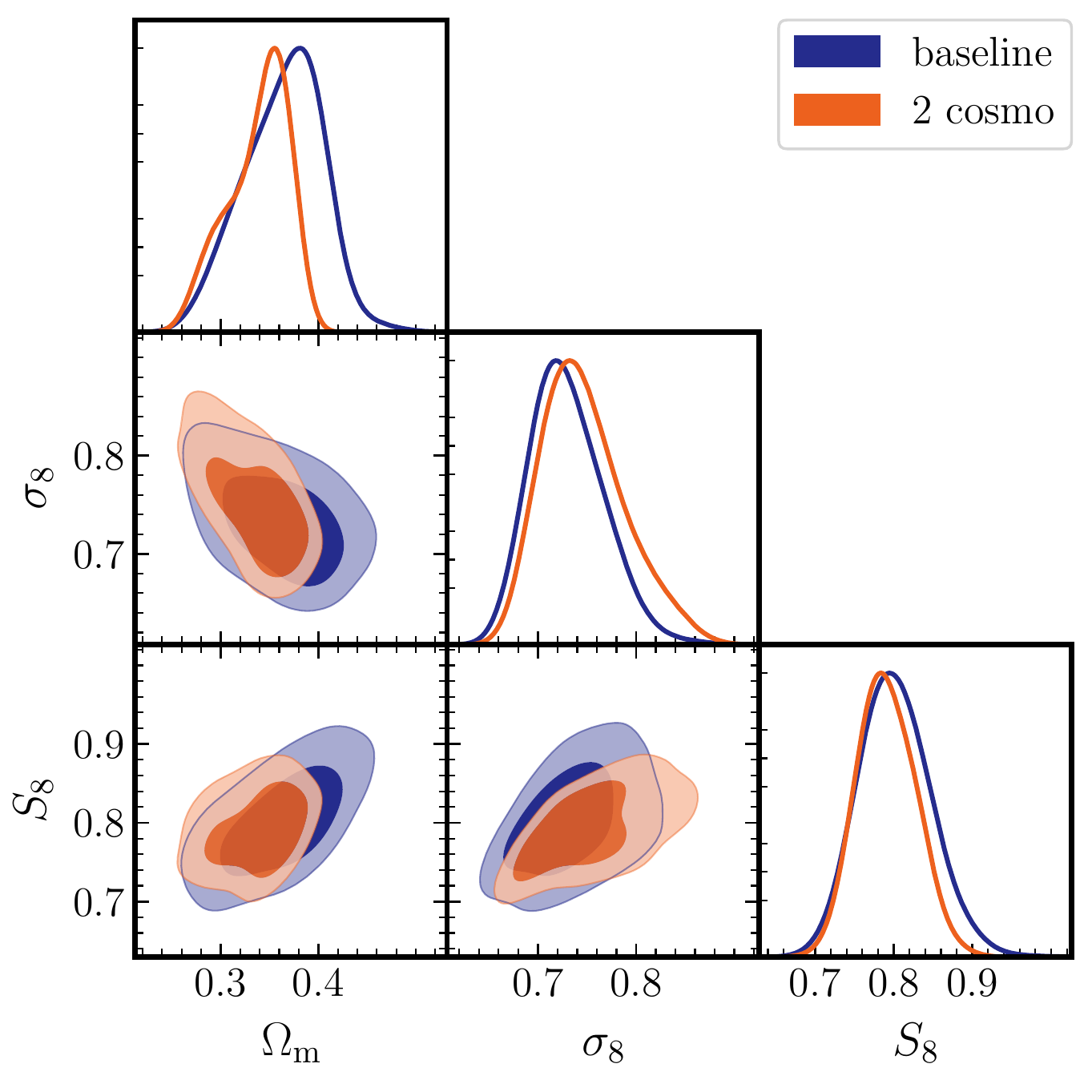}
\end{center}
\caption{Similar to Fig.~\ref{fig:contour_baseline}, but this figure show the posterior distributions when only $\ln(10^{10} A_{\rm s})$ and $\Omega_{\rm de}$ among the five cosmological parameters are varied. }
\label{fig:contour_5vs2cosmo_lens}
\end{figure}
\section{Systematic tests with different analysis setups}
\label{sec:test_analysis_setup}

In this appendix, we demonstrate the robustness of the results with regards to different analysis setups, as listed in Table~\ref{tab:analysis_setups}. These results are also summarized in Table~\ref{tab:summary} and Fig.~\ref{fig:summary}. In this appendix we show the results for the 1- and 2-d posteriors of the parameters for the different setups.

Fig.~\ref{fig:contour_minus_lens} shows the posteriors if we remove one of the lens samples in the parameter inference (also see Fig.~\ref{fig:summary} and Table~\ref{tab:summary}). All the results for $S_8$ are consistent with the baseline result to within the 68\% credible interval, but the result removing the LOWZ sample displays a sizable shift in the cosmological parameters. This might be due to the statistical scatters of the SDSS galaxies because our constraints are mainly from $\wgg$ and the $\wgg$ information for the different samples are considered independent. In fact, such a shift in the cosmological parameters for the different samples were also seen from the cosmological analysis of redshift-space galaxy clustering \citep[e.g. see Fig.~2 in][]{2020JCAP...05..042I,Kobayashi:2021}. Since the shift in the parameters are not large compared to the current statistical errors, we leave this to our future work using a larger dataset from the ongoing HSC survey. 

One notable feature of this paper is a conservative treatment of the nuisance parameters, i.e. the photo-$z$ biases and the multiplicative shear biases, which are among the most important systematic errors in weak lensing. Throughout this paper, as a conservative approach, we employed rather broad priors for these nuisance parameters, $\Delta z_{\rm ph}$ and $\Delta m_\gamma$, and derive cosmological constraints after marginalization over the nuisance parameters. On the other hand, most previous weak lensing based studies employ tight priors on these parameters, typically a few per cent in the amplitudes of $\Delta z_{\rm ph}$ and $\Delta m_\gamma$. Here we study the impact of the nuisance parameters on our results. As one extreme case, Fig.~\ref{fig:contour_nuisance_cosmo} shows how the cosmological constraints are changed if either of $\Delta z_{\rm ph}$ or $\Delta m_\gamma$ is fixed to the central value (i.e. $\Delta z_{\rm ph}=0$ or $\Delta m_\gamma=0$ as implied by the fiducial photo-$z$ code or the shear calibration). The posterior distributions remain almost unchanged, meaning that our constraints are robust against these nuisance parameters to within the prior width. For completeness, the figure also shows how the results are changed if we ignore the magnification bias in the template of $\dSigma$. Again it is clear that the magnification bias does not have a large impact on the results. 

As another extreme case, we study how the broader prior widths of $\Delta z_{\rm ph}$ and $\Delta m_\gamma$ change the results. In doing this, we employ the prior width of $\sigma(\Delta z_{\rm ph})=0.2$ or $\sigma(\Delta m_\gamma)=0.1$, compared to our fiducial choices of $\sigma(\Delta z_{\rm ph})=0.1$ or $\sigma(\Delta m_\gamma)=0.01$. These widened priors are quite conservative, and the results would be basically equivalent to the case using these nuisance parameters as free parameters. Figs.~\ref{fig:contour_nuisance_ldm} and \ref{fig:contour_nuisance_ldpz} show the results. Encouragingly these conservative choices only moderately enlarge the size of the credible intervals for $S_8$, and at the same time constrain each of the nuisance parameters by the credible interval smaller than the prior width. That is, these joint-probe cosmology method enables us to perform, to some extent, a self-calibration of these nuisance parameters. It is intriguing to find that both results prefer a slightly smaller value of $S_8$, meaning that the HSC-Y1 and SDSS data prefer such a value or the central values inferred from the photo-$z$ calibration or the shear calibration might involve unknown systematic errors under the assumption of a flat $\Lambda$CDM model. This is definitely an interesting direction to further explore with upcoming large HSC datasets.

In Fig.~\ref{fig:contour_model_variants} we show the results for the extended halo model, where the off-centering effect or the incompleteness effect for central galaxies is included. To model the effect we need to introduce two additional model parameters for each of the effects as can be found from Table~\ref{tab:analysis_setups}. The figure clearly shows that the cosmological constraints are robust against these variants in the theoretical templates. 

Fig.~\ref{fig:contour_different_photoz} shows the posterior distributions inferred from data vectors computed with different photo-$z$ methods. We do not find any significant shift in the parameter constraints.

Fig.~\ref{fig:contour_5vs2cosmo_lens} compares the posterior distributions between the baseline analysis and the analysis with only two cosmological parameters $(\Omega_{\rm m}, \ln{10^{10}A_{\rm s}})$ varied while other cosmological parameters fixed to the {\it Planck} 2015 ``TT,TE,EE+lowP'' constraints \cite{PlanckCosmology:16}. We do not find any significant shift in $S_8$, while the statistical uncertainty shrinks by $\sim20$\%.

In summary, we investigate how the cosmological constraints are changed for different model templates or different combinations of data vector, which are shown in Table~\ref{tab:summary}, Fig.~\ref{fig:summary} and Figs.~\ref{fig:contour_minus_lens}--\ref{fig:contour_5vs2cosmo_lens}. All results for $S_8$ are consistent with the baseline result to within the 68\% credible interval. Hence we conclude that none of the tests indicates an unknown systematic error compared to the current statistical error, and our results, especially the result for $S_8$, are robust against possible residual systematic effects.


\end{document}